\documentclass[10pt]{article}
\pdfoutput=1

\usepackage{amsmath,amsfonts,amssymb}
\usepackage{enumerate}
\usepackage{epsfig}
\usepackage{textcomp}
\usepackage{wasysym}
\usepackage{cite}
\usepackage{url}

\usepackage{subfigure}
\usepackage{amsthm}

\usepackage{chngpage} 

\usepackage[T1]{fontenc}

\usepackage{etoolbox} 

\usepackage{stmaryrd} 

\newcommand{\layerindex}[1][]{ %
\ifstrequal{#1}{}{\ensuremath{\alpha} }{}
\ifstrequal{#1}{1}{\ensuremath{\alpha} }{}
\ifstrequal{#1}{2}{\ensuremath{\beta} }{}
\ifstrequal{#1}{3}{\ensuremath{\gamma} }{}
}
\newcommand{\layerindexvector}[1][]{ %
\ifstrequal{#1}{}{\ensuremath{\boldsymbol\alpha} }{}
\ifstrequal{#1}{1}{\ensuremath{\boldsymbol\alpha} }{}
\ifstrequal{#1}{2}{\ensuremath{\boldsymbol\beta} }{}
\ifstrequal{#1}{3}{\ensuremath{\boldsymbol\gamma} }{}
}

\newcommand{\nodeindex}[1][]{ %
\ifstrequal{#1}{}{\ensuremath{u} }{}
\ifstrequal{#1}{1}{\ensuremath{u} }{}
\ifstrequal{#1}{2}{\ensuremath{v} }{}
}

\newcommand{\aspects}{d}
\newcommand{\aspectindex}{a}
\newcommand{\nnodes}{n}
\newcommand{\nlayers}{b}
\newcommand{\layers}{L}
\newcommand{\layervector}{ {\bf L} }

\newcommand{\ctreduced}{\hat{C}}

\newcommand{\aff}[1]{\textit{  #1 }}

\usepackage[font=footnotesize]{caption}

\newcommand{\comment}[2]{} 

\begin{document}

\title{\bf Multilayer Networks}

\author{Mikko Kivel\"a \\ \aff{Oxford Centre for Industrial and Applied Mathematics, Mathematical Institute,}\\ \aff{University of Oxford, Oxford OX2 6GG, UK}
\and
Alexandre Arenas\\ \aff{Departament d'Enginyeria Inform\'atica i Matem\'atiques,}\\ \aff{Universitat Rovira I Virgili, 43007 Tarragona, Spain }
\and
Marc Barthelemy\\
\aff{Institut de Physique Th\'eorique, CEA, CNRS-URA 2306, F-91191,}\\ \aff{Gif-sur-Yvette, France}\\ and \\
\aff{Centre d'Analyse et de Math\'ematiques Sociales, EHESS, 190-198 avenue de France,}\\ \aff{75244 Paris, France }
\and
James P. Gleeson\\
\aff{MACSI, Department of Mathematics \& Statistics, University of Limerick, Ireland}
\and
Yamir Moreno\\
\aff{Institute for Biocomputation and Physics of Complex Systems (BIFI),}\\ \aff{University of Zaragoza, Zaragoza 50018, Spain}\\ and \\
\aff{Department of Theoretical Physics, University of Zaragoza,}\\ \aff{Zaragoza 50009, Spain}\\
\and
Mason A. Porter\thanks{Corresponding author: porterm@maths.ox.ac.uk}\\
\aff{Oxford Centre for Industrial and Applied Mathematics,} \\ \aff{Mathematical Institute, University of Oxford, Oxford OX2 6GG, UK}\\ and \\ \aff{CABDyN Complexity Centre, University of Oxford, Oxford OX1 1HP, UK}}

\maketitle

\section*{Abstract}

In most natural and engineered systems, a set of entities interact with each other in complicated patterns that can encompass multiple types of relationships, change in time, and include other types of complications.  Such systems include multiple subsystems and layers of connectivity, and it is important to take such ``multilayer'' features into account to try to improve our understanding of complex systems. Consequently, it is necessary to generalize ``traditional'' network theory by developing (and validating) a framework and associated tools to study multilayer systems in a comprehensive fashion.  The origins of such efforts date back several decades and arose in multiple disciplines, and now the study of multilayer networks has become one of the most important directions in network science. In this paper, we discuss the history of multilayer networks (and related concepts) and review the exploding body of work on such networks.  To unify the disparate terminology in the large body of recent work, we discuss a general framework for multilayer networks, construct a dictionary of terminology to relate the numerous existing concepts to each other, and provide a thorough discussion that compares, contrasts, and translates between related notions such as multilayer networks, multiplex networks, interdependent networks, networks of networks, and many others. We also survey and discuss existing data sets that can be represented as multilayer networks. We review attempts to generalize single-layer-network diagnostics to multilayer networks.  We also discuss the rapidly expanding research on multilayer-network models and notions like community structure, connected components, tensor decompositions, and various types of dynamical processes on multilayer networks.  We conclude with a summary and an outlook.

\section{Introduction} \label{sec:intro}

Network theory is an important tool for describing and analyzing complex systems throughout the social, biological, physical, information, and engineering sciences \cite{Newmanbook,Wasserman1994Social,bocca2006}.  Originally, almost all studies of networks employed an abstraction in which systems are represented as ordinary graphs \cite{bollobasbook}: the ``nodes'' (or ``vertices'') of the graphs represent some entity or agent, and a tie between a pair of nodes is represented using a single, static, unweighted ``edge'' (or ``link'').  Self-edges and multi-edges were also typically ignored.  Although this approach is naive in many respects, it has been extremely successful.  For example, it has been used to illustrate that many real networks possess a heavy-tailed degree distribution~\cite{Barabasi1999Emergence,Clauset2009Powerlaw}, exhibit the small-world property~\cite{Watts1998Collective,smallworld-scholarpedia}, contain nodes that play central roles \cite{Wasserman1994Social,Newmanbook}, and/or have modular structures~\cite{Porter2009,Fortunato2010Community,Lancichinetti2010Characterizing}.

As research on complex systems has matured, it has become increasingly essential to move beyond simple graphs and investigate more complicated but more realistic frameworks.  For example, edges often exhibit heterogeneous features: they can be directed~\cite{digraphbook,Newmanbook,Wasserman1994Social}, have different strengths (i.e., ``weights'')~\cite{Wasserman1994Social,Barrat2004Architecture,Newman2004Analysis}, exist only between nodes that belong to different sets (e.g., bipartite networks) \cite{Newmanbook,Wasserman1994Social,breiger1974}, or be active only at certain times~\cite{Holme2012Temporal,Holme2013Temporal}.
Most recently, there have been increasingly intense efforts to investigate networks with
multiple types of connections (see Section~\ref{multiplex}) and so-called ``networks of networks''~\cite{DAgostino2014Networks} (see Section~\ref{nodecolored}).  Such systems were examined decades ago in disciplines like sociology and engineering, but the explosive attempt to develop frameworks to study multilayer complex systems and to generalize a large body of familiar tools from network science is a recent phenomenon.

In social networks, one can categorize edges based on the nature of the relationships (i.e., ties) or actions that they represent~\cite{Wasserman1994Social,Krackhardt1987Cognitive,Padgett1993Robust}. Reducing a social system to a network in which actors are connected in a pairwise fashion by only a single type of relationship is often an extremely crude approximation of reality.  As a result, sociologists recognized decades ago that it is crucial to study social systems by constructing multiple social networks using different types of ties among the same set of individuals~\cite{Wasserman1994Social,Scott2012Social}.\footnote{Research in anthropology has also stated the need to consider multiple layers of social connectivity \cite{wolfe1961,wolfe1963}.} For example, consider the sociograms\footnote{A ``sociogram'' is one name for the usual type of network visualization that contains a collection of dots with pairwise connections drawn as lines between them.} that were drawn in the 1930s to represent social networks in a bank-wiring room \cite{Roethlisberger1939Management}. These sociograms depicted relations between 14 individuals via 6 different types of social interactions (see Fig.~\ref{fig:visuals} in Section \ref{data}).  
In the sociology literature, networks in which each edge is categorized by its type are 
called ``multiplex networks''~\cite{Gluckman55,Verbrugge1979Multiplexity} or ``multirelational networks''~\cite{Wasserman1994Social}.  (Such networks are also said to possess ``multi-stranded'' relationships \cite{mitchell1969}.) Social networks also often include several types of nodes (e.g., males and females) or hierarchical structures (e.g., individuals are part of organizations), which have been studied using ``multilevel networks'' (see Section~\ref{othernets}). 
The tools that have been developed to investigate multilayer social networks include exponential random graph models (ERGMs)~\cite{Pattison1999Logit,robins2013}, meta-networks and meta-matrices \cite{Krackhardt1998PCANS,CarleyStructural2001}, and methods for identifying social roles using blockmodeling and relational algebras~\cite{Lorrain1971Structural,boormanwhite1976,Breiger1986Cumulated,doreian,White1976Social,Winship1984Roles,Pattison2009Social}.

In the computer-science and computational linear-algebra communities, tensor-decomposition methods~\cite{Dunlavy2006Multilinear,Kolda2009Tensor} and multiway data analysis \cite{Acar2009Unsupervised} have been used to study various types of multilayer networks (see Sections \ref{centrality} and \ref{tensorcommunities}). These types of methods are based on representing multilayer networks as adjacency tensors of ``rank'' higher than 2 (i.e., of ``order'' higher than 2) and then applying machinery that has been developed for tensor decompositions. Perhaps the most widespread methods that use this approach are generalizations of the singular value decomposition (SVD) \cite{carla2012}, and these and other tools have been extremely successful in many applications \cite{Kolda2009Tensor}. For example, tensor-decomposition and multiway-data-analysis methods can be used to extract communities (i.e., sets of nodes that are connected densely to each other)~\cite{Dunlavy2006Multilinear} or to rank nodes~\cite{Kolda2006TOPHITS,Kolda2005HigherOrder} in multilayer networks. A clear benefit of a tensor representation is that one can directly apply methods from the tensor-analysis literature to multilayer networks --- e.g., by using dynamic tensor analysis~\cite{Sun2006Beyond} to study multiplex networks that change in time.

Networked systems that cannot be represented as traditional graphs have also been studied from a data-mining perspective. For example, heterogeneous (information) networks were developed as a general framework to take into account multiple types of nodes and edges~\cite{Zhou2007CoRanking,Cai2005Community,Sun2013Mining}. Similarly, one can use meta-matrices to conduct a dynamic network analysis~\cite{Carley2003Dynamic} that incorporates temporal and spatial information, node attributes and types, and other types of data about social networks in the same framework. Meta-matrices have been employed in the context of ``organizational theory'', as organizations, people, resources, and other types of entities are all interconnected~\cite{Krackhardt1998PCANS,CarleyStructural2001}.

Interconnected systems have been examined in the engineering literature as a source of cascading failures~\cite{Chang1996Estimation,Little2002Controlling,Rosato2008Modelling}.  Analogous to the notion of ``systemic risk'' in financial systems, increasing connectivity --- including the interconnectedness of different systems in an infrastructure --- has the potential to increase large-scale events.  In the last several years, these ideas have been formalized using interacting networks (and interdependent networks) \cite{Leicht2009Percolation,Buldyrev2010Catastrophic}. For example, it has been shown (especially using percolation processes) that interconnected systems can react to random failures in a manner that is different from ``monoplex'' (i.e., single-layer) networks. For some types of cascading-failure processes, an interdependent system can exhibit a ``first-order'' (i.e., discontinuous) phase transition instead of the ``second-order'' (i.e., continuous) phase transitions that are typical for monoplex systems \cite{Buldyrev2010Catastrophic,vespignani2010-comment}.  We will discuss this in more detail later.

In the last couple of years, it has suddenly become very fashionable to study networks with multiple layers (or multiple types of edges) and networks of networks.\footnote{As we discussed above, the sociology and engineering communities have been studying various types of multilayer networks for a long time.  Much more recently, the physics community also produced pioneering work on notions like multilayer networks~\cite{Jo2006Immunization,Kurant2006Layered,Kurant2007Error}, networks of networks \cite{kurths2006,kurths2007}, and node-colored networks~\cite{Newman2003Mixing,Vazquez2006Spreading} several years ago (i.e., before it became as popular as it is now).}  Unfortunately, the sudden and immense explosion of papers on multilayer networks has produced an equally immense explosion of disparate terminology, and the lack of a consensus (or even generally accepted) set of terminology and mathematical framework for studying multilayer networks is extremely problematic.\footnote{For example, there are numerous uses of different terminology for the same mathematical object (or, even more confusingly, for similar objects that are almost but not quite the same) as well as uses of the same words to describe different mathematical objects.}
 Additionally, research on generalizing monoplex-network concepts such as degree, transitivity, centrality, and diffusion is only in its infancy.  We also expect that it will be necessary to define many concepts that are intrinsic to multilayer networks.

In this paper, we present a general definition of multilayer networks that can be used to represent most types of complex systems that consist of multiple networks or include disparate and/or multiple interactions between entities.  We review the existing literature and find a natural mapping from each type of network to this multilayer network representation, and we classify the numerous existing notions of multilayer networks based on the types of constraints that they impose on this multilayer network representation (see Table~\ref{table:dictionary}). The framework that we advocate is able to handle networks with multiple modes of multiplexity (e.g., networks that are both multiplex and temporal~\cite{Barigozzi2011Identifying,DeDomenico2013Mathematical}), although it is important to note that the vast majority of the scholarly literature has been concerned thus far with networks that possess only a single ``dimension'' (i.e., what we call an ``aspect'') of layers.  As discussed in Ref.~\cite{DeDomenico2013Mathematical}, a multilayer network with a single aspect \emph{already} yields the inherent ``new physics'' of multiplexity, though writing down a framework that explicitly enumerates more aspects is convenient for expository purposes (in particular, to connect with terminology that has already been introduced into the literature) and to draw additional connections with applications. In this review, we concentrate primarily on multiplex networks that are represented as edge-colored multigraphs or sequences of networks, but we also give some attention to structures such as interdependent networks and networks of networks. One can also view a temporal network as a type of multilayer network, but we almost always leave them out of our discussion because they are reviewed elsewhere~\cite{Holme2012Temporal,Holme2013Temporal}. 
We also briefly discuss connections between multilayer networks and network structures such as hypergraphs, multipartite networks, and networks that are both node-colored and edge-colored.

The remainder of this paper is organized as follows.  In Section \ref{newform}, we present a general formulation for multilayer networks and map all of the reviewed types of multilayer networks to this general framework. We are thereby able to translate between the many existing notions for studying multilayer networks.  In Section \ref{data}, we review existing data sets that have been examined using multilayer-network frameworks and discuss other types of data that would be useful to study from such a perspective. In Section \ref{models}, we examine models of multilayer networks, methods and diagnostics that have been introduced (or generalized from single-layer networks) to analyze and measure the properties of multilayer networks, and dynamical processes on multilayer networks.  We conclude in Section \ref{conc}. We also include a glossary of important terms in Section~\ref{glossary}.

\section{Multilayer Networks}\label{newform}

\begin{figure}[!htp]
\begin{center}
\includegraphics[width=1.0\linewidth]{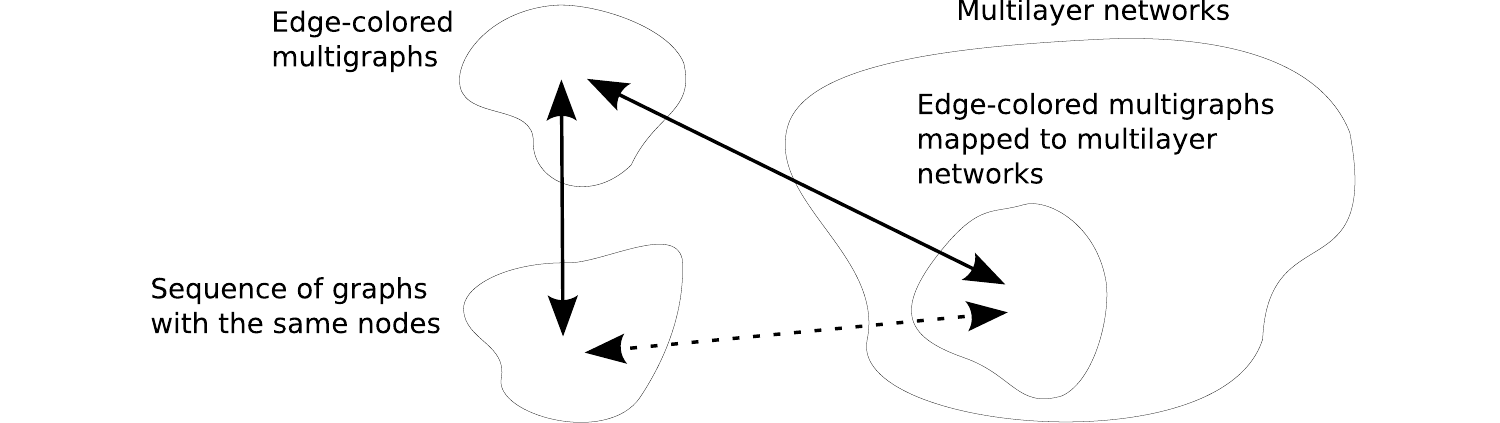}
\end{center}
\caption{Illustration of the idea of characterizing different types of network structures by finding ``natural'' injective maps from one set of network structures to another.   We do not define the notion of ``natural'' in a rigorous way, but we say that two network structures are related when there is a natural bijective map between them.  In such a scenario, one can think of the two structures as two different ways of representing the same object. For example, as we illustrate in the figure, there is a natural bijective mapping between edge-colored multigraphs and sequences of graphs that contain the same nodes, so one can construe these structures as equivalent (i.e., different representations of the same mathematical object). Further, the subset of multilayer networks that we label as ``Edge-colored multigraphs mapped to multilayer networks'' is the image of the bijective map from the set of edge-colored multigraphs to the set of multilayer networks. Each of the two solid lines indicates an explicit mapping, and the dashed line indicates a mapping that is implied by transitivity. It might be possible to find both a natural injective mapping (which is not surjective) from some structure $A$ to some other structure $B$ and another such mapping from $B$ to $A$.   (For any given structure, there is no upper bound to the number of layers, aspects, or other features that we can consider.) In other words, to argue that one network structure is a ``special case'' of another, one would need to find a natural, injective, non-surjective map in one direction and prove that no such map exists in the other direction.  It is not our aim to make such arguments; instead, we indicate mappings between different structures to illustrate different ways of representing the same object.
}
\label{fig:map_illustration}
\end{figure}

We start by presenting the most general notion of a multilayer network structure that we will use in this article and by defining various constraints for that structure. We then show how this structure can be represented as an adjacency tensor~\cite{DeDomenico2013Mathematical} and how one can reduce the rank (i.e., order) of such a tensor by constraining the space of possible multilayer networks or by ``flattening'' the tensor.  Taken to its extreme, such a flattening process yields ``supra-adjacency matrices'' (i.e., ``super-adjacency matrices'')~\cite{Gomez2013Diffusion,DeDomenico2013Random,Cozzo2013Clustering}, which have the advantage over tensors of being able to represent missing nodes in a convenient way. (When implementing methods for computation, most people are also much more familiar with working with matrices than with tensors.)
We then discuss numerous multilayer-network structures --- multiplex networks, networks of networks, etc. --- that have been formulated in the literature, and we show how they can be represented using our formulation of multilayer networks. Most of the ways of representing these network structures as general multilayer networks only cover a subset of all of the possible multilayer networks (see Fig.~\ref{fig:map_illustration}), and these subsets can be characterized by the constraints that they satisfy (i.e., by the properties of these constructions). In Table \ref{table:dictionary}, we summarize the properties of various multilayer-network structures from the literature have when they are represented using the general multilayer network structure. Finally, we discuss the relationship between multilayer networks and hypergraphs, temporal networks, and certain other types of networks.

\subsection{General Form}\label{general}

A graph (i.e., a single-layer network) is a tuple $G=(V,E)$, where $V$ is the set of nodes and $E \subseteq V\times V$ is the set of edges that connect pairs of nodes~\cite{bollobasbook}.  If there is an edge between a pair of nodes, then those nodes are \emph{adjacent} to each other.  This edge is \emph{incident} to each of the two nodes, and two edges that are incident to the same node are also said to be ``incident'' to each other.

To represent systems that consist of networks at multiple levels or with multiple types of edges (or with other similar features), we consider structures that have \emph{layers} in addition to nodes and edges. Our goal is to start from the most general structure of this kind and to yield existing notions of multilayer (and multiplex, etc.) networks by introducing relevant limitations and constraints. In our most general multilayer-network framework, we allow each node to belong to any subset of the layers, and we are able to consider edges that encompass pairwise connections between all possible combinations of nodes and layers.  (One can further generalize this framework to consider hyperedges that connect more than two nodes.) For example, a node $\nodeindex[1]$ in layer $\layerindexvector[1]$ can be connected to any node $\nodeindex[2]$ in any layer $\layerindexvector[2]$. Moreover, we want to consider ``multidimensional'' layer structures in order to include every type of multilayer-network construction that we have encountered in the literature. For example, one ``dimension'', which we henceforth call an \emph{aspect} (but for which the term ``feature'' might also be reasonable), of a layer might be the type of an edge and another aspect might be the time at which an edge is present.  The above use of the word ``dimension'' amounts to a standard English meaning of the word to mean aspect or feature, but the standard use of the monicker ``dimension'' as jargon in mathematics and physics compels us to use different terminology.  Additionally, in the social-networks literature, one might discuss different ``dimensions'' of interactions between people (friendship, family, etc.), so that a dimension would then correspond to a layer in a multilayer network. We wish to avoid this terminology clash as well.

\begin{figure}[!htp]
\includegraphics[width=1.0\linewidth]{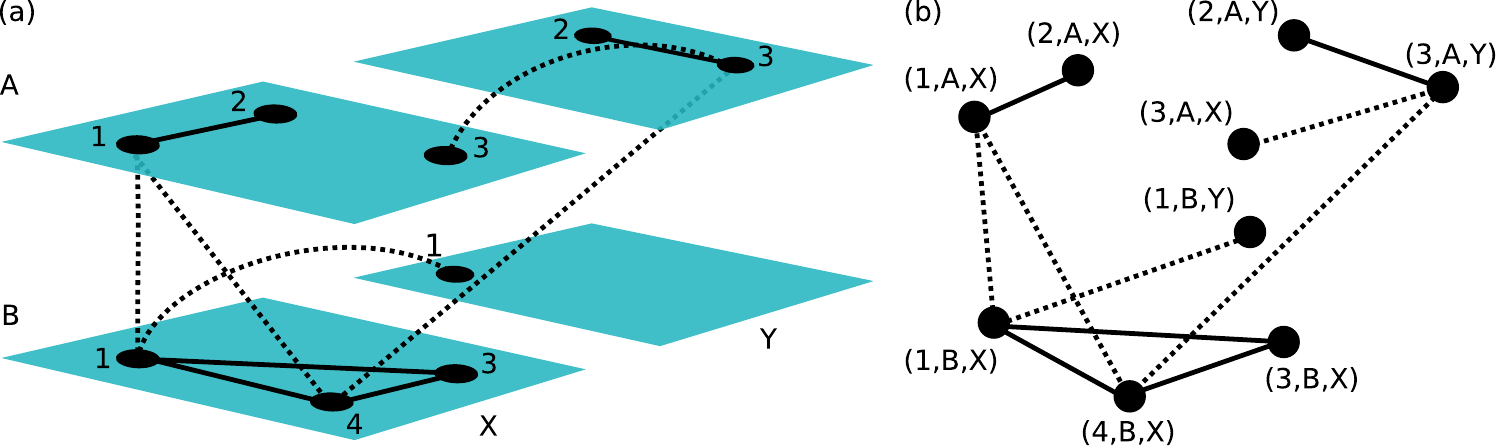}
\caption{(a) An example of the most general type of multilayer network, $M=(V_M,E_M,V,\layervector)$, that we consider in this article. The network $M$ has a total of four nodes, so $V=\{ 1,2,3,4 \}$, and two aspects, which have corresponding elementary-layer sets $L_1=\{ A, B\}$ and $L_2=\{ X, Y\}$.  There are thus a total of 4 different layers: $(A,X)$, $(A,Y)$, $(B,X)$, and $(B,Y)$. Each layer contains some subset of the node set $V$; for this example, the set of node-layer tuples is $V_M=\{ (1,A,X), (2,A,X), (3,A,X), (2,A,Y), (3,A,Y), (1,B,X), (3,B,X), (4,B,X), (1,B,Y) \} \subseteq V \times L_1 \times L_2$. The nodes can be connected to each other in a pairwise manner both within the layers and across the layers. We show the edges that remain inside of a layer (i.e., intra-layer edges) as solid lines and the edges that cross layers (i.e., inter-layer edges) as dotted lines. (b) The underlying graph $G_M=(V_M,E_M)$ of the same multilayer network.  We again show intra-layer edges as solid lines and inter-layer edges as dashed lines.  The adjacency matrix of this graph (or ``supra-graph'') is the multilayer network's supra-adjacency matrix.  
}
\label{fig:general}
\end{figure}

We will now give a precise definition of a multilayer network based on our above description. (See Ref.~\cite{mikko-viz} for software for the investigation and visualization of multilayer networks.) A multilayer network has a set of nodes $V$ just like a normal network (i.e., a graph). In addition, we need to have layers. However, because we want to be able to include multiple aspects in a multilayer network, we cannot restrict ourselves to having a single set of layers. For example, in a network in which the first aspect is interaction type and the second one is time, we need one set of layers for for interaction types and a second set of layers for time (e.g., for time stamps). To avoid confusion, we use the term ``elementary layer'' (see Fig.~\ref{fig:general}) for an element of \emph{one} of these sets and the term ``layer'' to refer to a combination of elementary layers from all aspects. In the previous example, an interaction type and a time stamp are both examples of an elementary layer, and a combination of an interaction type and a time stamp constitutes a layer. A multilayer network can have any number $\aspects$ of aspects, and we need to define a sequence $\layervector=\{ \layers_\aspectindex \}_{\aspectindex=1}^\aspects$ of sets of elementary layers such that there is one set of elementary layers\footnote{Note that a set of elementary layers need not be finite, and it could even be continuous. We will discuss this issue briefly in Section~\ref{temporal}, but we otherwise assume in this article that all of the elementary-layer sets are finite.  We also assume in this article that the number of aspects is finite.} $\layers_\aspectindex$ for each aspect $\aspectindex$. 

Using the sequence of sets of elementary layers, we can construct a set of layers in a multilayer network by assembling a set of all of the combinations of elementary layers using a Cartesian product $\layers_1 \times \dots \times \layers_\aspects$. We want to allow nodes to be absent in some of the layers. That is, for each choice of a node and layer, we need to indicate whether the node is present in that layer. To do so, we first construct a set $V \times \layers_1 \times \dots \times \layers_\aspects$ of all of these combinations and then define a subset $V_M \subseteq V \times \layers_1 \times \dots \times \layers_\aspects$ that contains only the node-layer combinations in which a node is present in the corresponding layer.\footnote{It is also convenient to require that there do not exist nodes that are not present in any of the layers. Therefore, each node appears in at least one layer, so $\{ u | (u,\layerindex[1]_1, \dots, \layerindex[1]_d) \in V_M \} = V $.)}  We will often use the term \emph{node-layer tuple} (or simply \emph{node-layer}) to indicate a node that exists on a specific layer. Thus, the node-layer $(\nodeindex,\layerindex_1,\dots,\layerindex_\aspects) $ represents node $\nodeindex[1]$ on layer $(\layerindex_1,\dots,\layerindex_\aspects)$.

In a multilayer network, we also need to define connections between pairs of node-layer tuples. As with monoplex networks, we will use the term \emph{adjacency} to describe a direct connection via an edge between a pair of node-layers and the term \emph{incidence} to describe the connection between a node-layer and an edge.  Two edges that are incident to the same node-layer are also ``incident'' to each other. We want to allow all of the possible types of edges that can occur between any pair of node-layers --- including ones in which a node is adjacent to a copy of itself in some other layer as well as ones in which a node is adjacent to some other node from another layer. In normal networks (i.e., graphs), the adjacencies are defined by an edge set $E \subseteq V \times V$, in which the first element in each edge is the starting node and the second element is the ending node. In multilayer networks, we also need to specify the starting and ending layers for each edge. We thus define an edge set $E_M$ of a multilayer network as a set of pairs of possible combinations of nodes and elementary layers.  That is, $E_M \subseteq V_M \times V_M$. 

Using the components that we set up above, we define a \emph{multilayer network} as a quadruplet $M=(V_M,E_M,V,\layervector)$. See Fig.~\ref{fig:general}a for an illustrative example. Note that if the number of aspects is zero (i.e., if $\aspects=0$), then the multilayer network $M$ reduces to a \emph{monoplex} (i.e., single-layer) network.  In that case, $V_M = V$, so the set $V_M$ becomes redundant. (By convention, the product term in the set $V \times \layers_1 \times \dots \times \layers_\aspects$ does not exist if $\aspects = 0$.)

The first two elements in a multilayer network $M$ yield a graph $G_M=(V_M,E_M)$, so one can interpret a multilayer network as a graph whose nodes are labeled in a certain way (see Fig.~\ref{fig:general}b). This observation makes it easy to generalize some of the basic concepts from monoplex networks to multilayer networks. For example, we define a \emph{weighted} multilayer network $M$ by defining weights for edges in the underlying graph $G_M$ (i.e., by mapping edges of a network to a real number using the function $w: E_M \rightarrow \mathbb{R}$), and we say that a multilayer network $M$ is undirected (or directed) if the underlying graph $G_M$ is undirected (or directed). We also employ the usual convention of disallowing self-edges in the multilayer network by disallowing self-edges in the underlying graph, but this can of course be relaxed. Expressing these concepts directly in terms of the multilayer-network formalism is trivial but a bit cumbersome when allowing an arbitrary number of aspects. To alleviate this problem, we denote the array of elementary layers using a bold typeface: $(\nodeindex,\layerindexvector) \equiv (\nodeindex,\layerindex_1,\dots,\layerindex_\aspects)$. With this notation, we write, for example, that a multilayer network is undirected if $((\nodeindex[1],\layerindexvector[1]),(\nodeindex[2],\layerindexvector[2])) \in E_M \implies ((\nodeindex[2],\layerindexvector[2]),(\nodeindex[1],\layerindexvector[1])) \in E_M$; and we disallow self-edges by requiring that $((\nodeindex[1],\layerindexvector),(\nodeindex[1],\layerindexvector)) \notin E_M$.

\begin{figure}[!htp]
\begin{center}
\includegraphics[width=1.0\linewidth]{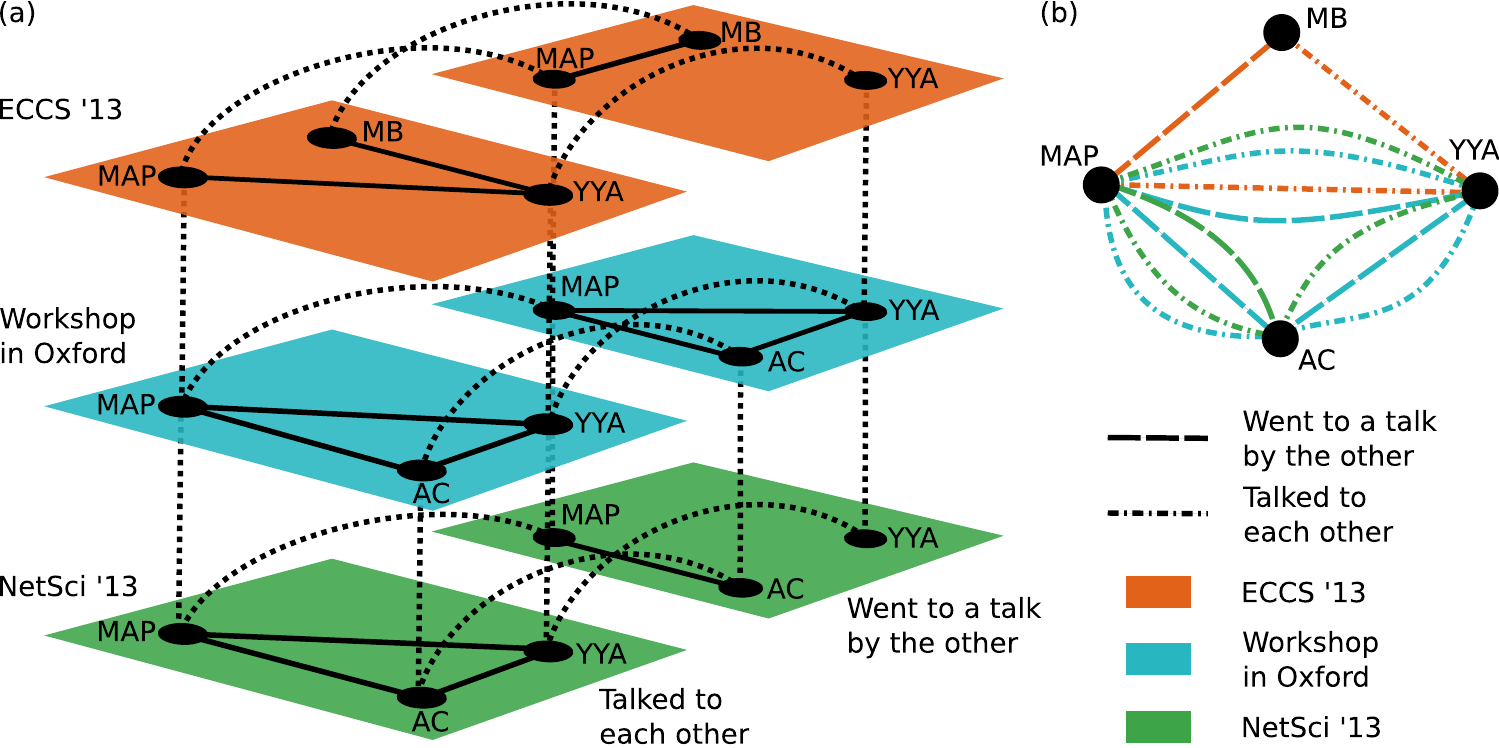}
\end{center}
\caption{[Color] (a) Visualization of the Zachary Karate Club Club (ZKCC) network as a multilayer network. Nodes (i.e., elements of $V$) in the network are the four network scientists who have held the coveted karate trophy for a period of time and have been awarded the associated membership in the ZKCC~\protect\cite{clubclub}.  The current members of the ZKCC are Cris Moore (CM), Mason A. Porter (MAP), Yong-Yeol Ahn (YYA), and Mari\'{a}n Bogu\~{n}\'{a} (MB).  In the figure, Aaron Clauset (AC) is standing in for CM, as the former awarded the karate trophy to MAP at NetSci '13 on behalf of the latter (who did not attend any of the conferences in the figure). The ZKCC network has two \emph{aspects}: the first one is the type of relationship between the scientists (talked to each other, went to a talk by the other) and the second one represents a conference in which the trophy was awarded (and thereby passed from one recipient to the next).  We assume for simplicity that all of the edges are undirected, even though the relationship of attending a talk obviously need not be a reciprocal one. Each \emph{layer} includes one \emph{elementary layer} from each of the two aspects. We represent \emph{intra-layer edges} using solid curves and \emph{inter-layer edges} using dotted curves. All of the inter-layer edges are \emph{coupling edges} because nodes are adjacent only to themselves (and not to other nodes) in other layers, and the inter-layer edges are therefore \emph{diagonal}. The coupling edges in the first aspect are \emph{categorical} (because such edges exist between corresponding nodes in all possible pairs of layers), and the coupling edges in the second aspect are \emph{ordinal} (see Section~\protect\ref{temporal}) because there exist such edges only in conferences that are contiguous to each other in time. Additionally, there is a coupling edge between a node and its counterpart in two different layers only when the layers differ from each other in exactly one aspect. [For example, MAP in layer (``ECCS '13'', ``Talked to each other'') is not adjacent to MAP in layer (``Workshop in Oxford'', ``Went to a talk by the other'').]  That is, the ZKCC network does not include \emph{inter-aspect coupling} (see Section~\protect\ref{constrain}). Additionally, the ZKCC network is not node-aligned because some of the nodes are missing from some of the layers. Finally, the ZKCC network is an example of a \emph{multiplex network} (see Section~\protect\ref{multiplex}). (b) Visualization of ZKCC network as an edge-colored multigraph (see Section~\protect\ref{multiplex}). 
}
\label{fig:kc2}
\end{figure}

\begin{table}[!p]
\begin{adjustwidth}{-5cm}{-5cm}
\setlength{\tabcolsep}{4pt}
\begin{center}
\begin{tabular}{lccccccccc}
Name                          & Aligned   & Disj.      & Eq. Size    & Diag.     & Lcoup.      & Cat.       & $|\layers|$ & $\aspects$ &Example refs. \\
\hline
Multilayer network            &            &            &            &\checkmark &\checkmark  & \checkmark &Any      &1   &\cite{Cardillo2013Modeling} \\
                              & \checkmark$\dagger$ &            & \checkmark$\dagger$ &           &            &            &Any      &1   &\cite{DeDomenico2013Mathematical}\\ 
Multiplex network             & \checkmark$\dagger$ &            & \checkmark$\dagger$ &\checkmark &            &            &Any      &1   &\cite{DeDomenico2013Random,DeDomenico2013Mathematical}\\
                              & \checkmark &            & \checkmark &\checkmark & \checkmark & \checkmark &Any      &1   &\cite{Yagan2012Analysis,Brummitt2012Multiplexityfacilitated,Nicosia2013Growing,Bianconi2013Statistical,Cellai2013Percolation,Battiston2013Metrics,Horvat2012Onemode}\\
                              & \checkmark &            & \checkmark &\checkmark & \checkmark & \checkmark &2        &1   &\cite{Min2013Layercrossing,Lee2012Correlated,Min2013Network}\\
                              &            &            &            &\checkmark & \checkmark & \checkmark &Any      &1   &\cite{Cozzo2012Stability}\\
                              & \checkmark &            & \checkmark &\checkmark & \checkmark &            &Any      &1   &\cite{SoleRibalta2013Spectral,Cozzo2013Contactbased,Sola2013Centrality}\\
Multivariate network          & \checkmark &            & \checkmark &\checkmark & \checkmark & \checkmark &Any      &1   &\cite{Pattison1999Logit}\\
Multinetwork                  & \checkmark &            & \checkmark &\checkmark & \checkmark & \checkmark &Any      &1   &\cite{Barigozzi2010Multinetwork}\\
                              & \checkmark &            & \checkmark &\checkmark & \checkmark & \checkmark &Any      &2   &\cite{Barigozzi2011Identifying}\\
Multirelational network       & \checkmark &            & \checkmark & \checkmark& \checkmark & \checkmark &Any      &1   &\cite{Wasserman1994Social,Cai2005Community,Harrer2012Approach,Stroele2009Mining} \\
Multirelational data          & \checkmark &            & \checkmark &\checkmark & \checkmark & \checkmark &Any      &1   &\cite{Li2012HAR,MichaekKwokPoNg2011MultiRank}\\
Multilayered network          & \checkmark &            & \checkmark & \checkmark& \checkmark & \checkmark &Any      &1   &\cite{Brodka2010Method,Brodka2011Shortest,Brodka2012Analysis,Sola2013Centrality} \\
Multidimensional network      & \checkmark &            & \checkmark & \checkmark& \checkmark & \checkmark &Any      &1   &\cite{Berlingerio2011Pursuit,Berlingerio2013ABACUS,Berlingerio2012Multidimensional,Tang2012Community,Barrett2012Taking,Kazienko2011MultidimensionalB,Coscia2013You} \\
                              & \checkmark &            & \checkmark & \checkmark& \checkmark & \checkmark &Any      &3   &\cite{Kazienko2011Multidimensional}\\
Multislice network            & \checkmark$\dagger$ &            & \checkmark$\dagger$ & \checkmark&            &            &Any      &1   &\cite{Mucha2010Community,Mucha2010Communities,Carchiolo2011Communities,Bassett2013Robust} \\
Multiplex of interdep. networks& \checkmark &            & \checkmark & \checkmark& \checkmark & \checkmark &Any      &1   &\cite{GomezGardenes2012Evolution} \\
Hypernetwork                  & \checkmark &            & \checkmark &\checkmark & \checkmark & \checkmark &Any      &1   &\cite{Irving2012Synchronization,Sorrentino2012Synchronization} \\
Overlay network               & \checkmark &            & \checkmark &\checkmark & \checkmark & \checkmark &2        &1   &\cite{Funk2010Interacting,Marceau2011Modeling}\\
Composite network             & \checkmark &            & \checkmark &\checkmark & \checkmark & \checkmark &2        &1   &\cite{Wei2012Competing}\\
Multilevel network**          &            & \checkmark &            &           &            &            &Any      &1   &\cite{Wang2013Exponential,Lazega2008Catching} \\   
                              &            &            &            &\checkmark & \checkmark & \checkmark &Any      &1   &\cite{Cozzo2012Stability,Criado2011Mathematical}\\    
Multiweighted graph           & \checkmark &            & \checkmark &\checkmark &\checkmark  &\checkmark  &Any      &1   &\cite{Rocklin2013Clustering}\\
Heterogeneous network         &            & \checkmark &            &           &            &            &2        &1   &\cite{Zhou2007CoRanking,Cai2005Community}\\         
Multitype network             &            & \checkmark &            &           &            &            &Any      &1   &\cite{Hindes2013Epidemic,Allard2009Heterogeneous,Vazquez2006Spreading}\\         
Interconnected networks       &            & \checkmark & \checkmark &           &            &            &2        &1   &\cite{Dickison2012Epidemics,Louzada2013Breathing} \\
                              &            & \checkmark &            &           &            &            &2        &1   &\cite{SaumellMendiola2012Epidemic,Sahneh2012Effect}\\
Interdependent networks*      &            & \checkmark & \checkmark &           &            &            &2        &1   &\cite{Buldyrev2010Catastrophic} \\
*                             &            & \checkmark &            &           &            &            &2        &1   &\cite{Parshani2010Interdependent} \\ 
                              &            &            & \checkmark &           &            &            &2        &1   &\cite{MartinHernandez2013Synchronization}\\
                              &            & \checkmark &            &           &            &            &2        &1   &\cite{Brummitt2012Suppressing,Bashan2013Extreme} \\
                              & \checkmark &            & \checkmark & \checkmark& \checkmark & \checkmark &Any      &1   &\cite{Baxter2012Avalanche} \\
Partially interdep. networks* &            & \checkmark &            &           &            &            &2        &1   &\cite{Zhou2013Percolation} \\
Network of networks*          &            &            & \checkmark &           &            &            &Any      &1   &\cite{Gao2011Robustness} \\
Coupled networks              &            &            &            &\checkmark & \checkmark & \checkmark &Any      &1   &\cite{Yagan2012Conjoining}\\
Interconnecting networks      &            &            &            &\checkmark & \checkmark & \checkmark &2        &1   &\cite{Xu2011Interconnecting}\\
Interacting networks          &            & \checkmark &            &           &            &            &Any      &1   &\cite{Leicht2009Percolation,Donges2011Investigating}\\
                              &            & \checkmark &            &           &            &            &2        &1   &\cite{Brummitt2012Suppressing} \\
Heterogenous information net**&            &            &            &           &            &            &Any      &2   &\cite{Davis2011Multirelational,Sun2011PathSim,Sun2012Thesis,Sun2013Mining}\\
                              &            & \checkmark &            &           &            &            &Any      &1   &\cite{Sun2009Ranking}\\
Meta-matrix, meta-network     &            &            &            &           &            &            &Any      &2   &\cite{CarleyStructural2001,Carley2007Toward,Tsvetovat2004DyNetML}\\
\end{tabular}
\end{center}
\end{adjustwidth}
\caption{Network types that can be represented using our general multilayer-network framework and the properties that these networks have in this representation. 
\emph{Aligned}: Is the network node-aligned? \emph{Disj.}: Is the network layer-disjoint? \emph{Eq. Size}: Do all of the layers have the same number of nodes? \emph{Diag.}: Are the couplings diagonal? \emph{Lcoup.}: Do the inter-layer couplings consist of layer couplings? \emph{Cat.}: Are the inter-layer couplings categorical? Additionally, $|L_i|$ denotes the number of possible layers, and $d$ denotes the number of ``aspects''.  (We define these terms in the main text in Sections \protect\ref{general} and discuss them further in Section \protect\ref{reps}.  Additionally, we summarize several of them in the glossary in Section \protect\ref{glossary}.)  
The symbol ``*'' indicates the additional constraints that the inter-layer edges are undirected and that a node in one layer can only be adjacent to a single node in any other layer (see the last paragraph of Section~\protect\ref{multiplex}).
The networks that are marked with the symbol ``**'' have additional constraints, we discuss in Section~\ref{othernets}.  (Each case of the symbols ``*'' and ``**'' applies to exactly one row in the table.)
Some network structures can be represented in multiple different ways using our multilayer-network framework. For example, a network structure that consists of a sequence of graphs that share the same set of nodes can be represented using either categorical or ordinal couplings. In our discussions, we have always assumed that such a network has implicit categorical couplings if the order of the graphs in the sequence is not used in the article in which the terminology was introduced.  
We mark ``multislice''~\protect\cite{Mucha2010Community}, ``multilayer''~\protect\cite{DeDomenico2013Mathematical}, and some types of multiplex networks with the symbol $\dagger$ to illustrate an important point: these frameworks technically require a network to be node-aligned, but there is a way to use them (which has been exploited successfully in various papers) to consider networks in which it is \emph{not} the case that all nodes are shared among all layers.  See the discussion in Section \protect\ref{reps}.
}
\label{table:dictionary}
\end{table}

It is both typical and convenient to use different semantics for edges that cross layers than for edges that stay within a single layer. It is thus often useful to partition the set of edges into \emph{intra-layer edges} $E_A=\{ ((\nodeindex[1],\layerindexvector[1]),(\nodeindex[2],\layerindexvector[2])) \in E_M  | \layerindexvector[1]=\layerindexvector[2] \}$ and \emph{inter-layer edges} $E_{C}=E_M \setminus E_A$ (see, e.g., Refs.~\cite{Cozzo2013Contactbased,Cozzo2013Clustering,Estrada2013Communicability}, and Fig.~\ref{fig:general} for an example). We also define \emph{coupling edges} $E_{\tilde{C}} \subseteq E_{C}$ as edges for which the two nodes represent the same entity in different layers: $E_{\tilde{C}}=\{ ((\nodeindex[1],\layerindexvector[1]),(\nodeindex[1],\layerindexvector[2])) \in E_C \}$. We thereby define an \emph{intra-layer graph} $G_A=(V_M,E_A)$, an \emph{inter-layer graph} $G_{C}=(V_M,E_C)$, and a \emph{coupling graph} $G_{\tilde{C}}=(V_M,E_{\tilde{C}})$.\footnote{We use a naming convention that is motivated by the block structure of supra-adjacency matrices. We use $A$ to denote intra-layer blocks (which are diagonal blocks in a matrix), and we use $C$ to denote other (off-block-diagional) matrix elements. Note that the letter $C$ has been used in the literature to represent both coupling edges and more general inter-layer edges (see, e.g., Refs.~\cite{Radicchi2013Driving,Estrada2013Communicability}), but we want to differentiate between these two ideas in the present paper.}

As we discuss later, we can obtain existing notions of multilayer networks (and similar objects) from the literature by applying various constraints to our general framework.  We now give names to these constraints. We say that a multilayer network is \emph{node-aligned}\footnote{The term ``node-aligned'' should not be confused with the concept of \emph{graph alignment}, which refers to the problem of renaming nodes of two or more graphs such that the graphs become as similar to each other as possible.} (or ``fully interconnected'') if all of the layers contain all nodes: $V_M = V \times L_1 \times \dots \times L_\aspects$. A multilayer network is \emph{layer-disjoint} if each node exists in at most one layer: $(\nodeindex[1],\layerindexvector[1]), (\nodeindex[1],\layerindexvector[2]) \in V_M \implies \layerindexvector[1]=\layerindexvector[2]$. We say that couplings are \emph{diagonal} if all of the inter-layer edges are between nodes and their counterparts in another layers: $E_{\tilde{C}}=E_C$. We say that a diagonal multilayer network is \emph{layer-coupled} if the coupling edges and their weights are independent of the nodes~\cite{Sola2013Centrality}: if $((\nodeindex[1],\layerindexvector[1]),(\nodeindex[1],\layerindexvector[2])) \in E_C$ and $(\nodeindex[2],\layerindexvector[1]),(\nodeindex[2],\layerindexvector[2]) \in V_M$, then $((\nodeindex[2],\layerindexvector[1]),(\nodeindex[2],\layerindexvector[2])) \in E_C$ and $w(((\nodeindex[1],\layerindexvector[1]),(\nodeindex[1],\layerindexvector[2])))=w( ((\nodeindex[2],\layerindexvector[1]),(\nodeindex[2],\layerindexvector[2])) )$ for all $\nodeindex[1],\nodeindex[2],\layerindexvector[1],\layerindexvector[2]$. That is, for any two layers, the coupling is the same for all nodes (so it depends only on the layers). The couplings are \emph{categorical} if each node is adjacent to all of its counterparts in the other layers: $(\nodeindex[1],\layerindexvector[1]),(\nodeindex[1],\layerindexvector[2])  \in V_M \implies ((\nodeindex[1],\layerindexvector[1]),(\nodeindex[1],\layerindexvector[2])) \in E_M$.

We say that couplings are categorical with respect to a single aspect if each node is adjacent to all of its counterparts in layers that only differ in that aspect. When a network is node-aligned, we denote the number of entities (which are represented by nodes) by $\nnodes=|V|$. Note that the number of nodes in each layer can be equal to each other (i.e., the layers have the same size) even if a network is not node-aligned. In this case, the node identities in the different layers are different. When $\aspects=1$, there is a single additional aspect beyond an ordinary monoplex network, so we denote the set of layers by $\layers=\layers_1$ and the number of layers by $\nlayers=|\layers|$. Note that categorically-coupled multilayer networks are a subset of layer-coupled networks, which are in turn a subset of diagonal networks. We summarize the constraints that we have just discussed in the Glossary (see Section~\ref{glossary}), and we illustrate these constraints with an example multilayer network in Fig.~\ref{fig:kc2}a.

\subsection{Tensor Representations}\label{reps}

It is often convenient to represent ordinary graphs (i.e., monoplex networks) using adjacency matrices. For a node-aligned multilayer network, we represent the analogous structure using a rank-$2(\aspects+1)$ [i.e., order-$2(\aspects+1)$] adjacency tensor\footnote{In the social-networks literature, such multidimensional arrays are sometimes called ``super-sociomatrices''~\cite{Wasserman1994Social} (Similarly, adjacency matrices are often called ``sociomatrices'' in that literature.}~\cite{DeDomenico2013Mathematical} $\mathcal{A} \in \{ 0,1 \} ^{|V| \times |V| \times |\layervector_1| \times |\layervector_1| \times \dots \times |\layervector_\aspects| \times |\layervector_\aspects| }$. We use the simplified notation\footnote{In the rest of the present article, we assume (without loss of generality) that the nodes $\nodeindex[1] \in V$ and layers $\layerindex[1] \in \layers_\aspectindex$ are represented by consecutive integers that start from $1$.} $\mathcal{A}_{\nodeindex[1]\nodeindex[2]\layerindexvector[1]\layerindexvector[2]}=\mathcal{A}_{\nodeindex[1]\nodeindex[2]\layerindex[1]_1\layerindex[2]_1 \dots \layerindex[1]_\aspects\layerindex[2]_\aspects}$, where the tensor element $\mathcal{A}_{\nodeindex[1]\nodeindex[2]\layerindexvector[1]\layerindexvector[2]}$ has a value of $1$ if and only $((\nodeindex[1],\layerindexvector[1]),(\nodeindex[2],\layerindexvector[2])) \in E_M$; otherwise, $\mathcal{A}_{\nodeindex[1]\nodeindex[2]\layerindexvector[1]\layerindexvector[2]}$ has a value of $0$.  As a convention, we group indices according to layer type (i.e., according to aspect). Analogous to weighted adjacency matrices in monoplex networks, we can define a weighted adjacency tensor $\mathcal{W}$ in which the value of each element corresponds to the weight of an edge (and to $0$ when there is no edge).

Technically, the above tensor representation is only appropriate for a multilayer network that is node-aligned. However, as discussed in Ref.~\cite{DeDomenico2013Mathematical}, many of the tensor-based methods for network analysis can still be employed by adding extraneous nodes (which we will call \emph{empty nodes}) that are not adjacent to any other nodes.  In Table \ref{table:dictionary}, we mark ``multislice''~\cite{Mucha2010Community}, ``multilayer''~\cite{DeDomenico2013Mathematical}, and some types of multiplex networks with the symbol $\dagger$ to illustrate this important point: although these frameworks technically require a network to be node-aligned, they have been used explicitly (by ``padding'' them with empty nodes) to consider networks in which it is \emph{not} the case that all nodes are shared between all layers.\footnote{Such a ``padding'' process is also viable for other types of multilayer networks in the table.  In Table \ref{table:dictionary}, we only use the notation \checkmark in situations for which such padding is not necessary.} Such ``padding'' yields a structure that is node-aligned from a mathematical perspective, but one has to be very careful in practice when studying the resulting tensors.  From a technical perspective, appending empty nodes makes it possible to use these adjacency tensors as inputs in other frameworks --- e.g., this has been very successful for studies of community structure in the type of multilayer network known as ``multislice networks'' \cite{Mucha2010Community,Mucha2010Communities} (and sometimes also called ``multilayer networks'' in subsequent papers \cite{Bassett2013Robust}) --- but it can lead to highly misleading results when computing network diagnostics, such as mean degree or clustering coefficients, unless one accounts for the presence of empty nodes in an appropriate way.  One must be cautious. In theory, one could also represent layer-disjoint networks using adjacency tensors by renaming (or enumerating) the nodes that start from 1 at each layer and padding using empty nodes if the layers are not of equal size. We discuss this type of renaming in Section~\ref{nodecolored} and illustrate it in Fig.~\ref{fig:nodecolored}.

\begin{figure}[!htp]
\includegraphics[width=1.0\linewidth]{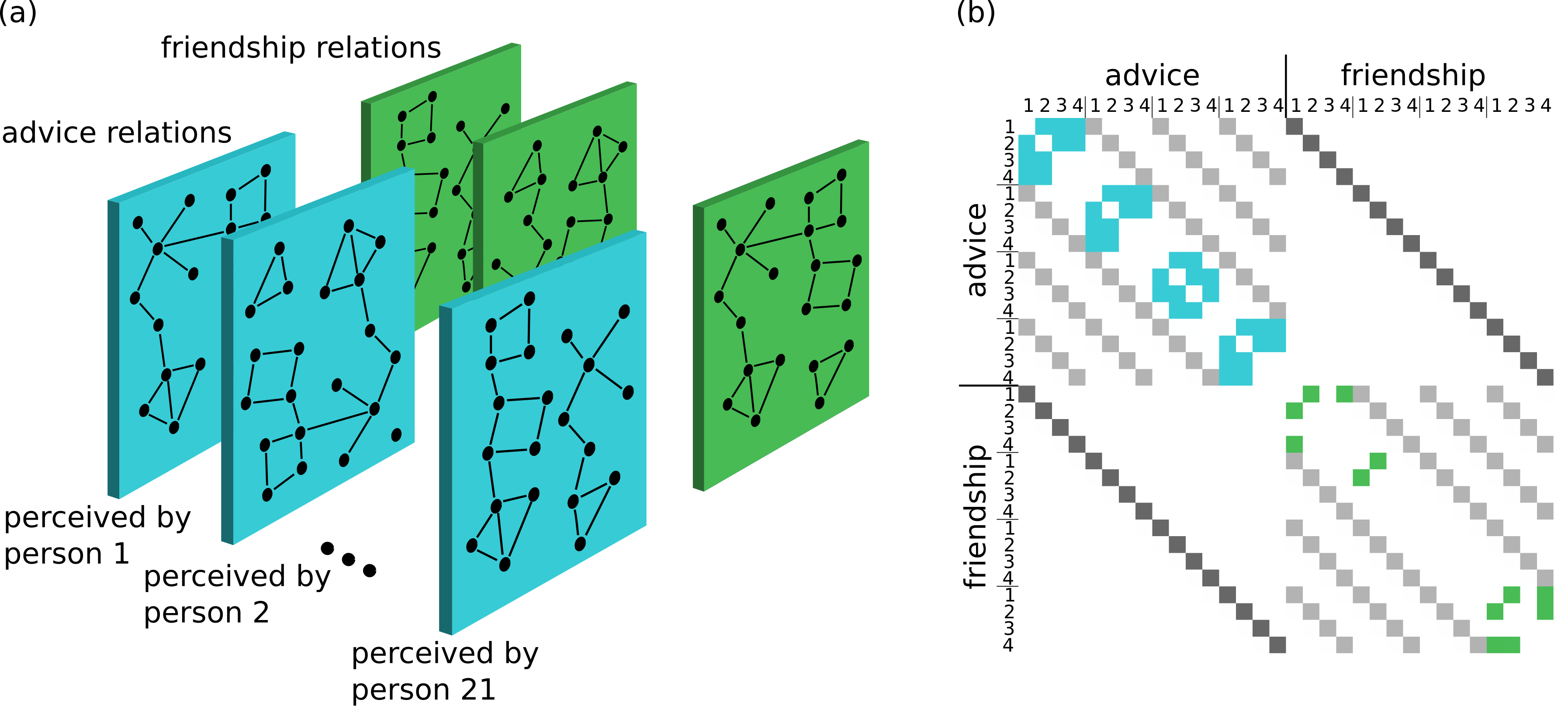}
\caption{(a) Schematic of a cognitive social structure~\protect\cite{Krackhardt1987Cognitive} represented as a multilayer network (though note that we have not drawn the coupling edges). The ``perceiver'' aspect has a size of 21, and the ``edge-type'' aspect has a size of 2.  This yields a total of 42 network layers. In the general tensorial framework, this structure corresponds to a 6th-order (i.e., rank-6) tensor with components $\mathcal{A}_{\nodeindex[1]\nodeindex[2]\layerindex[1]_1\layerindex[2]_1\layerindex[1]_2\layerindex[2]_2}$. By using implicit couplings, one can instead use a 4th-order tensor $\mathcal{A}$, where the element $\mathcal{A}_{\nodeindex[1]\nodeindex[2]\layerindex[1]_1\layerindex[1]_2}$ corresponds to an edge of type $\layerindex[1]_1$ between $\nodeindex[1]$ and $\nodeindex[2]$ that is perceived by node $\layerindex[1]_2$. (b) Illustration of (a subset of) the supra-adjacency matrix that represents a cognitive social structure. In the depicted setup, each aspect is categorically coupled and there is no inter-aspect coupling (see Section~\protect\ref{constrain}). This yields a nested block-diagonal structure, where the interior block-diagonal structures (in blue and green) correspond to intra-layer adjacency matrices and the two large blocks correspond to the couplings between different perceivers when the edge-type (i.e., friendship or advice) aspect is fixed to one of the two alternatives. The multiplex network that we show in this figure is a subset of the full data set. (We show nodes and layers only for persons 1--4.)
}
\label{fig:cogsoc}
\end{figure}

\subsubsection{Constraints.}\label{constrain}

We now formulate some of the multilayer network constraints that we defined in the previous section using the above tensorial framework. We assume that the multilayer network is node-aligned. Constraints force some of the tensor elements to be $0$ and makes it possible to represent constrained multilayer networks using tensors with a lower rank than would otherwise be necessary. If a multilayer network has diagonal couplings --- i.e., if inter-layer edges are only allowed between two representations of the same node --- then the adjacency-tensor element $\mathcal{A}_{\nodeindex[1]\nodeindex[2] \layerindexvector[1]\layerindexvector[2]}=0$ if the two node indices differ from each other (i.e., if $\nodeindex[1] \neq \nodeindex[2]$) and the layers are not the same (i.e., if $\layerindexvector[1] \neq \layerindexvector[2]$). With this restriction, one can express a multilayer network as a combination of an intra-layer adjacency tensor with elements $\mathcal{A}_{\nodeindex[1]\nodeindex[2] \layerindexvector[1]}=\mathcal{A}_{\nodeindex[1]\nodeindex[2] \layerindexvector[1]\layerindexvector[1]}$ and a coupling tensor with elements $\mathcal{C}_{\nodeindex[1] \layerindexvector[1]\layerindexvector[2]}=\mathcal{A}_{\nodeindex[1]\nodeindex[1] \layerindexvector[1]\layerindexvector[2]}$. This is equivalent to the ``multislice'' formulation in Ref.~\cite{Mucha2010Community} and its sequels.

If one disallows inter-aspect couplings in a diagonal multilayer network, then the adjacency tensor also has elements equal to $0$ when the layer indices differ in more than one aspect; that is, $\mathcal{A}_{\nodeindex[1]\nodeindex[2] \layerindexvector[1]\layerindexvector[2]}=0$ if either $\nodeindex[1] \neq \nodeindex[2]$ and $\layerindexvector[1] \neq \layerindexvector[2]$, or if $\nodeindex[1] = \nodeindex[2]$ but $\layerindexvector[1]$ and $\layerindexvector[2]$ differ in more than one element. (Recall that an element of $\layerindexvector[1]$ corresponds to one aspect.) One can now represent the multilayer network using an intra-layer adjacency tensor with elements $\mathcal{A}_{\nodeindex[1]\nodeindex[2] \layerindexvector[1]}$ and a sequence of ``reduced'' coupling tensors (one for each aspect $\aspectindex$) with elements $C_{\nodeindex[1]\layerindexvector[1]\layerindex[2]}^\aspectindex=\mathcal{A}_{\nodeindex[1]\nodeindex[1]\layerindex[1]_1 \layerindex[1]_1 \dots \layerindex[1]_{\aspectindex-1}\layerindex[1]_{\aspectindex-1}\layerindex[1]_a\layerindex[2]\layerindex[1]_{\aspectindex+1}\layerindex[1]_{\aspectindex+1} \dots \layerindex[1]_\aspects \layerindex[1]_\aspects}$. If $\aspects=1$, then there is no difference between the reduced coupling tensor and the original coupling tensor (i.e., $\mathcal{C}_{\nodeindex[1] \layerindex[1]_1\layerindex[2]_1}=C_{\nodeindex[1]\layerindex[1]_1\layerindex[2]_1}^1$).

If there are no inter-aspect couplings and the coupling strengths only depend on the two elementary layers for each aspect, then the coupling tensor reduces even further to $\ctreduced_{\nodeindex[1]\layerindex[1]\layerindex[2]}^\aspectindex=\mathcal{A}_{\nodeindex[1]\nodeindex[1]\layerindex[3]_1 \layerindex[3]_1 \dots \layerindex[3]_{\aspectindex-1}\layerindex[3]_{\aspectindex-1}\layerindex[1]\layerindex[2]\layerindex[3]_{\aspectindex+1}\layerindex[3]_{\aspectindex+1} \dots \layerindex[3]_\aspects \layerindex[3]_\aspects}$ for any $\layerindexvector[3] = (\layerindex[3]_1, \ldots, \layerindex[3]_\aspects)$. In layer-coupled multilayer networks (which are also necessarily diagonal, by definition), the couplings are independent of the node: $\mathcal{A}_{\nodeindex[1]\nodeindex[1] \layerindexvector[1]\layerindexvector[2]}=\mathcal{A}_{\nodeindex[2]\nodeindex[2] \layerindexvector[1]\layerindexvector[2]}$ for all $\nodeindex[1],\nodeindex[2],\layerindexvector[1],\layerindexvector[2]$. This makes it possible to drop the node indices in the coupling tensors and write $\mathcal{C}_{\layerindexvector[1]\layerindexvector[2]}=\mathcal{C}_{\nodeindex[1] \layerindexvector[1]\layerindexvector[2]}$, $C_{\layerindexvector[1]\layerindex[2]}^\aspectindex=C_{\nodeindex[1]\layerindexvector[1]\layerindex[2]}^\aspectindex$, and $\ctreduced_{\layerindex[1]\layerindex[2]}^\aspectindex=\ctreduced_{\nodeindex[1]\layerindex[1]\layerindex[2]}^\aspectindex$ for any $\nodeindex[1]$. If such a multilayer network has no couplings (respectively, categorical couplings), then all of the coupling-tensor elements have the value $\mathcal{C}_{\layerindexvector[1]\layerindexvector[2]}=0$ (respectively, $\mathcal{C}_{\layerindexvector[1]\layerindexvector[2]}=1$). The network is then completely defined by the intra-layer adjacency tensor, which has elements $\mathcal{A}_{\nodeindex[1]\nodeindex[2] \layerindexvector[1]}$. If $\aspects=1$, this yields a rank-3 tensor, which has been used previously to represent this type
of multilayer network~\cite{Li2012HAR,MichaekKwokPoNg2011MultiRank,Lin2009MetaFac,Li2011Integrative}.

\subsubsection{Tensor Flattening.} \label{flat}

One can reduce the number of aspects of an adjacency tensor by combining aspects $i$ and $j$ into a new aspect $h$. That is, we map a node-aligned multilayer network $M=(V_M,E_M,V,\layervector)$ with $\aspects$ aspects into a new multilayer network $M^\prime=(V_{M^\prime},E_{M^\prime},V,\layervector^\prime)$ with $\aspects -1$ aspects, such that the total number of elements in the corresponding tensors is retained: $\layers_h^\prime=\layers_i \times \layers_j$, so $|\layers_h^\prime|=|\layers_i| |\layers_j|$. There is a bijective mapping between the elements of the old and new tensors, which is why we have noted that the essential ``new physics'' in multilayer networks already occurs when there is a single aspect.  The above mapping process is often called \emph{flattening} and is sometimes also known as ``unfolding'' or ``matricization''. Without loss of generality, suppose that one flattens aspects $\aspects-1$ and $\aspects$ of an adjacency tensor $\mathcal{A}$ to obtain a new tensor $\mathcal{A}^\prime$.  Again without loss of generality, we denote the layers using integers starting from $1$. With this labeling convention, the corresponding mapping is $\mathcal{A}_{\nodeindex[1]\nodeindex[2]\layerindex[1]_1\layerindex[2]_1 \dots \layerindex[1]_\aspects \layerindex[2]_\aspects}=\mathcal{A}^\prime_{\nodeindex[1]\nodeindex[2] \layerindex[1]_1 \layerindex[2]_1 \dots ((\layerindex[1]_{\aspects-1}-1)|\layers_\aspects|+\layerindex[1]_\aspects)((\layerindex[2]_{\aspects-1}-1)|\layers_\aspects|+\layerindex[2]_\aspects)}$. To reduce the number of aspects further, one can again apply such a mapping (and it can be desirable to continue this process until one obtains a tensor with $\aspects = 1$ aspects).  

Flattening tensors can be useful conceptually, and it is also often very convenient when writing software implementations of algorithms \cite{Kolda2009Tensor,Jutla2012}. The representation of multilayer networks using supra-adjacencies (see Refs.~\cite{Gomez2013Diffusion,SoleRibalta2013Spectral,Cozzo2013Clustering,SanchezGarcia2013Dimensionality} and Section \ref{supra}) is an extreme case of tensor flattening in which one constructs a network with $\aspects = 0$ aspects by combining all of the layer aspects and node indices to obtain additional node indices. See Fig.~\ref{fig:cogsoc} for an example of a supra-adjacency matrix.

\subsection{Supra-adjacency Representation} \label{supra}

The adjacency-matrix representation for monoplex networks is powerful because one can exploit the numerous tools, methods, and theoretical results that have been developed for matrices. To get access to these tools for investigations of multilayer networks, one can represent such networks using supra-adjacency matrices~\cite{Gomez2013Diffusion,SoleRibalta2013Spectral,Cozzo2013Clustering,SanchezGarcia2013Dimensionality}. As illustrated in Fig.~\ref{fig:general}, the adjacency matrix of a ``supra-graph'' corresponds to the supra-adjacency matrix of a multilayer network. (Other ``supra-matrices'' are defined analogously.) A supra-adjacency representation has already yielded interesting insights for processes such as diffusion~\cite{Gomez2013Diffusion,SoleRibalta2013Spectral}, epidemic spreading~\cite{Sahneh2012Effect,Wang2013Effect}, and synchronizability~\cite{SoleRibalta2013Spectral}.  In such studies, it is often helpful to investigate the properties of so-called ``supra-Laplacian matrices''.  Supra-adjacency matrices are also convenient for describing walks on multilayer networks~\cite{Cozzo2013Clustering}. An additional advantage of supra-adjacency matrices over adjacency tensors is that they provide a natural way to represent multilayer networks that are not node-aligned without having to append empty nodes. However, this boon comes with a cost: one must flatten a multilayer network to obtain a supra-adjacency matrix and one thereby loses some of the information about the aspects. Partitioning a network's edge set into intra-layer edges, inter-layer edges, and coupling edges makes it possible to retain some of this information. Supra-adjacency matrices, intra-layer supra-adjacency matrices, inter-layer supra-adjacency matrices, and coupling supra-adjacency matrices are the adjacency matrices that correspond, respectively, to the graphs $G_M$, $G_A$, $G_C$, and $G_{\tilde{C}}$ that we defined in Section~\ref{general}. Alternatively, for node-aligned multilayer networks, one can start from the tensor $\mathcal{A}$ and then use the flattening process discussed in Section \ref{flat} to transform it into a matrix.

One can derive a supra-Laplacian matrix from a supra-adjacency matrix in a manner that is analogous to the way that one derives a Laplacian matrix from the adjacency matrix of a monoplex graph. For example, the combinatorial supra-Laplacian matrix is ${\bf L_M} = {\bf D_M} - {\bf A_M}$, where ${\bf D_M}$ is the diagonal supra-matrix that has node-layer strengths (i.e., weighted degrees) along the diagonal and ${\bf A_M}$ denotes the supra-adjacency matrix that corresponds to the graph $G_M$.  Hence, each diagonal entry of the supra-Laplacian ${\bf L_M}$ consists of the sum of the corresponding row in the supra-adjacency matrix ${\bf A_M}$, and each non-diagonal element of ${\bf L_M}$ consists of the corresponding element of ${\bf A_M}$ multiplied by $-1$. The eigenvalues and eigenvectors of this supra-Laplacian are important indicators of several structural features of the corresponding network, and they also give crucial insights into dynamical processes that evolve on top of it~\cite{Radicchi2013Abrupt,Gomez2013Diffusion,Radicchi2013Driving} (see Section~\ref{spreading}). The second smallest eigenvalue and the eigenvector associated to it, which are sometimes called (respectively) the ``algebraic connectivity'' and ``Fiedler vector'' of the corresponding network, are very important diagnostics for the structure of a network. For example, the algebraic connectivity of a multilayer network with categorical coupling\footnote{Reference~\cite{Radicchi2013Abrupt} considered interconnected networks with certain restrictions. See Sections \ref{nodecolored} and \ref{multiplex} for a description of how one can map such networks into categorical multilayer networks.} has two distinct regimes when examined as a function of the relative strengths of the inter-layer edges and the intra-layer edges~\cite{Radicchi2013Abrupt}. Additionally, there is a discontinuous (i.e., first-order) phase transition --- a so-called ``structural transition'' --- between the two regimes. In one regime, the algebraic connectivity is independent of the intra-layer adjacency structure, so it is determined by the inter-layer edges. In the other, the algebraic connectivity of the multilayer network is bounded above by a constant multiplied by the algebraic connectivity of the unweighted superposition (see Section~\ref{agg}) of the layers. Combinatorial supra-Laplacian matrices have also been used to study a diffusion process on multiplex networks~\cite{Gomez2013Diffusion}. (See Section~\ref{spreading} for more on diffusion processes on multiplex networks.) Radicchi~\cite{Radicchi2013Driving} studied the spectrum of the normalized supra-Laplacian matrix ${\bf \hat{L}_M} = {\bf I} - {\bf D_M}^{-1/2} {\bf A_M} {\bf D_M}^{-1/2}$ (where ${\bf I}$ is an identity matrix) on two-layer interconnected networks that were generated using a generalized configuration model that includes correlations between intra-layer degrees and inter-layer degrees (see Section~\ref{interconnectedmodels}). Similar to Ref.~\cite{Radicchi2013Abrupt}, Radicchi varied the relative strengths of the inter-layer edges and the intra-layer edges, and he observed qualitatively different behavior for the eigen-spectrum of the normalized supra-Laplacian for different values of the relative strengths.

All calculations that one does using supra-adjacency matrices can also be done using adjacency tensors by defining a tensor multiplication that mimics the supra-adjacency multiplication. To see this, we define a \emph{flattening function} $f: \mathbb{R}^{|V| \times |V| \times |\layervector_1| \times |\layervector_1| \times \dots \times |\layervector_\aspects| \times |\layervector_\aspects|} \rightarrow \mathbb{R}^{|V| \prod_{\aspectindex=1}^\aspects |\layervector_\aspectindex|} \times \mathbb{R}^{|V| \prod_{\aspectindex=1}^\aspects |\layervector_\aspectindex|}$ that maps weighted adjacency tensors to weighted supra-adjacency matrices. The function $f$ is bijective, its inverse $f^{-1}$ is well-defined [$\mathcal{A}=f^{-1}(f(\mathcal{A}))$], and both $f$ and $f^{-1}$ are linear. We can now define a tensor multiplication $\times_f$ for the class of tensors that are spanned by the adjacency tensors of the multilayer networks: $\mathcal{A} \times_f \mathcal{B} = f^{-1}(f(\mathcal{A}) \cdot f(\mathcal{B}))$, where $\cdot$ is ordinary matrix multiplication. One can trivially (and usefully) extend the flattening function $f$ and the tensor multiplication $\times_f$ to include vectors in the vector space in which the supra-adjacency matrices operate.  This makes it possible, for example, to calculate the number of walks of length $N$ that start from node-layer $(\nodeindex[1], \layerindexvector[1])$ and end at node-layer $(\nodeindex[2]$,$\layerindexvector[2])$ in the multilayer network using the formula $(\underbrace{\mathcal{A} \times_f \cdots \times_f \mathcal{A}}_{N})_{\nodeindex[1]\nodeindex[2]\layerindexvector[1]\layerindexvector[2]}$.

\subsection{Node-Colored Networks, Interconnected Networks, Interdependent Networks, and Networks of Networks} \label{nodecolored}

Discussions of node-colored networks and similar structures have concentrated mostly on spreading processes, cascading failures, and network models (see Section \ref{models}).  In this section, we present the basic structural notions for these types of multilayer networks.

In \emph{interdependent networks}, nodes in two or more monoplex networks are adjacent to each other via edges that are called \emph{dependency edges}. For example, one can construe a electrical grid and a computer network as a pair of interdependent networks, as the proper function of a router in the computer network can depend on a power station~\cite{Buldyrev2010Catastrophic} and vice versa. Similarly, \emph{interconnected networks}, \emph{interacting networks}, and \emph{networks of networks} are sets of networks in which some of the nodes from the various networks are adjacent to each other, but the edges that connect different networks need not indicate dependency relations~\cite{SaumellMendiola2012Epidemic,Sahneh2012Effect,Leicht2009Percolation,Donges2011Investigating,Gao2011Robustness}. If the connections in interdependent networks and similar structures are limited in a certain way, then there is a relationship between them and multiplex networks (see the discussion at the end of Section~\ref{multiplex}). In \emph{multitype networks} and \emph{heterogeneous networks}~\cite{Zhou2007CoRanking,Cai2005Community}, all of the nodes are labeled with some ``type'' and they can be adjacent to nodes that are labeled with either the same or a different type. For example, the nodes in social multitype networks might be labeled with demographic characteristics such as sex, age, and ethnic group~\cite{Newman2003Mixing,Allard2009Heterogeneous}. In Ref.~\cite{Sun2009Ranking}, \emph{heterogeneous information networks} were defined as graphs in which each node has a distinct type.\footnote{In other articles, such as Ref.~\cite{Sun2011PathSim}, heterogeneous information networks were defined as possessing both colored nodes and colored edges.  This is a different type of structure from the others that we discuss in this paragraph.  See the discussion in Section \ref{othernets}.} In the types of multilayer networks that we have discussed in this paragraph, one can also think of each layer as a module (i.e., a community). This can be convenient for purposes like defining random-graph ensembles \cite{melnik-hetero}.

\begin{figure}[!htp]
\includegraphics[width=1.0\linewidth]{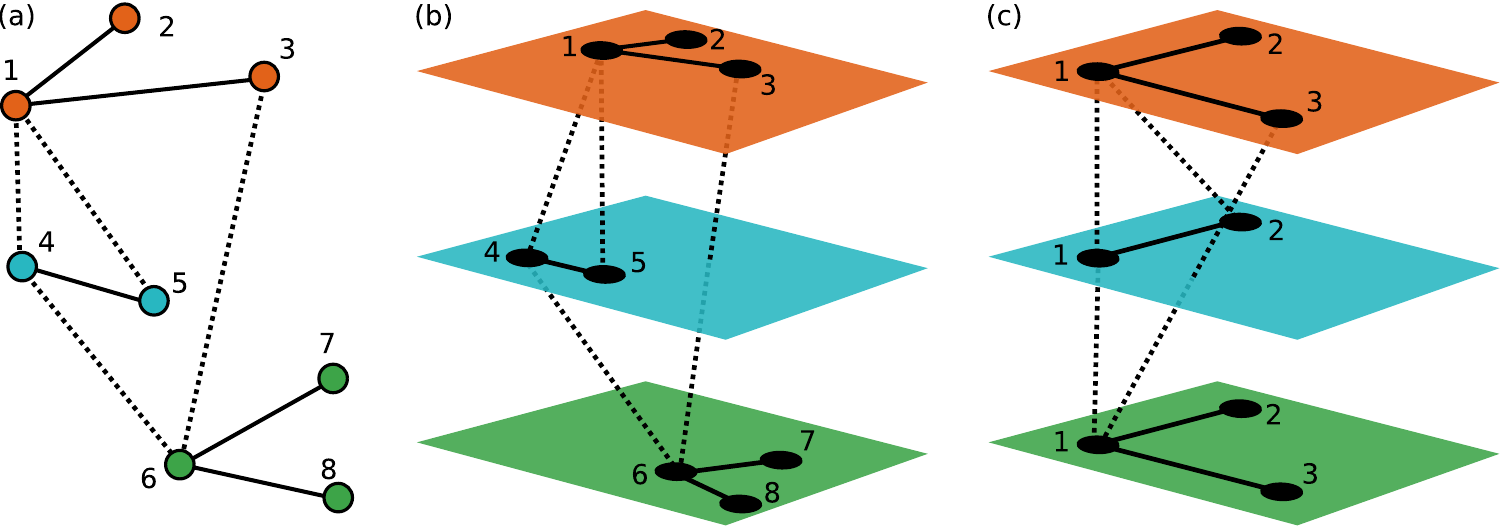}
\caption{(a) An example of a node-colored network (i.e., an interconnected network, a network of networks, etc.). (b) Representation of the same node-colored network using our multilayer-network formalism. We keep the node names from the original network. (c) Alternative representation of the same node-colored network in our multilayer-network formalism. This time, we use consecutive integers starting from $1$ to name the nodes in each layer, so we also need to include the identity of the layer to uniquely specify each node.
}
\label{fig:nodecolored}
\end{figure}

The multilayer networks that we have discussed above are equivalent to \emph{node-colored networks} \cite{Allard2009Heterogeneous,Vazquez2006Spreading}, although these various frameworks were formulated with different ideas in mind. (Note that we are using the word ``color'' in a very general sense; in particular, two nodes of the same color \emph{are} allowed to be adjacent.  This type of ``color'' is really a label, which is a more general usage along the lines of role-assignment problems \cite{roleequiv}.) Node-colored networks are graphs in which each node has exactly one color: $G_c=(V_c,E_c,C,\chi)$, where $V_c$ and $E_c$ are the nodes and edges, $C$ is the set of possible ``colors'' (where each color is a possible categorical label for the nodes), and $\chi: V_c \to C$ is a function that indicates the color of each node. For multitype networks and heterogeneous networks, the mapping to node-colored networks is obvious, as each type is now called a ``color''. For interdependent networks and networks of networks (and related frameworks), one needs to map the networks into a flattened graph and then assign colors to nodes according to the subnetwork to which each node belongs.

One can represent node-colored graphs using our multilayer-network framework with $\aspects=1$ by considering each layer as a color. That is, we let $V=V_c$, $\layers=C$, $V_M=\{ (\nodeindex[1],c) \in V \times \layers | \chi(\nodeindex[1])=c  \} $, and $E_M=\{ ((\nodeindex[1],c_1),(\nodeindex[2],c_2)) \in V_M \times V_M | (\nodeindex[1],\nodeindex[2]) \in E_c  \} $. See Fig.~\ref{fig:nodecolored} for an example of such a mapping. Because each node has only a single color, this multilayer network is disjoint, but it does not have any further restrictions (i.e., constraints).  Alternatively, if it is not important to preserve the node names, one can choose to rename the nodes such that the nodes in each layer start from 1. To do this, we define a function $\Upsilon : V_c \rightarrow \{ 1, \dots ,n_c \}$ that names each node using an integer between 1 and the maximum number of nodes $n_c$ with the same color in such a way that two nodes of the same color do not share the same name [i.e., $\Upsilon(\nodeindex[1])=\Upsilon(\nodeindex[2]) \implies \chi(\nodeindex[1]) \neq \chi(\nodeindex[2])$)]. It then follows that $V=\{ 1, \dots ,n_c \} $, $\layers=C$, $V_M=\{ (\Upsilon(\nodeindex[1]),c) \in V \times L | \chi(\nodeindex[1])=c, \nodeindex[1] \in V_c  \} $, and $E_M=\{ ((\Upsilon(\nodeindex[1]),c_1),(\Upsilon(\nodeindex[2]),c_2)) \in V_M \times V_M | (\nodeindex[1],\nodeindex[2]) \in E_c  \} $. This mapping is illustrated in Fig.~\ref{fig:nodecolored}c. For node-colored network that are constrained in certain way, note that it is possible to do such a mapping in a way that guarantees that the resulting multilayer network has diagonal couplings. (This is useful, for example, for studies of interdependent networks~\cite{Buldyrev2010Catastrophic,Gao2011Robustness}.) See the discussion at the end of Section~\ref{multiplex}.

\subsection{Multiplex Networks and Multirelational Networks} \label{multiplex}

One can define a \emph{multiplex network}, a \emph{multirelational network}, and similar types of multilayer networks as a sequence of graphs~\cite{Wasserman1994Social,Sola2013Centrality,Nicosia2013Growing,Bianconi2013Statistical,Battiston2013Metrics}: $\{ G_{\layerindex[1]} \}_{\layerindex[1]=1}^\nlayers = \{ (V_{\layerindex[1]},E_{\layerindex[1]})\}_{\layerindex[1]=1}^\nlayers$, where $E_{\layerindex[1]} \subset V_{\layerindex[1]} \times V_{\layerindex[1]}$ is the set of edges and $\layerindex[1]$ indexes the graphs. Usually, the node sets are the same across the different layers (i.e., $V_{\layerindex[1]}=V_{\layerindex[2]}$ for all $\layerindex[1],\layerindex[2]$), or they at least share some nodes (i.e., $\bigcap_{\layerindex[1]=1}^\nlayers V_{\layerindex[1]} \neq \emptyset$). Alternatively, one can define multiplex networks as \emph{edge-colored multigraphs}, which are networks with multiple types of edges. One defines an edge-colored multigraph as a triple $G_e=(V,E,C)$, where $V$ is the node set, $C$ is the color set (which is used for labeling the type of edge), and $E \subset V \times V \times C$ is the edge set. We use a general definition of a ``color'' as a label, so edges that are incident to the same node are allowed to have the same color. In this definition of edge-colored multigraphs, a pair of nodes cannot be adjacent to each other via multiple edges of the same color. One can use edge-colored multigraphs to represent a set of multiple networks that have the same set of nodes in each layer by associating each layer with a unique color~\cite{Berlingerio2011Foundations,Coscia2013You}.

One can map structures that amount to sequences of graphs to our multilayer-network framework by mapping each graph in the sequence to a single intra-layer network. For an edge-colored multigraph, each edge color corresponds to a layer in a multilayer network. See Fig.~\ref{fig:kc2} for an example of a multilayer network representation and an edge-colored multigraph of the same network. This type of mapping only gives the intra-layer edges. However, in the literature, it is usually assumed implicitly that nodes are somehow coupled to their counterparts in different layers.  Our multilayer framework can incorporate explicit adjacencies of a node with itself across multiple layers, and we use such inter-layer edges to represent the coupling structure of layers (similar to Ref.~\cite{Mucha2010Communities}). Thus far, it has been very common to assume that nodes are adjacent in an identical manner to each of their counterparts in the different layers, which corresponds to categorical inter-layer couplings in the general multilayer-network framework. The other well-known type of coupling is \emph{ordinal} coupling, in which the layers are ordered and nodes are adjacent only to their counterparts in consecutive (``adjacent'') layers.  Ordinal coupling structure arises, for example, when temporal networks are mapped into a multilayer-network framework. This was discussed several years ago in Ref.~\cite{Mucha2010Communities} (also see the discussion in Ref.~\cite{DeDomenico2013Mathematical}), and a similar idea was used very recently in Ref.~\cite{wehmuth2014}. Although categorical and ordinal couplings are the most common situations, they are not the only ones. For example, for $d = 1$ aspect, one can also represent situations with more general inter-layer couplings either by using 4th-order tensors \cite{DeDomenico2013Mathematical} or by using both block-diagonal and off-block-diagonal 3rd-order tensors~\cite{Mucha2010Communities,Bassett2013Robust} (see Section~\ref{reps}).  By employing higher-order tensors, similar representations are possible for any number of aspects.
 
In the rest of this paper, we will use the term ``multiplex network'' in a manner that is similar to Ref.~\cite{DeDomenico2013Mathematical}: we consider all diagonally coupled multilayer networks in which each layer shares at least one node with some other layer in the network to be multiplex networks. That is, we include multilayer networks that are not node-aligned in our definition of multiplex networks, but we leave out layer-disjoint networks. In several respects, our use of the term ``multiplex'' is thus more general than its typical usage in the literature thus far (see Table~\ref{table:dictionary}). For example, we allow temporal networks (which have ordinal couplings) to be construed as multiplex networks.  Importantly, we also do not require all nodes to exist on every layer to have a multiplex network. Additionally, this definition of multiplex networks also includes multilayer networks with more than one aspect, because it is natural to represent certain multiplex structures using multilayer networks with more than one type (i.e., aspect) of layer. For example, in \emph{cognitive social structures}~\cite{Krackhardt1987Cognitive}, one aspect can be used for the set of layers that contain each person's perception of a social network and the other can be used to represent different types of relationships in these social networks. See Fig.~\ref{fig:cogsoc} for an illustration. Furthermore, as discussed in Section~\ref{ordinalcategorical}, temporal networks with multiple types of edges can be considered as multiplex networks with two aspects.

An archetypical example of a multiplex network is a social network in which the different layers (i.e., different edge labels) represent different types of social relationships~\cite{Wasserman1994Social,TTT,Szell2010Multirelational,moodyjimi2013}.  For example, one can place friendship ties, family ties, and coworker ties in different layers. Other examples include gene co-expression networks, protein interaction networks, and transportation networks.  In a gene co-expression network, each layer can represent a different tissue type or environment~\cite{Li2011Integrative}.  In a protein interaction network, each layer can include interactions from one of the many possible experimental protocols.  There are many types of multiplex transportation networks. For example, one can construct a multiplex air-transportation network whose individual layers contain routes from a single airline \cite{Cardillo2013Emergence,Cardillo2013Modeling}, a shipping network with different types of vessels in different layers \cite{blasius2010}, or a ground-transportation network in which each layer includes edges from a single mode of transportation.  In other types of multiplex transportation networks, such as a city's metropolitan system, each layer corresponds to a different ``line'' (e.g., the Tube in London has the Circle Line, the District Line, and many others) \cite{puck-cp,Cozzo2013Clustering}, though this example includes very different sets of nodes in different layers. See Section~\ref{data} for more discussion and examples of real-world multiplex networks.

In theory, one can also map the structures that are represented naturally as multiplex networks to the node-colored networks that we discussed in Section~\ref{nodecolored}. For example, consider a transportation network in which cities are adjacent via railroad lines, airplane routes, and shipping routes.  For simplicity, let's assume that each city has a single airport, a single train station, and a single port. When representing this situation using a node-colored network, the colors correspond to the three modes of transportation, and the nodes represent the airports, train stations, and ports (which are adjacent to each other if they are in the same city). In the multiplex-network representation of this same network, the nodes are the cities, and the three layers represent the three transportation modes. In our multilayer-network representation, these two ways of representing the data are almost but not quite equivalent. The difference is that the multiplex-network representation has node sets that are same across the layers, whereas the node-colored-network representation has disjoint node sets. In this type of node-colored network, any path\footnote{We use the convention of defining a ``walk'' as a sequence of adjacent nodes (and the associated edges). A ``path'' is a walk that has no repeated nodes, with the exception that the starting and ending node are allowed to be the same.  An \emph{intra-layer walk} is a walk that occurs only within a single layer, and an \emph{intra-layer path} is defined similarly.} whose edges are between nodes of a different color (i.e., inter-layer edges, which correspond to intra-city edges in the above example) cannot contain more than one node of any given color. That is, in the above example, it is not possible to (for example) start from an airport and go to another airport by following only intra-city edges. In undirected multiplex networks with only two layers, this condition is equivalent to enforcing both ``uniqueness'' and ``no-feedback'' conditions~\cite{Gao2011Robustness} for networks of networks (i.e., interdependent networks). Several authors have exploited this connection between the two representations to map dynamics on networks of networks to dynamics on multiplex networks~\cite{Son2011Percolation,Son2012Percolation,Baxter2012Avalanche,Bianconi2014Mutually} (see Section~\ref{cascades}). Note that some interdependent networks that include nodes without any inter-layer edges --- these are sometimes called ``partially-interdependent networks'' to distinguish them from ``fully-interdependent networks'' (see, e.g., Ref.~\cite{Son2011Percolation}) --- and one can use a similar argument to map them to multiplex networks that are not node-aligned.

\subsection{Hypergraphs} \label{hyper}

In a \emph{hypergraph} $H=(\mathcal{V},\mathcal{E})$, each \emph{hyperedge} in the hyperedge set $\mathcal{E}$ can connect any number of nodes (rather than being restricted to connecting exactly two nodes, as in a graph without self-edges), so $\mathcal{E} \subset \mathcal{P}^\prime(\mathcal{V}) \subset \mathcal{P}(\mathcal{V})$, where $\mathcal{P}(\mathcal{V})$ denotes the power set (i.e., set of subsets) of $\mathcal{V}$ and $\mathcal{P}^\prime(\mathcal{V})$ denotes the power set with subsets of cardinality larger than $1$.  (As with graphs, we are excluding self-edges, which correspond to subsets of cardinality $1$.  One can, of course, include self-edges if one wants.) One constructs a weighted hypergraph by associating a non-negative number to each hyperedge, and one constructs a directed hypergraph by associating each hyperedge with a sequence of nodes rather than a set of nodes. Any undirected hypergraph can be mapped to a categorically-coupled multilayer network: each hyperedge corresponds to a single layer in which all of the nodes in the hyperedge form a clique~\cite{Criado2011Mathematical}. Additionally, note that there exists a mapping between hypergraphs (with the possibility for multiedges) and bipartite graphs (see Section~\ref{othernets}): one set of nodes in the bipartite graph represents the nodes of the hypergraph, and the other set represents the edges.

A \emph{$k$-uniform hypergraph} is a hypergraph in which the cardinality of each hyperedge is exactly $k$, so each hyperedge represents a connection among exactly $k$ nodes.  This is relevant for folksonomies \cite{ghosh1,ghosh2} and other applications (e.g., finding interologs in protein interactions)~\cite{Michoel2012Alignment}. One can represent weighted, directed, $k$-uniform hypergraphs using adjacency tensors. Let the element $\mathcal{W}_{\nodeindex[1]_1 \dots \nodeindex[1]_k}^H$ be the weight of any existing hyperedge $(\nodeindex[1]_1, \dots, \nodeindex[1]_k) \in \mathcal{E}$, and assign a value of $0$ to any hyperedge that is not in the hyperedge set $\mathcal{E}$. This connection is useful, for example, for investigating spectral theory in adjacency tensors of hypergraphs~\cite{Pearson2013Spectral,RotaBulo2009New,Cooper2012Spectra}.

Multiplex networks have been studied by mapping them into directed $3$-uniform hypergraphs~\cite{Michoel2012Alignment}. A node-aligned multiplex network with node set $V$ and layer set $\layers$ is mapped to a 3-uniform hypergraph $H$ such that the node set in the hypergraph is $\mathcal{V}=V \cup \layers$ and $(\nodeindex[1],\nodeindex[2],\layerindex[1]) \in \mathcal{E}$ if and only if there is an edge between node-layers $(\nodeindex[1],\layerindex[1])$ and $(\nodeindex[2],\layerindex[1])$.  (That is, there needs to be an intra-layer edge between nodes $\nodeindex[1]$ and $\nodeindex[2]$ in layer $\layerindex[1]$.) It is also possible to define a mapping in the reverse direction (i.e., from $k$-uniform hypergraphs to node-aligned multiplex networks). Given $H$, one defines a multiplex network with $V=\mathcal{V}$ and $\layers_\aspectindex=\mathcal{V}$ for all $\aspectindex \in \{ 1 ,\dots , k \}$. One can construct similar maps for multiplex networks with an arbitrary number of aspects.

\subsection{Ordinal Couplings and Temporal Networks} \label{temporal}

One can represent a temporal network as a set of events or an ordered sequence of graphs~\cite{Holme2012Temporal,Holme2013Temporal}.\footnote{How many temporal networks can be defined in this way depends, e.g., on how restrictively one wants to define the notion of an ``event''.} This is also the case for temporal networks that are constructed from similarities in coupled time series \cite{Bassett2013Robust}, which can arise either from experimental data or from the output of a model. In the case of event sets, suppose that events are triples $e=(\nodeindex[1],\nodeindex[2],t)$, where $\nodeindex[1],\nodeindex[2] \in V$ are nodes and $t \in T$ is a time stamp of an event.  An event-based temporal network is equivalent to an edge-colored graph in which the set of colors is the set of possible time stamps (i.e., $C=T$) and the edges $E$ are the events in the network. The only difference between this structure and the edge-colored multigraphs that we defined in Section~\ref{multiplex} is that the set of ``colors'' is ordered instead of categorical. 

When using our general multilayer-network framework, one can be explicit about the fact that events or intra-layer graphs are ordered \cite{Mucha2010Community}. In this case, two identical nodes from different layers are adjacent via an inter-layer edge only if the layers are next to each other in the sequence. Furthermore, time's arrow can be incorporated into the network structure by using directed edges between corresponding nodes in different layers.  One can also allow a generalized ordinal coupling that includes a horizon $h$ by considering not only neighboring layers but all layers that are within $h$ steps.  In the context of temporal networks, one can view this construction as a time horizon.

The multilayer-network framework that we discussed in Section \ref{newform} allows non-diagonal couplings. In the context of temporal networks, one can use the edges that are associated with such couplings to represent delays in events. For example, consider an airline network in which a flight from airport $\nodeindex[1]$ leaves at time $t$ and arrives at airport $\nodeindex[2]$ at time $t+t_d$. In an event-based temporal-network representation, this flight is a quadruplet $(\nodeindex[1],\nodeindex[2],t,t_d)$, where the additional element $t_d$ represents a delay. In our multilayer-network framework, one can represent the associated edge as an inter-layer edge $((\nodeindex[1],t),(\nodeindex[2],t+t_d))$.

Our general multilayer-network framework also allows continuous time, whereas the tensor and the supra-adjacency matrix representations assume that there are a finite number of layers. (See Ref.~\cite{grindrod2013} for a discussion about using continuous versus discrete time for studying temporal networks.) Moreover, in some contexts, it might be desirable to distinguish between events that have a long duration from several consecutive events. For example, consider a mobile-phone calling network in which an event represents a call between two people, and edges in a layer correspond to all of the calls that take place within one minute. Clearly, two consecutive calls that each last one minute is a different situation from a single call that lasts two minutes, but one can represent each of these cases by using two edges in two consecutive layers. If there is a need to distinguish between these two cases, then one should add an intervening ``connecting'' layers between each pair of time points. In such a representation, the two consecutive calls would not include an edge in the connecting layer, whereas the continuous call would include such an edge.

\subsubsection{Networks with Both Ordinal and Categorical Aspects} \label{ordinalcategorical}

It is also possible to use our multilayer-network framework to represent multiplex networks that both have multiple types of edges and are temporal. In this case, $\aspects=2$, and one of the aspects is categorical and the other is ordinal. When using a tensor representation, this leads to 6th-order tensors (or 4th-order tensors if one chooses to use a lower-order representation; see Section~\ref{constrain}).

Examples of networks with both ordinal and categorical networks include transportation networks with multiple modes of travel (e.g., flights, trains, etc.) and different departure times. Additionally, social networks and communication networks are typically both multiplex and temporal. (See Fig.~\ref{fig:kc2} for an example of such social network.) There are currently very few studies that construct networks that explicitly incorporate both time stamps and edge types. One of them is the international trade networks in Refs.~\cite{Barigozzi2010Multinetwork,Barigozzi2011Identifying}. In this example, each layer corresponds to one year (out of 12 years) in one category (out of 97 categories).  Other studies have examined data sets with both temporal and multiplex features (e.g., the shipping data set in Ref.~\cite{blasius2010} as well as numerous social networks, such as the ones that were studied in Refs.~\cite{TTT,Szell2010Multirelational,andreacrime}), but such investigations typically have not assembled their data into a multilayer network that is both multiplex and temporal.  An old data set that is technically both multiplex and temporal is discussed in Ref.~\cite{Kapferer1972Strategy}, but it only has two time points and two types of edges.  

Some recent papers have explicitly incorporated both time-dependence and multiplexity. For example, Ref.~\cite{snijders2013} examined the dynamics of a stochastic actor-oriented model in a multiplex network in which a bipartite network (with actors adjacent to groups) coevolves with a unipartite multiplex network (which encapsulates interactions between the actors).  Additionally, Oselio et al. \cite{oselio2013} used stochastic blockmodeling to develop a method for inference in time-dependent multiplex networks.

\subsection{Other Types of Networks and Graphs} \label{othernets}

We now briefly discuss $k$-partite graphs, networks with both colored nodes and colored edges, multilevel networks, and some other networks structures. 

Define a \emph{$k$-partite network} as a tuple $G_k=(\mathcal{V}_k,E_k)$, where $\mathcal{V}_k=\{ V_i \}_{i=1}^k$ is a collection of $k$ pairwise disjoint sets of nodes (i.e., $V_i \cap V_j = \emptyset$ if $i \neq j$), such that each set $V_i$ represents nodes of a certain type; and $E_k \subset  \bigcup_{i=1}^k V_i \times \bigcup_{i=1}^k V_i $ is the set of edges, where edges are not allowed between nodes of the same type (i.e., $\nodeindex[1],\nodeindex[2] \in V_i \implies (\nodeindex[1],\nodeindex[2]) \notin E_k$ for any $i$).  Clearly, a $k$-partite graph is a special case of the node-colored graphs that we discussed in Section~\ref{nodecolored}. Each node type corresponds to a color, and the coloring is a proper node-coloring, so two nodes of the same color cannot be incident to the same edge. ``Bipartite'' (i.e., $2$-partite) networks have received considerable attention \cite{breiger1974,wilson1982,Newmanbook,Wasserman1994Social}, but tripartite and more general $k$-partite (aka, ``multi-mode'' or ``multipartite'') networks are also interesting, even though they have not been studied with close to the same intensity.  However, they have been used to investigate various social systems~\cite{Fararo1984Tripartite,Melamed2013Community,knoke2013}. Additionally, some scholars have examined multiplex bipartite networks and their unipartite projections to multiplex networks~\cite{Horvat2012Onemode,Horvat2013Fixed}.  In such a projected network, there are $\binom{\nlayers+1}{2}$ layers --- one for each combination $(\layerindex[1],\layerindex[2])$ of layers of the bipartite network (including combinations in which the same layer is included twice, for which $\layerindex[1]=\layerindex[2]$). There is an edge between a pair of nodes in layer $(\layerindex[1],\layerindex[2])$ if one node is adjacent to a second node in the bipartite network via an edge of type $\layerindex[1]$ and the other node is adjacent to that second node via an edge of type $\layerindex[2]$. Allard et al. \cite{Allard2012Bond} examined bond percolation (see Section \ref{connect}) on networks that are constructed via unipartite projections from node-colored bipartite graphs that are generated using a (generalized) configuration model. The projected networks include colors for both nodes and edges. Allard et al. also considered the projection of node-colored bipartite graphs to various other types of multilayer networks.

We discussed multilayer networks with multiple types of edges in Section~\ref{multiplex}, and we discussed multilayer networks with multiple types of nodes in Section~\ref{nodecolored}.  It is also possible for networks to include both of these features. A natural way to map such a network to our general multilayer-network framework is to consider node colors and edge colors as separate aspects. Such multilayer networks have been examined under the monickers of \emph{heterogeneous information networks}~\cite{Han2009Mining,Davis2011Multirelational,Sun2012Thesis,Sun2011PathSim} and \emph{coupled-cell networks with multiple arrows}~\cite{Stewart2003Symmetry,Golubitsky2005Patterns}.  Other authors have studied similar structures called \emph{meta-networks} (and associated \emph{meta-matrices}) \cite{CarleyStructural2001,Carley2007Toward,Tsvetovat2004DyNetML}  Additionally, see the research on topics such as ``semantic graphs''~\cite{Sowa1983Conceptual} and ``attributed relational graphs''~\cite{Brayer1975Web,Tsai1979Error}. Some of aforementioned multilayer networks possess an additional constraint: if two edges have the same color, then the pair of nodes that are incident to one of these edges must share the same color combination as the pair of nodes that are incident to the other edge \cite{Davis2011Multirelational,Sun2012Thesis,Sun2013Mining,Golubitsky2005Patterns}.  (For example, consider two red edges.  They can both be incident to one blue node and one green node, but it is not permissible for one edge to be incident to two blue nodes but the other to be incident to one blue node and one green node.)

\emph{Multilevel networks} are based on the application of ideas from ``multilevel analysis'' \cite{snijders2012book} to networks \cite{snijders2003,snijders1995}. One can use the general framework for multilevel networks that is described in Ref.~\cite{Wang2013Exponential} to represent networks in which nodes can have any finite number of types (i.e., ``levels'') and in which there can be edges between nodes of the same type or between nodes of ``adjacent'' types. Reference~\cite{Wang2013Exponential} included a mapping of all previous multilevel networks (see, e.g., Refs.\cite{hedstrom2000,snijders2003,Lazega2008Catching,Iacobucci1990Social}) to a two-level hierarchical network that is a special case of their general multilevel-network framework.  They used the term ``macro-level network'' for one level, ``micro-level network'' for the other level, and ``meso-level network'' for a network that consists of micro-level nodes, macro-level nodes, and exclusively inter-level edges. For example, a social network of researchers (micro-level) and a resource-exchange network between laboratories (macro-level) to which the researchers belong constitutes a multilevel network with two levels~\cite{Lazega2008Catching,Wang2013Exponential}. In many cases, the structure of multilevel networks is restricted even further --- for example, by only allowing only a single inter-layer edge from each micro-level node (see the discussion in Ref.~\cite{Wang2013Exponential} and references therein). Multilevel networks fit our multilayer-network framework naturally, as each level is a layer. The resulting structure amounts to a node-colored network in which only inter-layer edges between ``adjacent'' colors (i.e., consecutive layers) are allowed. Clearly, two-level networks are equivalent to node-colored networks with two colors.

A type of multilayer network of particular relevance for telecommunication networks (such as the internet) is a ``hierarchical multilayer network'', in which the bottom layer constitutes a ``physical'' network and the remaining layers are ``virtual layers'' that operate on top of the physical layer~\cite{pioro2004,Kurant2006Layered,Kurant2007Error,Pacharintanakul2009Effects,Mattia2013Polyhedral}. Such networks are typically similar to interdependent networks in that (1) a node in a given layer is dependent on a node in the layer below, and (2) a node in a given layer cannot be adjacent to another node in its own layer unless there is a path via the corresponding nodes in the layer below.

\section{Empirical Multilayer Networks}\label{data}

\begin{figure}[!htp]
\begin{center}
\includegraphics[width=1\linewidth]{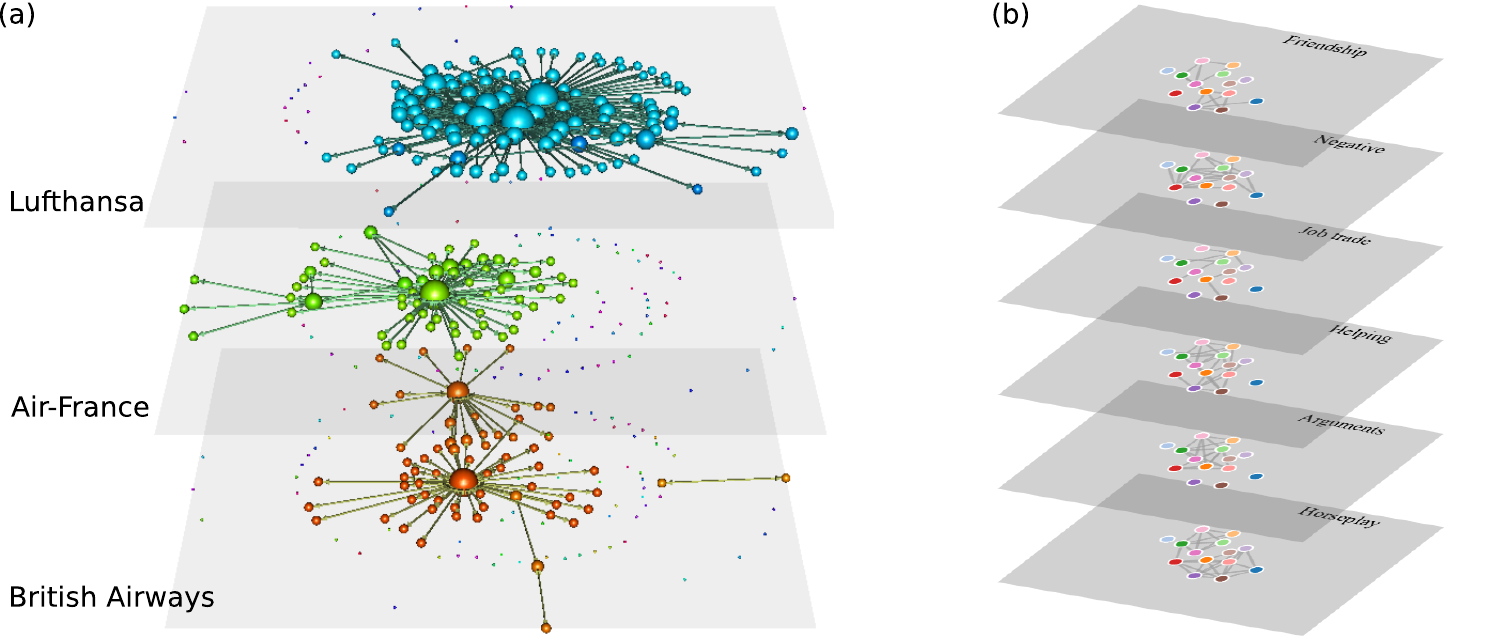}
\end{center}
\caption{Visualization of two multilayer data sets. (a) Air transportation network from Ref.~\protect\cite{Cardillo2013Emergence}.  Each layer contains the flights from a different airline.  We drew this network using a recently-developed visualization tool for multilayer networks \protect\cite{manlio-viz}. (b) Bank-wiring room network from Ref.~\protect\cite{Roethlisberger1939Management}. We represent each individual using a different node color, and each layer contains the ties from a different type of a relationship.  We drew this network using a recently-developed software library for the investigation (and visualization) of multilayer networks \protect\cite{mikko-viz}. In both visualizations, the layout of the nodes is calculated using an aggegrated network (which is monoplex). It is thus identical in all layers.
}
\label{fig:visuals}
\end{figure}

Developing representations and models for multilayer networks is useful for increasing the understanding of the structure and function of multilayer systems, and it can lead to discoveries of new phenomena that cannot be explained using a monoplex-network framework (see Section~\ref{models}). However, in order to understand how real-world multilayer networks behave and are organized, it is also necessary --- indeed, it is crucial --- to collect and study empirical data for which such frameworks are appropriate.  It is also helpful to develop new visualization tools, data structures,\footnote{It is important to use sparse data structures for general multilayer networks, because the number of possible edges grows as $\mathcal{O}(n^{2(d+1)})$. Additionally, it is best not to explicitly store the inter-layer edges even for single-aspect categorical multiplex networks because their number grows as $\mathcal{O}(n b^2)$.  Recall that $n = |V|$, the parameter $d$ is the number of aspects, and (for a single-aspect network) $b$ is the number of layers.} and computational methods \cite{mikko-viz} for multilayer networks. In Fig.~\ref{fig:visuals}, we show visualizations of two different multilayer networks.  In panel (a), we show the air transportation network from Ref.~\cite{Cardillo2013Emergence}; in panel (b), we show the bank-wiring room network of Ref.~\cite{Roethlisberger1939Management} (see Section \ref{sec:intro}).

By design, networks with multiple layers are able to encapsulate a much more detailed description of a system than monoplex networks.  This, in turn, yields significant new data-collection challenges, which need to be surmounted for ideas developed for multilayer networks to be genuinely useful for applications. For example, although it is clear that using a single type of edge between a pair of people does not always provide a suitable level of abstraction to study social networks~\cite{Wasserman1994Social}, data on empirical social networks is still primarily available in a format that is more suitable for monoplex networks than for multilayer ones. Small multiplex social-network data, which was predominantly collected manually either via questionnaires or by carefully observing a group of people, has been available for decades. See Ref.~\cite{UCINET1} for a several examples of such small data sets, including the well-studied Sampson monastery network \cite{sampson}).  More recently, large-scale multiplex social network data sets have been acquired automatically.  See Ref.~\cite{plexmath-data} for some example data sets (from air transportation \cite{Cardillo2013Emergence} and Twitter \cite{higgs-twitter2013,15M-2011}).  Methods for large-scale data collection include combining data from several social-networking sites~\cite{Magnani2013Formation}, observing several different types of interactions between players in a massive multiplayer online game (MMOG)~\cite{Szell2010Multirelational}, and using mobile-phone billing information to construct networks with calls and text messages considered as separate layers~\cite{Zignani2014Exploiting}. Most of the large-scale multiplex network data sets are node-aligned --- i.e., all of the nodes are present in all of the layers --- but, for example, there is also a large data set that covers relationships via multiple social-networking sites that yields layers that are not node-aligned~\cite{Buccafurri2013Bridge}.

Thus far, most empirical studies of multilayer networks have used data that can fit the multiplex-network framework.  In Table~\ref{table:multiplexdata}, we give a sample of data sets that have been used in the literature. Importantly, a small number of empirical studies have used other types of multilayer networks: four of the investigated examples of empirical interacting and interdependent networks are power stations and the internet servers on which they depend~\cite{Rosato2008Modelling,Buldyrev2010Catastrophic}, coupled power grids~\cite{Brummitt2012Suppressing}, coupled climate networks~\cite{Donges2011Investigating}, and interconnected transportation networks such as an airport and a railroad network~\cite{Halu2013Emergence} or an airport and port network~\cite{Parshani2010Intersimilarity}. There have also been empirical studies on hierarchical multilevel networks (see Section~\ref{othernets}), such as a social network between cancer researchers, their affiliations, and connections between the affiliations~\cite{Lazega2008Catching}. Nevertheless, the vast majority of studies on interconnected and interdependent networks have been theoretical.

We categorize the data sets in Table~\ref{table:multiplexdata} to facilitate exposition and to illustrate different approaches for generating multiplex data. (Obviously, our categorization is not definitive.) For example, one can map a network whose edges are labeled with text to a multiplex network by letting each layer correspond to a keyword that appears in its edge. Additionally, one can construe a system with nodes that belong to multiple different bipartite networks as a multiplex network by letting each layer be a unipartite projection of one of these bipartite networks. As discussed in Section~\ref{temporal}, most temporal networks can also be examined using a multiplex framework.  In Table~\ref{table:multiplexdata}, however, we only include temporal-network examples in which the authors of the articles that used the data explicitly employed a multilayer-network construction.

Most of the multiplex data that has been collected thus far has concentrated on intra-layer networks and has completely disregarded inter-layer connection strengths. Consider, for example, the problem of information diffusion in multiple social-networking websites, which together constitute a multiplex network in which each layer corresponds to a single site. In this type of network, it is typically straightforward to collect data on intra-layer edges, but describing diffusion of information in such systems requires somehow quantifying (ideally, in a direct way from real data) the relative rates at which information moves across layers versus within layers~\cite{Min2013Layercrossing}. For example, one can try to estimate transition probabilities between different layers by considering the relative amounts of time that are spent on activities in each layer. Similarly, in a transportation network that includes multiple modes of transportation (e.g., airline networks, railroad networks, and a single city's transportation system), the cost of changing modes needs to be quantified.  Such data is often measured --- e.g., Transport for London possesses excellent data on the amount of time that it takes to change lines in their metropolitan transportation system~\cite{duncanhorne} --- and it is important to collect or acquire such data to construct empirical estimates for inter-layer edge weights.

When constructing multilayer networks, it is essential to be creative about determining what constitutes the layers.  For example, a multilayer interbank network can include different types of credit relations in different layers \cite{bargigli2013}, a social network can include different types of social relations in different layers, a Twitter network can include different hashtags in different layers \cite{higgs-twitter2013,plexmath-data}, a communication network can include interactions that use different languages in different layers, a multilayer brain network can contain different layers for structural and functional networks (or different layers for brain activity that is measured using different modalities, different layers for different experimental subjects, and so on). Different layers in a multilayer network could also represent how much time people spend doing different activities (with transition probabilities between different activities to yield weights for inter-layer edges), different populations in a metapopulation model for biological epidemics, the different swinging platforms (where several oscillating metronomes are placed on each platform) in the experiments of Ref.~\cite{martens2013}, and so on.  For some applications, there might be some ``obvious'' way to identify layers in the construction of a multilayer network, but other times it might not be so obvious. We expect a multilayer-network framework to yield useful insights in both types of situations.  Moreover, it is also useful to develop methods to try --- given some data --- to identify different layers \cite{Chiang2007Layering,prescott2013}.

\begin{table}[!p]
\begin{adjustwidth}{-3cm}{-3cm}
\begin{center}
\begin{tabular}{|l|c|c|c|}
\hline
Name                          & Nodes                   & Layers                         & Refs.\\
\hline 
\multicolumn{3}{c}{Social networks with multiple types of ties} \\
\hline
Baboon social network         & Individuals (12)        & Interaction types (3)          & \cite{Barrett2012Taking} \\
MMOG social network           & Players (300000)        & Interaction types (6)          & \cite{Szell2010Multirelational,Szell2010Measuring} \\
Aarhus computer science department    & Employees (61)          & Social activities (5)          & \cite{Magnani2013Combinatorial} \\
Friendfeed network            & Users (7629)            & Social-networking services (3) & \cite{Magnani2013Formation} \\ 
Online forum social network   & Users (1899)            & Forum, instant messages (2)    & \cite{Halu2013Multiplex} \\
Top Noordin terrorist network & Terrorists (78)         & Relation type (4)              & \cite{Battiston2013Metrics} \\
Florentine families           & Families (16)           & Business, marriage ties (2)    & \cite{Breiger1986Cumulated,Cozzo2013Clustering}\\ 
Facebook (``Tastes, Ties, and Time'') & Users (1640)            & Friends, in same group, in same picture (3) & \cite{TTT}\\
\hline
\multicolumn{3}{c}{Coauthorship networks} \\
\hline
DBLP coauthorship (3)      & Authors (6771--558800)   & Conferences (6--2536)           & \cite{Berlingerio2012Multidimensional,Coscia2013You,Berlingerio2013ABACUS}\\
DBLP coauthorship          & Authors (10305)         & Publication categories (617)   & \cite{MichaekKwokPoNg2011MultiRank} \\
\hline
\multicolumn{3}{c}{Unipartite projections of bipartite graphs} \\
\hline
Youtube users              & Users (15088)           & Shared activities (5)          & \cite{Tang2012Community}\\ 
{\tt Extrarandom.pl}             & Users (4404)            & Shared activities (11)         & \cite{Brodka2012Analysis}\\
Netflix                    & Movies (13581)          & Rating category pairs (3)      & \cite{Horvat2012Onemode,Horvat2013Fixed}\\
\hline
\multicolumn{3}{c}{Temporal networks considered as multilayer networks} \\
\hline
Enron e-mail                & Users (184)             & Months (44)                    & \cite{Bader2007Temporal}\\ 
World trade                & Countries (18)          & Years (10)                     & \cite{Bader2007Temporal}\\ 
US Senate                  & Senators (1884)         & Congresses (110)               & \cite{Mucha2010Communities,Mucha2010Community} \\
IMDB                       & Actors (28042)         & Years (10)                     & \cite{Berlingerio2011Finding}\\
DBLP Coauthorship          & Authors (582179)        & Years (55)                     & \cite{Berlingerio2012Multidimensional}\\
\hline
\multicolumn{3}{c}{Layers based on keywords of text on edges} \\
\hline
Enron e-mail                & Users (500)             & Keywords (500)                 & \cite{Sun2009Multivis}\\
IBM social network             & Users (3679)             & Keywords (1000)                & \cite{Sun2009Multivis}\\
Webpages, anchor text (3)   & Web pages (560--$10^5$)& Anchor terms (533--39255)       & \cite{Kolda2006TOPHITS,Li2012HAR,Kolda2005HigherOrder}\\
\hline
\multicolumn{3}{c}{Layers based on different aspects of node similarities} \\
\hline
SIAM journals               & Articles (5022)         & Similarity types (5)          & \cite{Dunlavy2006Multilinear}\\
{\tt ArXiv} articles        & Articles (30000)        & Similarity types (4)           & \cite{Rocklin2013Clustering} \\
Corporate Governance network& Companies (273)         & Similarity types (3)          & \cite{Bonacina2014Multiple}\\
\hline
\multicolumn{3}{c}{Transportation networks} \\
\hline
Air transportation nets (3)& Airports (308--3108)     & Airlines (15--530)              & \cite{Cardillo2013Emergence,Cardillo2013Modeling,Cozzo2013Clustering} \\
London underground         & Stations (314)          & Lines (14)                     & \cite{puck-cp,Cozzo2013Clustering} \\
Global cargo ship network      & Ports (951)             & Ship types (3)                 & \cite{blasius2010} \\
\hline
\multicolumn{3}{c}{Other types and mixed types of networks} \\
\hline
Web search queries (2)     & Words (131268, 184760)   & Rank of clicked result (5, 6)   & \cite{Berlingerio2011Pursuit,Berlingerio2012Multidimensional,Berlingerio2011Foundations,Berlingerio2013ABACUS}\\
Flickr (2)                 & Users (1000, 1186895)    & Contacts, shared activities (11, 4) & \cite{Brodka2010Method,Kazienko2011MultidimensionalB,Berlingerio2011Pursuit,Berlingerio2012Multidimensional}\\
DBLP citations (2)         & Authors (6848, 10305)    & Publication categories (617)   & \cite{MichaekKwokPoNg2011MultiRank,Li2012HAR} \\
DBLP proceedings           & Authors                 & Conferences (70)               & \cite{Cai2005Community}\\ 
DBLP author network        & Authors (424455)        & Coauthorship, co-citations (3)  & \cite{Brodka2011Shortest} \\
Gene co-expression         & Genes                   & Experimental conditions (130)   & \cite{Li2011Integrative} \\ 
Cognitive social structures (2) & People (21, 48)          & People, perceptions (21, 47)    & \cite{Krackhardt1987Cognitive,Harrer2012Approach,Cozzo2013Clustering}\\
Global terrorist network       & Terrorist groups (2509) & Target country (124)           & \cite{Berlingerio2011Finding}\\
International trade network (2)   & Countries (15, 162)         & Years (3, 12), commodities (15, 97)   & \cite{Barigozzi2011Identifying,Barigozzi2010Multinetwork,Mahutga2013MultiRelational} \\ 
\hline
\end{tabular}
\end{center}
\end{adjustwidth}
\caption{Examples of multiplex data sets that have been used in the literature. We categorize the networks to make the table easier to read. A number in parentheses next to a network name indicates the number of different networks (when it is larger than 1). The numbers in parentheses next to the ``nodes'' and ``layers'' fields indicate the number of nodes and layers in those networks (if it is indicated in the cited articles). We separate the sizes of two networks by a comma, and we write the size range when there are three or more networks. For the international trade network, we use a comma to separate its two different aspects. This table only includes temporal networks that have been considered explicitly using a multilayer formalism. The temporal networks in this table have ordinal coupling, whereas all other networks in this table have categorical coupling (see Section~\protect\ref{temporal}). 
}
\label{table:multiplexdata}
\end{table}

\section{Models, Methods, and Dynamics}\label{models}

There has been considerable interest in generalizing concepts from monoplex networks 
to multilayer networks, and it is very important to do so. Ideas that have been examined include (but are not limited to) node degrees and neighborhoods, walks, clustering coefficients and transitivity, centrality measures, community structure, network models, and component connectivity.  Multilayer structures can have important effects on dynamical systems on networks, and examples such as percolation and spreading processes have been studied in detail to illustrate some of these effects. An increasingly large body of other dynamical processes have also been studied on multilayer networks. 

Thus far, almost all of the scientific literature on multilayer networks has concerned networks with only a single aspect. Accordingly, in this section, we assume that a multilayer network has only a single aspect unless we explicitly indicate otherwise. We also typically consider multiplex networks to be node-aligned unless we comment explicitly on this feature.

\subsection{Network Aggregation: From Multiplex Networks to Monoplex Networks}\label{agg}

A ``traditional'' way to examine systems with multiplexity is to construct a monoplex network by aggregating data from the different layers of a multiplex network and then study the resulting monoplex network \cite{DeDomenico2013Mathematical}.\footnote{Recently, Ref.~\cite{prescott2013} discussed the inverse problem: one can imagine starting with a monoplex network and trying to assign layers to nodes to obtain a multilayer-network representation.  Additionally, applying a ``filtration'' to a weighted monoplex network can, for example, yield a (hierarchical) multilayer network in which the first layer includes only the strongest edge, the second layer includes only the two strongest edges, and so on \cite{Lee2012weighted}.} One way to construct an aggregated network (which is also known as a superposition network~\cite{Gomez2013Diffusion}, overlapping network~\cite{Battiston2013Metrics}, or overlay network~\cite{DeDomenico2013Mathematical}) is to define edge weights between two nodes in the resulting monoplex network as a linear combination of the weights between those same nodes from each of the layers. This yields a weighted adjacency matrix ${\bf W}$ whose components are ${W}_{\nodeindex[1]\nodeindex[2]}=\sum_{\layerindex[1]=1}^\nlayers m_{\layerindex[1]} \mathcal{W}_{\nodeindex[1]\nodeindex[2]\layerindex[1]}$, where $\mathcal{W}$ is the 3rd-order weighted adjacency tensor of the multiplex network and ${\bf m} \in \mathbb{R}^\nlayers$ is a vector that weights the importances of different layers~\cite{Rocklin2013Clustering}.  If information about the relative importances of the layers is not available, then it is typical to set each of the weights to $1$ by taking ${\bf m} = (1,\dots,1)$. If desired, one can also normalize after the aggregation process.  For the choice ${\bf m} = (1,\dots,1)$ and an unweighted multiplex network, the weights in the aggregated network equal the number of different types of edges between pairs of nodes and the weighted network corresponds to a multigraph in which the edge layers are discarded.~\footnote{This aggregation scheme is equivalent to removing the edge colors in an edge-colored multigraph.} It is sometimes possible to indirectly infer the relative importances of layers; for example, one can do do by exploiting known community structure of a multilayer network~\cite{Cai2005Community,Rocklin2013Clustering}. Aggregation of temporal networks~\cite{Holme2012Temporal,Holme2013Temporal} is a special case of multilayer-network aggregation, and one needs to consider event-time statistics in that case \cite{hoffmann2012}.

Typically, aggregations of multiplex networks into monoplex networks tend to be performed manually rather than through an explicit mathematical computation. For example, the weighted version of the well-known Zachary Karate Club network~\cite{zachary_1977_model} was constructed by considering eight different contexts for which there exist relationships between pairs of nodes, although precise representations (e.g., in the form of adjacency matrices) for these eight networks were not reported. In some cases, it is desirable to disregard the weights of a monoplex network that one obtains by aggregating multiplex networks and to instead examine an unweighted monoplex network~\cite{Cardillo2013Emergence,Sola2013Centrality,Cardillo2013Modeling,Battiston2013Metrics} (References \cite{Criado2011Mathematical,Sola2013Centrality} used the term ``projected network'' to describe this situation, though it is preferable to use words like ``projection'' in a more mathematical sense --- such as in a one-mode projection of a bipartite network that is obtained by multiplying an adjacency matrix with its transpose \cite{Newmanbook}.) In this case, ${\bf m}$ is a binary vector, and the projection results in an unweighted adjacency matrix: $m_{\layerindex[1]}=1$ if layer $\layerindex[1]$ is included in the aggregation (i.e., ``projection'') and $m_{\layerindex[1]}=0$ if it is not. In this approach, there are $2^b-1$ nontrivial ways ($1$ for each combination of layers) of performing an aggregation, and the aggregation process can be construed as taking a union of the edges of a selected set of layers. A complementary approach was taken by Ref.~\cite{CorominasMurtra2013Detection}, who instead considered intersections of intra-layer edge sets with different combinations of layers.

An aggregation process can discard a lot of valuable information about an inherently multiplex system. For example, in the international trade network, the aggregated network that uses the total amount of trade between countries as edge weights cannot capture the richness of the structures in the multiplex network, for which each layer represents trading of different categories of products~\cite{Barigozzi2010Multinetwork,Barigozzi2011Identifying}. Another interesting example is an air-transportation multiplex network, for which each layer contains edges that represent connections between airports that arise from the flights of a single airline.  In this example, the layers that correspond to low-cost airlines have qualitatively different properties from those that correspond to major airlines ~\cite{Cardillo2013Emergence}.  It was also demonstrated recently that social networks and transportation networks include multiple types of transitivity, which cannot be inferred by exclusively studying a weighted network obtained from aggregation~\cite{Cozzo2013Clustering}. As these examples illustrate, aggregation can distort the properties of multiplex networks. In addition, in multilayer networks with nontrivial inter-layer coupling, the information related to transitions between layers (i.e., the inter-layer edges) disappears as a result of aggregation processes because self-edges cannot account for such transitions.\footnote{Among other phenomena, this implies that processes that are Markovian for a multilayer network need not yield Markovian processes on a monoplex network that one obtains via aggregation. There is a similar effect when considering a temporal network: a sequence of symmetric pairwise interactions nevertheless generally includes asymmetries as to which nodes can pass information (or a disease, etc.) to which other nodes \cite{Holme2005Network,Holme2012Temporal}.} However, despite the loss of information inherent in an aggregation process, there can be interesting relationships between inherently multiplex diagnostics and corresponding diagnostics calculated for aggregated networks. For example, the eigenvalues of the supra-adjacency matrix and combinatorial supra-Laplacian matrix of a node-aligned multiplex network (see Section~\ref{supra}) interlace with the eigenvalues of the weighted adjacency matrix and combinatorial Laplacian matrix\footnote{Note that the adjacency matrix is weighted, so the diagonal matrix ${\bf D}$ needs to be constructed using weighted degrees.} of the aggregated network that one constructs by counting the number of edges between pairs of nodes and dividing by the number of layers~\cite{SanchezGarcia2013Dimensionality}. (Also see Ref.~\cite{tsubakino2012} from the control-theory literature.) Similar results also hold for multiplex networks that are not node-aligned, but there are several ways of doing the averaging. Furthermore, a multiplex clustering coefficient (see Section~\ref{clustering}) with certain parameter values reduces to what one would obtain by using a weighted clustering coefficient on a weighted monoplex network~\cite{Cozzo2013Clustering}.

\subsection{Diagnostics for Multilayer Networks}

We now examine some of the diagnostics that have been developed for multilayer networks.  All of them are defined for multilayer networks with only a single aspect (i.e., $\aspects=1$), and most of them have been defined in the context of multiplex networks.

\subsubsection{Node Degree and Neighborhood.} \label{degree}

In monoplex networks that are undirected and unweighted, a node's ``degree'' (i.e., degree centrality) gives the number of nodes that are adjacent to it (i.e., the number of its immediate neighbors). Equivalently, a node's degree is the number of edges that are incident to it, and a focal node's ``neighborhood'' is the set of nodes that are reached by following those incident edges. One can generalize the notion of degree for directed networks to obtain in-degree and out-degree.  These indicate, respectively, the number of incoming and outgoing edges.  The weighted degree, or strength, of a weighted network is given by the sum of the weights of all edges that are incident to a node \cite{Newmanbook,Barrat2004Architecture}. There are several ways to generalize the notions of degree and neighborhood for multiplex networks \cite{Criado2011Mathematical,Battiston2013Metrics}, though one of course obtains the usual definitions if one considers only a single intra-layer network at a time. 

The simplest way to generalize the concepts of degree and neighborhood for multiplex networks is to use network aggregation. One can then define the degree as the number of edges of any type that are incident to a node and the neighborhood as the set of nodes that can be reached from a focal node by following any of those edges. Alternatively, one can threshold an aggregated network such that two nodes are considered to be adjacent if and only if the number of edges that connect them in a multiplex network is larger than some threshold value. This approach was taken by Refs.~\cite{Brodka2010Method,brodka2011,Brodka2012Analysis}, whose definition of the neighborhood of a node in a directed multiplex network is based on counting the number of different types of edges and taking into account the directions of the edges. References \cite{Brodka2010Method,brodka2011,Brodka2012Analysis} also defined several versions of degree centrality that use different normalization factors.

Similar to aggregating multiplex networks themselves, for which one can use any combination of layers in an aggregation process, it is possible to define degree and neighborhood in terms of a focal node and any subset of the layers. This makes it possible to define an aggregated multiplex degree or neighborhood of a node in $2^{\nlayers}$ different ways. This approach was taken in Refs.~\cite{Berlingerio2011Foundations,Berlingerio2011Pursuit,Berlingerio2012Multidimensional}, who defined the \emph{neighbors} $\Gamma(\nodeindex[1],D)$ of a node $\nodeindex[1]$ given a subset $D \subseteq \layers$ of the layers as the set of nodes that can be reached by following any edge that starts from node $\nodeindex[1]$ in any of the layers in $D$. Additionally, the \emph{neighbors-XOR} of node $\nodeindex[1]$ is $\Gamma_{\mathrm{XOR}}(\nodeindex[1],D)=\Gamma(\nodeindex[1],D) \setminus \Gamma(\nodeindex[1],\layers \setminus D)$, which gives the set of nodes that one can reach by following any edge that is incident to node $u$ in any of the layers in $D$ but which are unreachable if one starts in any layer that is not in the set $D$. These definitions give a starting point for developing other measures to quantify, for example, the level of redundancy in the layers of a multiplex network. Bianconi defined a notion of a ``multidegree'' starting from the concept of a ``multi-edge''~\cite{Bianconi2013Statistical}. One defines a \emph{multi-edge} using the binary vector ${\bf m} \in \{ 0,1 \}^{\nlayers}$ (which plays a similar role as ${\bf m} \in \mathbb{R}^{\nlayers}$ above), which gives the set of \emph{node pairs} $(\nodeindex[1],\nodeindex[2])$ (i.e., the set of multi-edges) for which a pair of nodes $\nodeindex[1]$ and $\nodeindex[2]$ are adjacent (when ${m}_{\layerindex[1]}=1$) and not adjacent (when ${m}_{\layerindex[1]}=0$) on layer $\layerindex[1]$. The \emph{multidegree} $k_{\nodeindex[1]}^{\bf m}$ of node $\nodeindex[1]$ is then the number of multi-edges that node $\nodeindex[1]$ has with vector ${\bf m}$, and the \emph{multistrength} $s_{\nodeindex[1],\layerindex[1]}^{\bf m}$ is the sum of the weights of those edges in the intra-layer network of layer $\layerindex[1]$~\cite{Menichetti2013Weighted}. The number of different vectors ${\bf m}$ (and, equivalently, the number of possible subsets $D \subseteq \layers$ grows exponentially with the number of layers.  If the number of layers is large, such growth can cause both computational problems and difficulties with interpretation of results. To help alleviate these problems, it is useful to define the concept of \emph{overlap multiplicity} $\nu({\bf m})$ \cite{Menichetti2013Weighted}, which is given by the number of $1$ entries in the vector ${\bf m}$. One can then average multiplex quantities defined for ${\bf m}$ over all ${\bf m}$ for which $\nu({\bf m})=\nu$ (e.g., for degree, one would define $k_{\nodeindex[1]}(\nu)=\binom{b}{\nu}^{-1}\sum_{\nu({\bf m})=\nu} k_{\nodeindex[1]}^{\bf m}$). For a given quantity, this entails taking $\nlayers +1$ different averages.

\subsubsection{Walks, Paths, and Distances.} \label{paths}

Walks and paths --- and their lengths --- are important concepts in both graph theory and network science. The ability to define generalizations of such concepts for multilayer networks yields natural extensions of many other measures for monoplex networks --- including graph distance, connected components, betweenness centralities, random walks, communicability~\cite{Estrada2008Communicability}, and clustering coefficients. One can then use some of these concepts, such as betweenness centralities and random walks, to define additional tools, such as methods for community detection or centrality measures. (See Ref.~\cite{DeDomenico2013Centrality} for an example of such an approach.) Furthermore, one can use the notion of a shortest path (i.e., a ``geodesic path'') to develop new measures that are intrinsic to multiplex networks.  One example of this is \emph{interdependence}, which is defined as the ratio of the number of shortest paths that traverse more than one layer to the total number of shortest paths~\cite{Morris2012Transport,Nicosia2013Growing}.
 
When defining a \emph{walk} on a multilayer network, one should be able to answer at least two basic questions: 
\begin{enumerate}
\item{Is changing layers considered to be a step in the multilayer network? In other words, are different copies of the same node in different layers considered to be a set of distinct objects such that it ``costs'' something to change layers~\cite{Cozzo2013Clustering}?}
\item{Is there a difference between taking intra-layer steps in different layers?
} 
\end{enumerate}
In many of the existing notions of multilayer-network walks, these decisions have been implicit, and they have depended on the types of systems that were considered. For example, the costs of inter-layer steps are often very small in social networks (though this is not always the case), but they might be substantial in transportation networks \cite{Cozzo2013Clustering}. Thus far, it has been extremely difficult to quantify these costs in empirical data (see Section \ref{data}), and whether one should explicitly consider the costs of inter-layer steps depends on their values relative to other scales in a system. Clearly, it is also desirable to try to find ways to estimate such values from data even if one cannot obtain them directly. 

If the answer to the first of the above questions is ``yes'', then a step and a walk are each defined as occurring between a pair of node-layer tuples. This approach has been used to generalize concepts like random walks~\cite{DeDomenico2013Random} (see Section~\ref{spreading}), clustering coefficients~\cite{Cozzo2013Clustering} (see Section~\ref{clustering}), several centrality measures~\cite{DeDomenico2013Centrality} (see Section~\ref{centrality}), and communicability~\cite{Estrada2013Communicability}. When using this approach, it is often natural to generalize concepts from monoplex network by simply replacing nodes with node-layer tuples. For example, one can calculate centrality measures and other node diagnostics for each node-layer tuple separately and then obtain a single value for a node via some method of aggregation (e.g., by summing the values). Additionally, if there is a cost when changing layers, then one needs to ask what values they take. In some types of networks, such as transportation networks, it is possible to estimate costs of inter-layer steps using empirical data in straightforward manner (see Section~\ref{data}).  In other situations, it can be difficult to determine reasonable values for the costs of changing layers.

\emph{Labeled walks}~\cite{Pattison1993Algebraic} (i.e., \emph{compound relations}~\cite{Lorrain1971Structural}) are walks in a multiplex network in which each walk is associated with a sequence of layer labels. In such a situation, one can define a \emph{walk length} that takes intra-layer steps into account in at least three different ways: each intra-layer step is of equal length (i.e., it does not depend on the layer), step lengths in different layers are comparable but can be weighted differently in different layers, or step lengths in different layers are incomparable. Reference \cite{Magnani2013Pareto} considered the last alternative and defined the length of a walk as a vector that counts how many steps are taken in each of the layers. They defined a path to be \emph{Pareto efficient} if there are no paths that are better in at least one of the vector elements and are equally good in all of the other elements. The \emph{Pareto distance} between two nodes is then given by a set of distance vectors (rather than a scalar value) corresponding to Pareto-efficient paths. 

As discussed in Section \ref{agg}, a straightforward way of generalizing any network diagnostic is to first aggregate a multiplex network to obtain a monoplex network and then apply tools that have already been developed for weighted networks. This approach was taken by Br\'{o}dka et al.~\cite{Brodka2011Shortest}, who formulated several ways of aggregating a network and only kept edges whose aggregated weight was above some threshold. They then defined an aggregated distance for each edge to calculate shortest paths in an aggregated network.

There are natural ways to define walks, paths, and path lengths in other types of multilayer networks besides multiplex networks. For example, Sahneh et al.~\cite{Sahneh2012Effect} decomposed all walks in an interconnected network (i.e., a node-colored network) into classes $(l_1,\dots,l_N)$: there are first $l_1$ intra-layer steps, then one inter-layer step, then $l_2$ intra-layer steps, and so on. They then defined the lengths of a walk to be the sum of the total number of intra-layer and inter-layer steps. Similarly, Sun et al.~\cite{Sun2011PathSim} defined ``metapaths'' (and one can similarly define ``metawalks'') of multilayer networks that have labels on both nodes and edges as sequences of node labels and edge labels. (See Section~\ref{othernets} for a discussion of the ``heterogeneous information networks'' that were examined by publications such as \cite{Sun2011PathSim}.)  An ``instance'' of a metawalk is then a walk in a network that respects those labelings.

\subsubsection{Clustering Coefficients, Transitivity, and Triangles.}\label{clustering}

A local clustering coefficient~\cite{Watts1998Collective,Newmanbook} measures transitivity in a monoplex network. One way to define it is by using the fraction of existing adjacencies versus all possible adjacencies in (i.e., the \emph{density} of) the neighborhood of a node. As discussed in Section \ref{degree}, the concept of neighborhood and the existence of pairwise connections (i.e., pairwise adjacencies) both become ambiguous in multilayer networks, as there are multiple ways to define these ideas in such settings. A second way to define a clustering coefficient is by using the ratio of closed triples (i.e., triangles) to connected triples. However, triangles also cannot be defined in a unique way in multilayer networks.  Indeed, as discussed in Ref.~\cite{Irving2012Synchronization}, there are many possible triples of nodes that contain 3 nodes and 2 layers. A third alternative is to define a monoplex clustering coefficient starting from the idea of 3-cycles (i.e., closed paths of length 3).  There are again multiple ways to define walks and paths in multiplex networks (see Section~\ref{paths}), so using this perspective also requires care. Consequently, attempting to define a multilayer clustering coefficient runs into even more significant complications than trying to define weighted~\cite{Saramaki2007Generalizations} or directed \cite{Fagiolo2007Clustering} clustering coefficients.

Despite the clear difficulties in defining clustering coefficients for multilayer networks (especially if one also considers directions and/or weights), there have been several attempts \cite{Brodka2010Method,Criado2011Mathematical,Brodka2012Analysis,Barrett2012Taking,Cozzo2013Clustering,Battiston2013Metrics,Donges2011Investigating,Parshani2010Intersimilarity,Podobnik2012Preferential} to define a notion of multilayer clustering coefficients (and, more generally, to develop notions of transitivity for multilayer networks). Most of these definitions are for multiplex networks. The definitions in Refs.~\cite{Brodka2010Method,Brodka2012Analysis} are not based on the standard definition of the local clustering coefficient~\cite{Watts1998Collective} for unweighted monoplex networks, and the definition in Ref.~\cite{Barrett2012Taking} is not normalized between $0$ and $1$.  Criado et al. \cite{Criado2011Mathematical} developed generalizations using the density-based perspective of a monoplex local clustering coefficient, and they defined a pair of local clustering coefficients  starting from two alternative ways of defining a neighborhood of a node in a multiplex network. Cozzo et al. \cite{Cozzo2013Clustering} formulated several definitions for a multiplex clustering coefficient based on different ways of defining a walk and a 3-cycle in a multiplex network.  They also compared the properties of their new notions and several existing notions.  As discussed in Ref.~\cite{Cozzo2013Clustering}, multiplexity induces novel notions of transitivity --- because, for example, one can close a triangle either on the layer in which a length-3 path starts or on a different layer --- and different notions of transitivity might be more appropriate in different situations (e.g., in social versus transportation networks).  Notions such as ``structural holes'' \cite{ronburt} that one can relate to local clustering coefficients \cite{borgatti97,Newmanbook} also need to be generalized to account for multilayer settings.  

There have been a few attempts to define clustering coefficients for node-colored graphs (and equivalent structures). Parshani et al.~\cite{Parshani2010Intersimilarity} defined an ``inter-clustering coefficient'' for interdependent networks in which each node has an inter-layer degree of $1$. Reference~\cite{Podobnik2012Preferential} defined a ``cross-clustering coefficient'' for arbitrary two-layer interdependent networks (i.e., for node-colored networks in wich all edges are allowed).

\subsubsection{Centrality Measures.}\label{centrality}

In the study of networks, a ``centrality'' measure attempts to measure the importance of a node, an edge, or some other subgraph \cite{Wasserman1994Social,Newmanbook}.  It is desirable to calculate centralities in multilayer networks \cite{jung2007}, and several monoplex-network centrality measures have now been generalized to multiplex networks. For example, one can generalize PageRank (i.e., PageRank centrality)~\cite{Page1999}, which is based on the stationary distribution of a random walker on a network (with ``teleportation'' \cite{smartteleport}, if the network is not strongly connected) by defining a random-walk process on a multilayer network. PageRank has been generalized for undirected multiplex networks so that a random walker can either take steps inside a layer or change layers, and both nodes and layers receive a rank~\cite{MichaekKwokPoNg2011MultiRank}. Halu et al.~\cite{Halu2013Multiplex} defined an alternative multiplex version of PageRank: random walks on one network layer are biased so that they take into account the normal PageRank values of some other layer. PageRank has also been generalized to node-colored networks with two distinct layers by considering a random walk in which intra-layer and inter-layer steps have different probabilities~\cite{Zhou2007CoRanking}.

Another popular centrality measure that has been generalized for multiplex networks is hyperlink-induced topic search (HITS)~\cite{Kleinberg1999Authoritative}, which produces both hub and authority scores for the nodes. The HITS algorithm can be expressed as a singular value decomposition (SVD) of the adjacency matrix of a network. This approach was utilized by Kolda et al.~\cite{Kolda2005HigherOrder,Kolda2006TOPHITS}, who developed a modified version of the HITS algorithm based on the tensor-decomposition method PARAFAC (which is a generalization of the matrix SVD). Li et al.~\cite{Li2012HAR} considered a random walker in directed multiplex networks that yields hub and authority scores for nodes via an algorithm that is similar to HITS. Myers et al.~\cite{myers2013} have also developed a multilayer generalization of hub and authority centrality.

Defining a random-walk process on a multilayer network yields several other centrality measures in addition to PageRank and HITS. De Domenico et al.~\cite{DeDomenico2013Centrality} developed generalizations of PageRank, random-walk occupation centrality, random-walk betweenness centrality, and random-walk closeness centrality based on a random walk process for a general single-aspect multilayer network.  In the same paper, they also used the tensorial framework for multilayer networks developed in Ref.~\cite{DeDomenico2013Mathematical} to define generalizations of eigenvector centrality, Katz centrality, and HITS centrality. Their methods yield a centrality measure for each node-layer tuple, and one can then calculate aggregate centralities for each node by summing (or aggregating in some other way) the node-layer centralities from each layer.

As we discussed in Section \ref{agg}, multiplex-network projections and aggregations make it possible to apply any monoplex methods to the resulting monoplex network. Sola et al.~\cite{Sola2013Centrality} took this approach and defined several generalizations of eigenvector centrality.  Some of their generalizations yield centrality values for node-layer tuples, whereas others only give aggregate centrality values for nodes. Sola et al. defined yet another eigenvector-centrality measure by constructing an $\nnodes \nlayers \times \nnodes \nlayers$ matrix based on the intra-layer adjacency matrices and then computing its eigenvectors.  (Recall that $\nnodes$ is the number of nodes in each layer and that $\nlayers$ is the number of layers.)

Other multiplex centrality measures that have been developed include multiplex versions of geodesic node betweenness centrality~\cite{Magnani2013Combinatorial,DeDomenico2013Centrality} and geodesic closeness centrality~\cite{DeDomenico2013Centrality}, which can each be calculated for any definition of shortest path (see Section \ref{paths}), and the centrality measure defined by Coscia et al.~\cite{Coscia2013You}, which ranks individuals in social networks based on how close they are to other people with some weighted ``skill'' set. Aguirre et al.~\cite{Aguirre2013Successful} studied how different inter-layer connection strategies in node-colored networks (i.e., networks of networks) affect the ratios of mean centralities of node-layer tuples from two different layers. Very recently, communicability was also generalized for multiplex networks \cite{Estrada2013Communicability}.

\subsubsection{Inter-Layer Diagnostics.} \label{interlayer}

Thus far, we have almost exclusively discussed generalizations of monoplex-network quantities for multilayer networks.  Indeed, most methods and diagnostics that have been developed thus far for multilayer networks are direct generalizations of monoplex quantities. However, we expect that the new ``degrees of freedom'' (via inter-layer relationships) that result from the introduction of multiple layers will yield interesting methods and diagnostics for multilayer networks that do not have counterparts in monoplex networks.  Although many papers have already illustrated various fascinating insights that arise from the ``new physics'' of multilayer network, it is also crucial to develop fundamentally new tools and techniques that take advantage of the structure of such networks.

One way to develop new quantities for multiplex networks is to compare the intra-layer networks of two or more layers. For example, one can compare two layers to each other by counting the number of edges that they share --- this quantity was called \emph{global overlap} in Ref.~\cite{Bianconi2013Statistical}, and a similar quantity called the \emph{global inter-clustering coefficient} was defined in Ref.~\cite{Parshani2010Intersimilarity} --- or by calculating the correlation between (weighted) adjacency-matrix elements of a pair of layers~\cite{Barigozzi2010Multinetwork}. One can calculate the \emph{degree of multiplexity} for a multiplex network by counting the number of node pairs that have multiple edge types between them divided by the total number of adjacent node pairs~\cite{kapferer1969}. (One can analogously calculate a node's degree of multiplexity by considering all pairs between a node and its neighbors.) One can also calculate correlations (e.g., via assortativity coefficients) between node diagnostics such as degree or local clustering coefficient~\cite{Barigozzi2010Multinetwork,Nicosia2013Growing,Parshani2010Intersimilarity,DeDomenico2013Centrality,Goh2013prl}. We discuss models of multiplex networks with built-in inter-layer correlations in Section~\ref{mplexmodels}.

Apart from examinations of correlations of network structures between layers, such as the ones that we just discussed, there has only been a little bit of work on genuinely multilayer diagnostics.  One example of an inherently multilayer diagnostic is the \emph{interdependence}~\cite{Morris2012Transport,Nicosia2013Growing}, which is the ratio of shortest paths in which two or more layers are used to the total number of shortest paths (see Section \ref{paths}). Moreover, most existing multilayer diagnostics have been designed for node-aligned multiplex networks (i.e., networks in which all nodes exist on every layers), though there are a few exceptions.  For example, Ref.~\cite{SanchezGarcia2013Dimensionality} defined the \emph{multiplexity degree} of a node as the number of layers in which the node exists, and Ref.~\cite{Buccafurri2013Bridge} studied online social networks that are not node-aligned by comparing nodes with a multiplexity degree of 1 to so-called ``bridge'' nodes, which have a multiplexity degree larger than $1$.

If one construes different communities in a network as belonging to different layers of a layer-disjoint multilayer network (e.g., interconnected networks, node-colored networks, etc.) \cite{melnik-hetero}, then one can interpret quantities like assortativity and modularity as inter-layer diagnostics.

\subsection{Models of Multiplex Networks}\label{mplexmodels}

A straightforward way to construct generative ensembles of synthetic multiplex networks is to consider any monoplex network model, such as an Erd\H{o}s-R\'enyi (ER) random-graph ensemble~\cite{Erdos59} or the configuration model~\cite{Bekessy72,Boll01,Newman01}, and use it to construct intra-layer networks that are independent of each other. One can then incorporate dependencies between layers by, for example, considering nodes that have joint degree distributions~\cite{Funk2010Interacting,Lee2012Correlated,Min2013Network} or by directly adding an arbitrary number of shared edges across layers~\cite{Funk2010Interacting,Marceau2011Modeling}. One can also add inter-layer correlations by starting from a multiplex network in which intra-layer networks are generated independently of each other and then changing the node identities in one of the layers (i.e., by relabeling the nodes in that layer) in order to introduce inter-layer correlations~\cite{DeDomenico2013Centrality}.

Some exponential random graph models (ERGMs)~\cite{Frank86,Robins07a,Robins07b} can handle multilayer networks such as multilevel networks \cite{Wang2013Exponential} (see Section~\ref{othernets}) and multiplex networks~\cite{Pattison1999Logit,lazega1999,robins2013}. For example, one can write the probability of any multiplex network $G_M$ to occur in an ERGM using the exponential form $P(G_M)=\exp\{{\boldsymbol\theta} \cdot {\bf f}(G_M)\}/Z({\boldsymbol\theta})$, where ${\bf f}(G_M)$ is a vector of network diagnostics (e.g., the number of triangles that contain different types of edges), ${\boldsymbol\theta}$ is a vector that represents the model parameters, and $Z({\boldsymbol\theta})$ is a normalization function.  One application of ERGMs to multiplex networks is in a recent study of the influence and reputation of interest groups \cite{heaney2014}.

Bianconi~\cite{Bianconi2013Statistical,Menichetti2013Weighted} introduced microcanonical and canonical network ensembles~\cite{Park2004Statistical,Squartini11} for multiplex networks. Microcanonical ensembles contain only networks that satisfy some strict set of constraints, whereas canonical network ensembles are based on maximizing Shannon entropy conditional on satisfying constraints only on average~\cite{Jaynes1957Information}. Deriving a canonical network ensemble thereby leads to exponential random graphs in single-layer networks~\cite{Park2004Statistical}.
Such ensembles were used recently to model interacting (i.e., node-colored) spatial networks and multiplex spatial networks. (A \emph{spatial network} is a network that is embedded in some space \cite{barth-review}.) The latter were reported to have higher values of edge overlap (see Section~\ref{interlayer}) than multiplex networks whose component layers are independent of each other (i.e., for which the intra-layer edges are independent across layers)~\cite{Halu2013Emergence}. In their examples, the authors of Ref.~\cite{Halu2013Emergence} embedded all nodes into the same space (independent of the layer).

It is also useful to define other types of generative models for multiplex networks.  For example, one can generalize monoplex-network attachment mechanisms such as preferential attachment~\cite{Price1976General,Barabasi1999Emergence} to multiplex networks.  Criado et al. \cite{Criado2011Mathematical} defined a set of models in which a multilayer network grows by adding new layers such that only a random subset of nodes participates in each layer. References~\cite{Goh2013prl,Nicosia2013Growing,Magnani2013Formation} examined multiplex-network models in which networks grow by addition of new nodes and new edges are created via preferential attachment.  In these models, the adjacencies between nodes are determined by applying a preferential-attachment rule such that the probability of a new node in a layer to form an intra-layer adjacency to any other node is proportional to a function of the intra-layer degrees of the node in all of the layers.  (Such a function is sometimes called an ``attachment kernel''.) Both Nicosia et al.~\cite{Nicosia2013Growing} and Kim and Goh~\cite{Goh2013prl} studied two-layer multiplex networks and attachment rules in which the attachment kernels were affine.\footnote{Recall that a function $f: \mathbb{R} \rightarrow \mathbb{R}$ is called ``affine'' if it is of the form $f(x) = c_1x+c_2$.  The standard ``linear preferential attachment'' model of de Solla Price \cite{Price1976General} has an attachment kernel that is an affine function of node degree. The kernel does not depend on anything else (such as an intrinsic node quality).} Additionally, Nicosia et al.~\cite{Nicosia2013Growing} allowed a node to be created at different times in different layers. Subsequently, Nicosia et al. considered the growth of multiplex networks via preferential attachment with nonlinear attachment kernels \cite{nicosia-growing2}. In the model introduced in Magnani et al.~\cite{Magnani2013Formation}, each intra-layer network is updated via an affine preferential attachment rule \cite{Bollobas2003Directed} that only considers intra-layer node degrees from that layer or by copying edges from some other layer. All of these studies~\cite{Goh2013prl,Nicosia2013Growing,Magnani2013Formation} found that their models can lead to positive correlations in intra-layer degrees across layers. Obviously, one can also construct multilayer versions of any other attachment mechanism.

\subsection{Models of Interconnected Networks}\label{interconnectedmodels}

One can also generalize generative models for monoplex networks for other multilayer settings, such as interconnected networks. Such models are very useful for studies of dynamical processes that occur on top of multilayer networks (see Section~\ref{dynamical}), because they can facilitate the derivation of (approximate) analytical results for important qualitative features of the dynamics. There is now a large body of research on network models from several flavors of multilayer networks that are very similar to each other (but which use different terminology for the layers). Terminology for an individual layer includes ``interacting network''~\cite{Leicht2009Percolation}, ``node color''~\cite{soderberg03,Soderberg2003Random,Soderberg2003Properties}, ``node type''\cite{Soderberg2002General,Newman2003Mixing,Allard2009Heterogeneous}, and ``module''~\cite{Dorogovtsev2008Organization,melnik2011}. One can also generate models of interconnected networks using notions such as block models \cite{doreian,airoldi2010} and mixture models \cite{mclachlan2000,leicht2007}.

A trivial way to construct interconnected (i.e., node-colored) networks is to start from a monoplex-network model and add edges uniformly at random in order to connect nodes from different layers. For example, this approach was taken to connect (via inter-layer edges) sets of regular lattices~\cite{Li2012Cascading,Bashan2013Extreme,Stippinger2013Enhancing}, networks produced by ER random graphs~\cite{Dong2013Robustness,Gao2013Percolation}, configuration-model networks~\cite{Dorogovtsev2008Organization}, and Barab\'asi-Albert (BA) networks~\cite{Tan2014Traffic}. In previous studies of interdependent-network models, it has been common to connect a pair of networks such that the resulting inter-layer network is undirected and the degree of each node in it is either $1$ or $0$. See Section~\ref{cascades} for details and Section~\ref{multiplex} and Refs.~\cite{Son2011Percolation,Baxter2012Avalanche,Son2012Percolation} for discussions of how one can also construe such models as models of multiplex networks. Importantly, note that inter-layer edges need not be added uniformly at random, as one can use other random (or deterministic) process to connect networks to each other. For example, Ref.~\cite{Aguirre2013Successful} studied how different inter-layer connection strategies affect the ratios of mean centralities of node-layer tuples from two different layers, and Ref.~\cite{Wang2011Effects} examined how they affect SIR spreading (see Section~\ref{spreading}). The Barab\'asi-Albert model~\cite{Barabasi1999Emergence}, which connects nodes via a preferential-attachment rule~\cite{Price1976General}, has also been generalized for interdependent networks~\cite{Podobnik2012Preferential,Jiang2014Effect}.

A natural extension of the monoplex configuration model to interconnected (i.e., node-colored) networks is to specify multiple degree distributions using multiple variables. As in the usual configuration model, the ends of edges (i.e., the stubs) are then connected to each other uniformly at random. References~\cite{Leicht2009Percolation,Allard2009Heterogeneous} defined a model in which $P_{\layerindex[1]}(k_{1}, \dots, k_{\nlayers})$ is the probability that a node on layer $\layerindex[1]$ has exactly $k_{\layerindex[2]}$ neighbors in layer $\layerindex[2]$. S\"{o}derberg~\cite{soderberg03,Soderberg2003Random,Soderberg2003Properties} started from a multi-degree distribution $P(k_1, \dots, k_b)$ that is independent of the layer and a mixing matrix ${\boldsymbol\tau}$ with elements $\tau_{\layerindex[1]\layerindex[2]}$ that gives the relative abundances of edges between layers $\layerindex[1]$ and $\layerindex[2]$. Similarly, Newman~\cite{Newman2003Mixing} defined a degree distribution $P_{\layerindex[1]}(k)$ for each layer and a mixing matrix similar to that in Refs.~\cite{soderberg03,Soderberg2003Random,Soderberg2003Properties}. Gleeson~\cite{gleeson2008} examined a connection-probability matrix ${\bf P}$ whose elements $P_{\layerindex[1]\layerindex[2]}(k)$ specify the probability that a node on layer $\layerindex[1]$ has $k$ edges to nodes on layer $\layerindex[2]$. Additionally, see Ref.~\cite{barbour2011} for an ER model for node-colored graphs (i.e., one connects a pair of nodes with colors $\layerindex[1]$ and $\layerindex[2]$ with probability $p_{\layerindex[1]\layerindex[2]}$) and Ref.~\cite{Allard2012Bond} for the definition of a configuration model for node-colored bipartite graphs.

One can also generate a synthetic ensemble of interconnected networks that incorporates both intra-layer and inter-layer degree-degree correlations by defining a model that is specified by the probability $P_{\layerindex[1]\layerindex[2]}(k,k')$ of selecting an edge that connects a node of degree $k$ in layer $\layerindex[1]$ to a node with degree $k'$ in layer $\layerindex[2]$~\cite{melnik-hetero}. Similarly, one can define a model with additional correlations whose input is the probability $P_{\layerindex[1]\layerindex[2]}(k_{\layerindex[1]\layerindex[1]},k_{\layerindex[1]\layerindex[2]},k'_{\layerindex[2]\layerindex[2]},k'_{\layerindex[1]\layerindex[2]})$ of the existence of an edge between nodes in layers $\layerindex[1]$ and $\layerindex[2]$ such that the node in layer $\layerindex[1]$ is adjacent to $k_{\layerindex[1]\layerindex[1]}$ nodes in layer $\layerindex[1]$ and $k_{\layerindex[1]\layerindex[2]}$ nodes in layer $\layerindex[2]$, and a node in layer $\layerindex[2]$ is adjacent to $k'_{\layerindex[2]\layerindex[2]}$ nodes in layer $\layerindex[2]$ and $k_{\layerindex[1]\layerindex[2]}$ nodes in layer $\layerindex[1]$~\cite{SaumellMendiola2012Epidemic}.

\subsection{Communities and Other Mesoscale Structures}

Community detection is one of the most popular topics in network science~\cite{Porter2009,Fortunato2010Community}. Roughly, the idea of community detection is to (algorithmically) find sets of nodes that are connected more densely to each other than they are to the rest of a network. Even in monoplex networks, there is no universally accepted definition of what constitutes a community (nor is it appropriate for there to be such a universal definition). Instead, there are numerous notions of a network ``community'' --- most of which are not actually definitions in a mathematical sense --- and associated computational heuristics to find them. This situation even messier in multilayer networks, for which there exist multiple viable generalizations even of concepts as simple as degree.

The earliest attempts to find structures similar to communities --- as well as more general types of mesoscale features --- in a multilayer network comes from the social-networks literature, where \emph{blockmodels} were used to find sets of nodes that all share similar connection patterns~\cite{Wasserman1994Social,doreian,Batagelj1997Notes,Harrer2012Approach}. Blockmodeling, which includes both deterministic and stochastic varieties, differs from community detection in that blockmodels do not necessarily seek densely-connected sets of nodes with sparse connections between sets but instead allow (in principle) any kind of connectivity pattern as long as most (or all) of the nodes in a block share a similar pattern.  One can also find mesoscale network features by trying to assign roles (i.e., colors or labels) to nodes in a network \cite{roleequiv}. One can also use multilayer ideas in the formulation of a stochastic blockmodel \cite{peixoto2013} or start with a monoplex network and try to assign layers to construct a multilayer-network representation \cite{Chiang2007Layering,prescott2013}. 

\subsubsection{Community Structure in Multilayer Networks}

Despite the plethora of research on community structure in monoplex networks, the development of community-detection methods for multilayer networks is in its infancy.  (Additionally, much more work is needed on mesoscale features other than community structure in both single-layer and multilayer networks.) Thus far, only a few community-detection methods have been generalized for multilayer networks. One of the best-known of these generalizations was formulated by Mucha et al.~\cite{Mucha2010Community,Mucha2010Communities}, who generalized the objective function known as ``modularity''~\cite{Newman2004Finding} to the ``multislice'' version of multilayer networks so that each node-layer tuple is assigned separately to a community. This makes it possible for the same entity to be in separate communities depending on the layer. Maximizing modularity is a computationally hard problem, and computational difficulties become even more acute in multilayer networks, as the system size now scales not only in the number of nodes but also in the number of layers. Higher-order tensors become large much faster than matrices as the number of nodes per layer increases, so computational tractability becomes increasingly challenging as one considers multilayer networks with a larger number of aspects. This challenge is especially prominent when representing a temporal network as a multilayer network, because the number of layers can be very large in such situations. To help address the computational challenges of optimizing modularity in multilayer networks, Ref.~\cite{Carchiolo2011Communities} developed a version of the multislice modularity-optimization method in Ref.~\cite{Arenas2007Size} that reduces the size of a network by grouping nodes such that the reduced network does not change the optimal partition and the modularity values of all of the partitions of the reduced network are the same as the corresponding partitions in the original network.  One can then apply modularity optimization to a smaller multilayer network and thereby consider larger multilayer networks than would otherwise be possible.  

Another challenge in optimizing multislice modularity is the construction of appropriate ``null models'' for multilayer networks.  When optimizing modularity in monoplex networks, the choice of null model quantifies what it means for a network to have sets of nodes (i.e., communities) that are connected to each other more densely than what would be expected ``at random''.  The null model is typically given by a random-graph ensemble, which specifies what it means for connections to have arisen by chance.  Naturally, a larger variety of null models are possible for multilayer networks than for monoplex networks, and the choice of null model has a large effect on the results of community detection. Reference \cite{Bassett2013Robust} developed several new ``null-model'' network ensembles for multislice modularity, though there remains much more work to do in this direction.  Optimization of multislice modularity has been used in multislice representations of temporal networks to study phenomena such as political-party reorganization \cite{Mucha2010Community,Mucha2010Communities}, the dynamics of behavioral \cite{wymbs2012} and functional brain networks \cite{bassett2011,bassett2013core}, and the qualitative behavior of time-series output of coupled nonlinear oscillators \cite{Bassett2013Robust,bassett2014}.  Multislice modularity has also been used to represent ``Kantian fractionalization'' in a multiplex network of countries, as there is a positive correlation between the value of multislice modularity and the future conflict rate between countries~\cite{Cranmer2014Kantian}.  It is possible to use the framework of multislice networks from Refs.~\cite{Mucha2010Community,Mucha2010Communities} to generalize community-detection methods other than modularity to multilayer networks. Reference.~\cite{cucu-multiplex2013} took this approach to examine \emph{group synchronization} (which has nothing to do with the concept of ``synchronization'' from dynamical systems), in which one tries to determine the elements of a (mathematical) group from noisy measurements of their pairwise ratios.\footnote{For the elements $g_i$ and $g_j$ of a group, a pairwise ratio is $g_{ij} = g_ig_j^{-1}$.}

Another important (and old \cite{Porter2009}) method for community detection in monoplex networks is spectral clustering. Michoel and Nachtergaele~\cite{Michoel2012Alignment} generalized spectral clustering and the Perron-Frobenius theorem to hypergraphs, and they applied their method to multiplex networks by mapping them to hypergraphs (see Section \ref{hyper}). Li et al.~\cite{Li2011Integrative} extended the framework of identifying ``heavy subgraphs'' (i.e., induced subgraphs of weighted networks with high values for the sum of the edge weights) for weighted multiplex networks by defining a \emph{recurrent heavy subgraph} as a multiplex subnetwork that is spanned by a subset of nodes and a subset of layers in a multiplex network such that the sum of edge weights in that subgraph is large. For computational reasons, Li et al. allowed fuzzy memberships of nodes and layers in such subgraphs.

One way to seek multiplex communities is to take advantage of monoplex community-detection methods and start by separately detecting communities in each intra-layer network. Barigozzi et al.~\cite{Barigozzi2011Identifying} took this approach and started by detecting communities in each layer via modularity optimization.  They examined an international trade network in which each layer corresponds to a different product category. They compared their results to communities that they found in an aggregated version of their multiplex network, and they observed considerable variation in the intra-layer communities across category layers.  It thus seems that much of the information about multiplex communities can be lost by aggregation a multilayer network, which underscores the importance of developing techniques for studying such networks without having to aggregate them into a monoplex network. Berlingerio et al.~\cite{Berlingerio2013ABACUS} also started by detecting intra-layer communities, and they defined a multiplex community as a ``closed frequent itemset''\footnote{An ``itemset'' is simply a set of items; see Refs.~\cite{Agrawal1993Mining,Agrawal1994Fast} for a precise definition of a ``closed frequent itemset''.} in which each node represents a single transaction and the items are tuples $(c,\layerindex[1])$, where $c$ is the community to which a node is assigned in layer $\layerindex[1]$. In network terms, they defined a multiplex community as a set of intra-layer communities such that at least some predefined number of nodes (the so-called ``support value'' of the community) are shared in all of those intra-layer communities. A multiplex community is only considered to be ``valid'' when there does not exist another multiplex community with the same support value that is a superset of the first community.

Obviously, one can exploit existing methods for monoplex community detection to find communities after aggregating multiplex networks into a single monoplex network~\cite{Berlingerio2011Finding}. Alternatively, one can examine community structure in multiplex networks by employing a network-aggregation process that considers all possible $2^{b}$ combinations of layers~\cite{Magnani2013Combinatorial}. Tang et al. \cite{Tang2012Community} identified two additional ways of using existing community-detection methods to study multiplex communities.  These methods interpolate between detecting communities in an aggregated network (``network integration'') and detecting them in individual intra-layer networks (``partition integration''). For example, Tang et al. defined ``utility integration'' as a method for calculating ``utility matrices'' of a community-detection method for each layer separately. They then optimized an objective function for a multilayer utility matrix that they obtained by summing all of the individual-layer utility matrices. Note that the definition of a single-layer utility matrix depends on the employed community-detection method. For modularity, for example, the utility matrix is the modularity matrix ${\bf Q}$ \cite{newman2006pre} of an intra-layer network. One then obtains the aggregate modularity value $Q$ as the sum $Q=\sum_{ij}{Q}_{ij}$. 

One can also apply ``inverse community detection'' to multiplex networks~\cite{Cai2005Community}. The basic idea behind inverse community detection is that one is given a ground-truth community structure for a multiplex network, and one then seeks an optimal linear combination of the weights ${\bf m}$ to use when aggregating the layers (see Section~\ref{agg}) so that the community structure in the aggregated network corresponds as closely as possible to the ground-truth communities.  Rocklin et al. \cite{Rocklin2013Clustering} formulated more complicated ways for defining the quality of network-aggregation weights when a ground-truth community structure is available. They also defined a ``metaclustering'' approach in which they first produced multiple different weighted networks with random aggregation weights, then clustered the resulting weighted networks, and finally calculated (using variation of information) a matrix of distances between the pairs of different clusterings.  They subsequently used a hierarchical-clustering method \cite{Porter2009} to obtain communities from this distance matrix.

Naturally, there are numerous other clustering techniques that can also be useful for examining communities in multilayer networks. For example, Str\"{o}ele and collaborators have taken various approaches from a data-mining perspective \cite{Stroele2009Mining,stroele2011,stroele2012} for clustering a multi-relational data set of scientific collaborations in Brazil.

\subsubsection{Methods Based on Tensor Decomposition.}\label{tensorcommunities}

Similar to using the SVD of an adjacency matrix to find communities in a monoplex network, one can employ tensor-decomposition methods~\cite{Kolda2009Tensor} to detect communities in a multiplex network~\cite{Kolda2005HigherOrder,Dunlavy2006Multilinear,Bonacina2014Multiple}. Several different tensor-decomposition methods amount to generalizations of the SVD to tensors. CANDECOM/PARAFAC (CP) is among the ones that can be used for multiplex community detection. It decomposes a tensor into $R$ factors, such that the tensor elements are approximated as $\mathcal{A}_{\nodeindex[1]\nodeindex[2]\layerindex[1]} \approx \sum_r^R x_{\nodeindex[1]r}y_{\nodeindex[2]r}z_{\layerindex[1]r}$, where ${\bf x},{\bf y} \in \mathbb{R}^{\nnodes \times R}$ and ${\bf z} \in \mathbb{R}^{\nlayers \times R}$. Each factor corresponds to a community, the nodes with the highest values in the columns of ${\bf x}$ and ${\bf y}$ correspond to nodes in the community, and the layers with the highest values in the columns of ${\bf z}$ correspond to layers in the community~\cite{Kolda2005HigherOrder,Dunlavy2006Multilinear}.

Other tensor decomposition methods, such as three-way DEDICOM or a Tucker decomposition, can also be used to detect communities in multiplex networks~\cite{Bader2007Temporal,Sun2009Multivis}. The three-way DEDICOM is similar to a blockmodeling approach because it finds classes of nodes that have similar connection patterns to other classes. (Importantly, nodes within a class may or may not be connected densely to each other.)  Gauvin et al. examined community structure in 3rd-order tensors using nonnegative tensor factorization \cite{gauvin2013}. Additionally, as discussed in Section~\ref{hyper}, one can map any node-aligned multiplex network to a $k$-regular hypergraph. One can then apply tensor-based clustering methods that have been developed for hypergraphs~\cite{Lin2009MetaFac} to study communities in multiplex networks.

\subsection{Dynamical Systems on Multilayer Networks} \label{dynamical}

One of the main reasons to study dynamical systems on networks is attempt to improve understanding of how nontrivial connectivity affects dynamical processes on networks (and conversely how dynamical processes affect network structure).  This can be very complicated, and there are interesting new wrinkles when studying dynamics on multilayer networks. It is very important to develop a deep understanding of such dynamics as well as how to design control strategies to achieve desired outcomes.

It has been illustrated repeatedly that multilayer variants of dynamical processes that have been studied on monoplex networks exhibit behavior that cannot be explained by examining one intra-layer network at a time or by aggregating layers.  Completely new phenomena can occur, and one of the primary challenges in the study of multilayer networks is to discern how features like multiplexity affect dynamical processes. Most of the multilayer dynamical processes that have been studied thus far have been examined using familiar mathematical machinery, such as generating functions and spectral theory (and such methods are, of course, subject to similar limitations as for monoplex networks), though we expect that new methods that incorporate more ideas from tensor algebra (and from geometry) will be necessary to obtain a complete understanding of dynamical systems on multilayer networks.

\subsubsection{Connected Components and Percolation.}\label{connect}

In an undirected monoplex network, a \emph{connected component} is a maximal set of nodes that are all connected to one another via some path. One can use the same definition for multilayer networks by allowing paths that are composed of all possible types of edges.  It is possible to use generating functions characterize the component-size distribution for the monoplex configuration model via a mean-field approximation~\cite{cohen2000,Callaway2000Network,Newmanbook,newman2007pre}.\footnote{The use of generating functions requires a network to be ``locally tree-like''~\cite{melnik2011}, so this calculation is only valid in that situation.}  One can take a similar approach for a configuration model defined on node-colored graphs, for which one also specifies a joint degree distribution that indicates the number of edges that connect nodes of different colors~\cite{soderberg03,Soderberg2003Random,Soderberg2003Properties,Newman2003Mixing,Vazquez2006Spreading,Allard2009Heterogeneous}. (Additionally, see Section~\ref{interconnectedmodels} and Ref.~\cite{barbour2011,Allard2012Bond} for a discussion of node-colored bipartite graphs and their unipartite projections.)  Therefore, one can also calculate the component-size distribution for equivalent structures such as interacting networks~\cite{Leicht2009Percolation}.

Percolation processes are among the simplest types of dynamics that can occur in a monoplex network \cite{Newmanbook}, so it is natural that they have already been investigated extensively in multilayer networks.  In \emph{site percolation} (i.e., \emph{node percolation}), one lets each node of a network be either occupied or unoccupied, and one construes occupied nodes as ``operational'' and unoccupied nodes as ``nonfunctional''.  In \emph{bond percolation} (i.e., \emph{edge percolation}), it is instead the edges that are either occupied (i.e., operational) or unoccupied (i.e., nonfunctional). As with the notion of a connected component, it is straightforward to generalize concepts from percolation theory --- such as the emergence (or destruction) of a giant connected component (GCC) as a function of the number of occupied nodes or edges --- to a multilayer-network framework. Note that it is typical to formulate percolation processes such that nodes or edges are removed from the network instead of labeling them as unoccupied. It is also often convenient to use a network diagnostic (e.g., mean degree) as a control parameter instead of the number of occupied nodes or edges.  This perspective is very useful for percolation processes in multilayer networks.

Many scholars have now studied percolation processes on multiplex networks. For example, Refs.~\cite{Cho2010Correlated,Lee2012Correlated} studied percolation in node-aligned multiplex networks that consist of two layers of ER networks whose nodes have intra-layer degrees that are correlated across the layers. They found that negative correlations in degrees increase the percolation threshold, whereas positive correlations decrease it. Moreover, for maximally positive correlations of intra-layer degrees --- i.e., when each node has exactly the same intra-layer degree --- the network contains a GCC for any nonzero edge density~\cite{Lee2012Correlated}. It was later demonstrated that multiplex networks with positively correlated intra-layer degrees are also more robust with respect to ``biconnectivity'' (i.e., by considering two nodes to be in the same component if and only if there are at least two paths between them), but that multiplex networks with negatively correlated intra-layer degrees are more resilient against attacks that are targeted based on the total degree (i.e., the sum of intra-layer degrees) of nodes \cite{Min2013Network}.  In this traditional type of targeted attack, nodes are removed in order from highest degree to lowest degree, though one can of course also target nodes based on other features.  Guha et al.~\cite{Guha2014Layered} considered percolation on multiplex networks in which all of the layers are subgraphs of some underlying network. They determined each subgraph by selecting nodes uniformly at random from the underlying network.

\subsubsection{Percolation Cascades.} \label{cascades}

\begin{figure}[!htp]
\begin{center}
\includegraphics[width=1\linewidth]{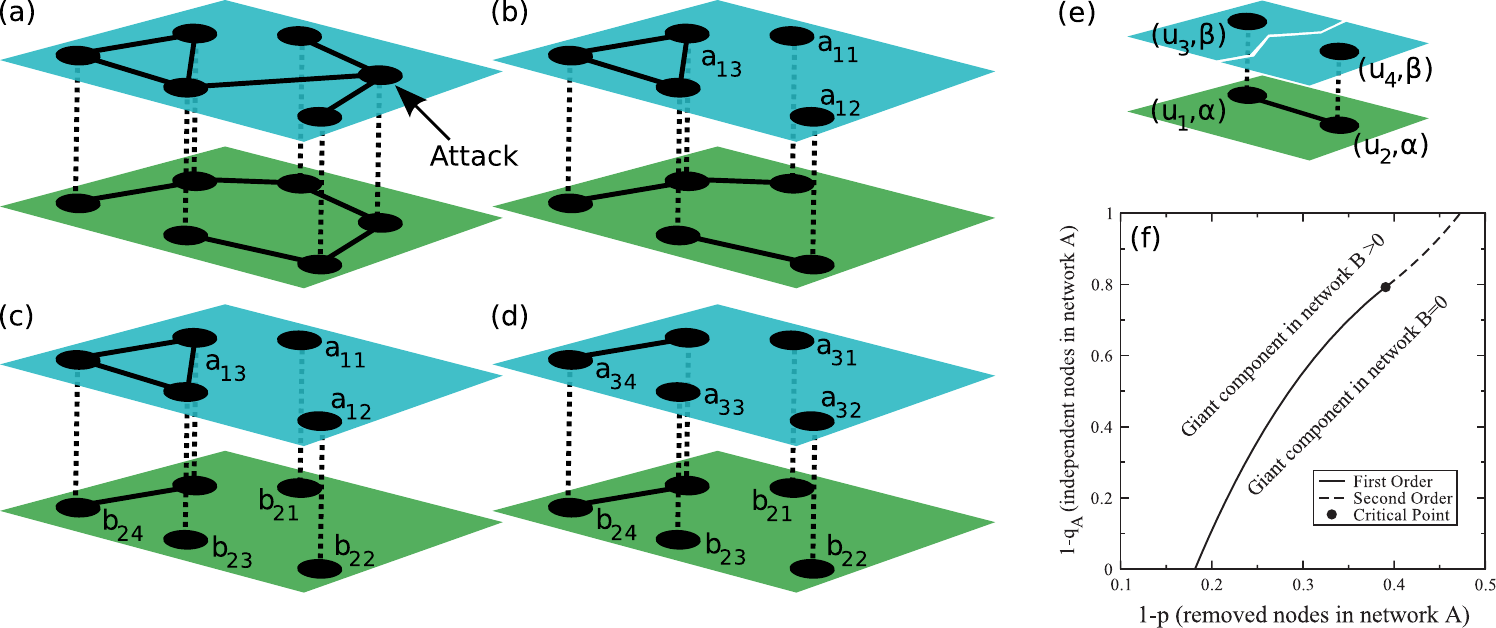}
\end{center}
\caption{(a--d) Schematic that illustrates, using a multilayer-network framework, the percolation process on interdependent networks that was studied in Ref.~\protect\cite{Buldyrev2010Catastrophic}.  We draw the interdependent networks using a layout that is similar to the one that we used for multiplex networks in Fig.~\protect\ref{fig:kc2}. In fact, one can map the network in the present figure to a multiplex network (see Refs.~\protect\cite{Son2011Percolation,Son2012Percolation,Baxter2012Avalanche,Bianconi2014Mutually} and Section~\protect\ref{multiplex}). (a) A node from the top layer is attacked.  (b) We then remove this node, along with all of its intra-layer edges, from both layers. (c)  From the bottom layer, we remove the intra-layer edges that are between nodes that are adjacent to nodes that are now in different components in the top layer. (d) We then remove dependency intra-layer edges from the top layer that are between nodes that are now in different components in the bottom layer. This process then continues --- alternating between the two layers --- and one divides the two networks into progressively smaller components until reaching a stationary state in which the nodes in connected components in each of the layers depend only on nodes that are in the same component in the other layer. [The inspiration for panels (a)--(d) is a figure in Ref.~\protect\cite{Buldyrev2010Catastrophic}, which illustrates the same mechanism but instead uses the language of interdependent networks.] (e) Schematic that illustrates the situation of an adjacent pair of nodes in one layer that are adjacent (e.g., via dependency edges) to nodes from different components of another layer. (f) A percolation phase transition for two interdependent networks, where the process that we described in panels (a)--(d) has been relaxed so that some fraction of the nodes in one layer are not dependent on any nodes in the other layer. The quantity $1-p$ is the fraction of nodes that have been removed from one network, and $1-q_A$ is the fraction of independent nodes in layer $A$ (and $q_B=1)$. The two intra-layer networks are ER graphs with average degree $3$. [In panel (f), we have reprinted a figure from Ref.~\protect\cite{Parshani2010Interdependent} with permission from the authors and the publisher. Copyright (2010) by the American Physical Society.]
}
\label{fig:cascade}
\end{figure}

In the percolation processes on node-colored networks that we discussed in Section~\ref{connect}, intra-layer edges and inter-layer edges are semantically equal, and a path that connects two nodes can include both types of edges. Buldyrev~et al.~\cite{Buldyrev2010Catastrophic} defined a cascade process in which intra-layer edges (called \emph{connectivity edges}) are defined in the same way as for monoplex networks, but inter-layer edges (which are called \emph{dependency edges}) encode dependencies between nodes. In panels (a)--(d) of Fig.~\ref{fig:cascade}, we illustrate their percolation process using a multilayer-network framework.  In Fig.~\ref{fig:cascade}e, we show a schematic that illustrates an intra-layer edge between a pair of nodes, which are adjacent (e.g., via dependency edges) to nodes from different components of another layer. Buldyrev~et al. studied multilayer networks with two layers, arbitrary intra-layer degree distributions, and inter-layer adjacencies that can exist between a node in one layer and its counterpart in the other layer. In their cascade process, one starts by removing a fraction $1-p$ (where $p \in [0,1]$) of the nodes uniformly at random [see panel (a)].  As we show in panel (b), one then divides the remaining nodes into disjoint sets according to their connected component in one specific layer (the proverbial ``first'' layer). One then updates the intra-layer network of the other layer (the equally proverbial ``second'' layer) by removing intra-layer edges between nodes that are adjacent to nodes from the first layer that are now in different components in that layer [see panel (c)] As we show in panel (d), the cascade then continues by removing intra-layer edges in the first layer that are between nodes that depend on nodes from different components in the second layer and by updating the components of the first layer accordingly. This process then continues --- alternating between the two layers --- and one divides the two networks into progressively smaller components until reaching a stationary state in which the nodes in connected components in each of the layers depend only on nodes that are in the same component in the other layer\footnote{If one is only interested in a giant component, then one can instead use a similar process, in which one removes nodes from a given layer if they are not in that layer's intra-layer GCC (which is guaranteed to be unique in the $n \rightarrow \infty$ limit). One then updates the other layer by removing all of the nodes from that second layer that were dependent on the nodes that have been removed from the first layer.  One then repeats this whole process for the second layer, and then does it again for the first layer, and so on.}. Buldyrev et al. studied this process in a mean-field setting by applying a generating-function formalism recursively until they reached a steady state. They reported that the system undergoes a first-order phase transition with respect to the control parameter $p$. From a big-picture perspective, an important result of Ref.~\cite{Buldyrev2010Catastrophic} is an illustration that interdependent networks with very heterogeneous intra-layer degree distributions can be less robust (for this type of percolation processes) than interdependent networks with more homogeneous distributions, which is the opposite of what has been observed for random failures (i.e., uniformly at random) using analogous network ensembles in monoplex networks \cite{Callaway2000Network,cohen2000,Albert2000Error,Newmanbook}. It was later suggested~\cite{Son2011Percolation} that the interdependencies between two networks might actually make the transition less steep than similar transitions in ordinary percolation for some networks, although the transition is discontinuous if the intra-layer networks are ER graphs.  See Refs.~\cite{Berezin2013Comment,Son2013Comment} for further discussion and debates on the issue of steepness, which has been controversial. The cascade process that we described above has received considerable attention in the last few years~\cite{Gao2012Networks}. For example, it has been extended to consider attacks on nodes that are targeted by degree (rather than the failures of nodes determined uniformly at random) \cite{Huang2011Robustness,Dong2012Percolation,Gao2013Percolation}.

The percolation mechanism that was studied in Ref.~\cite{Buldyrev2010Catastrophic} has been relaxed so that some fraction $1-q$ of the nodes in one layer are not dependent on any nodes (i.e., they are ``independent'' nodes) in the other layer~\cite{Parshani2010Interdependent}. In this situation, it was reported that there exists a critical fraction of dependency edges between nodes from different layers (i.e., a critical fraction of ``interdependent node pairs''), below which the percolation transition is second-order (i.e., continuous) and above which it is first-order (i.e., discontinuous) when the intra-layer networks are ER networks~\cite{Parshani2010Interdependent} (see Fig.~\ref{fig:cascade}f). If the intra-layer networks are produced from a configuration model with a power-law degree distribution, there is also a region of parameter space that exhibits interesting behavior that has been described as a ``hybrid'' transition~\cite{Zhou2013Percolation}.  In this regime, the size of the mutual GCC (which contains nodes from each layer) jumps discontinuously at a finite critical value of the percolation parameter $p$ to a very small but nonzero value and there is also a continuous transition in the size of the mutual GCC as $p \rightarrow 0$. Another situation in which percolation has been studied in interdependent networks includes a ``support-dependence relationship'', in which nodes are incident to more than one inter-layer edge and a node is removed only when all of its inter-layer neighbors are nonfunctional (i.e., have been removed in the percolation process)~\cite{Shao2011Cascade}. Reference~\cite{Hu2011Percolation} considered a two-layer node-aligned multiplex network in which one layer has connectivity edges and the other has dependency edges. Additionally, Refs.~\cite{Schneider2013Towards,Valdez2013Triple} examined strategies for choosing which nodes should not be interdependent (i.e., which nodes should be  ``autonomous'') for a network to be robust to failures. 

Reference~\cite{Buldyrev2010Catastrophic} included analytical calculations for two-layer networks in which both intra-layer networks were either ER networks or configuration-model networks with a power-law degree distribution. They considered interdependent networks by placing inter-layer edges uniformly at random between the two layers. In the past few years, there has been a great deal of interest in studying interdependent networks using a variety of assumptions about the structure of the intra-layer and inter-layer networks. It has been reported that abandoning the assumption of placing inter-layer edges uniformly at random can make interdependent networks more robust to random (i.e., uniformly at random) failures and that the percolation phase transition can be first-order instead of second-order. This is the case, for example, if the intra-layer degrees of interdependent node pairs are either exactly the same~\cite{Buldyrev2011Interdependent} or positively correlated with each other~\cite{Parshani2010Intersimilarity}, if numerous pairs of interdependent node pairs are adjacent to each other in both layers (so that there are many 4-cycles that consist of alternating inter-layer and intra-layer edges)~\cite{Parshani2010Intersimilarity}\footnote{If one interprets an interdependent network as a multiplex network, this feature guarantees that there is significant overlap between layers. See our discussion in Section~\ref{interlayer}.}, or if one includes a control parameter that creates some interdependent node pairs that are guaranteed to consist of nodes with high intra-layer degree (i.e., some pairs of interdependent nodes must include high-degree nodes from each layer)~\cite{Valdez2013Tricritical}. Reference~\cite{Zhou2012Assortativity} considered interdependent networks with degree correlations (i.e., degree assortativity) in intra-layer networks, Ref.~\cite{Watanabe2013Cavitybased} examined interdependent networks with both inter-layer and intra-layer degree correlations, and Refs.~\cite{Huang2013Robustness,shao2013} studied percolation cascades in interdependent networks in which intra-layer networks are produced using monoplex random-graph models with clustering (in particular, by using the degree-triangle model~\cite{Newman2009Random}). One can also study percolation cascades using any number of networks as well as with arbitrary types of coupling between component networks.  This yields a so-called \emph{network of networks}~\cite{Gao2011Robustness,Dong2013Robustness,Gao2013Percolation,Gao2012Robustness,Dong2013RobustnessB,Bianconi2014Mutually,Bianconi2014Multiple}. 

One of the motivations of the cascading-failure model in Ref.~\cite{Buldyrev2010Catastrophic} was to try to help explain failures on spatially-embedded infrastructure networks. Consequently, in subsequent studies of percolation on interdependent networks, there have been several attempts to take spatial constraints into account in the choices of the intra-layer networks.  For example, Li et al.~\cite{Li2012Cascading} studied square lattices with a tunable parameter to describe distances between nodes in an interdependent pair. They reported that there is a critical distance, such that the transition is second-order below this distance and first-order above it. (They measured the distance based on the lattice structure that is used to determine the intra-layer adjacencies; it depends only on the node coordinate in a node-layer tuple.) References~\cite{Bashan2013Extreme,Danziger2013Interdependent} extended the model by Li et al. by allowing nodes that are not interdependent.
For example, Bashan et al. \cite{Bashan2013Extreme} examined two-layer networks in which each layer is a two-dimensional spatial network (i.e., a network that is embedded in two dimensions).  They used equal-sized square lattices for their spatial networks and supposed for each layer that some fraction $q > 0$ of the nodes (i.e., the same fraction for each layer) are dependent on a node that is selected uniformly at random from the other layer. They reported that this situation yields a first-order phase transition for a giant component in the network.  This contrasts with the second-order transition that occurs when one considers analogous coupling when the layers are instead ER networks. Shekhtman et al. \cite{shekhtman2014} examined percolation on a network of interdependent spatially-embedded networks. Reference~\cite{Stippinger2013Enhancing} included the possibility for networks to ``heal'' (by adding new intra-network edges) after each cascade step, and Ref.~\cite{Kornbluth2013Cascading} studied two identical random regular graphs in which there is an upper limit to the intra-layer shortest path between a pair of interdependent nodes. Additionally, Berezin et al. \cite{berezin2013} examined a targeted-attack percolation problem (which they called ``localized attack'') in interdependent spatial networks. In their investigation, the initial node failures are localized geographically.

One can define a cascading-failure process in a multiplex network so that two nodes are considered to be in the same \emph{mutually-connected component} (i.e., they form a so-called \emph{viable cluster}) if there is an intra-layer path between them in all of the intra-layer networks~\cite{Son2011Percolation,Baxter2012Avalanche,Son2012Percolation}. This process is equivalent to a cascade process in an interdependent network in which nodes that are adjacent to each other via interdependency edges can be merged to create a single node in a multiplex network \cite{Baxter2012Avalanche,Son2012Percolation}. (This occurs, for example, when an interdependent network has two layers and each node has exactly one undirected inter-layer dependency edge.) One can also use a similar approach to map a cascading-failures process in an interdependent network in which only some fraction of nodes in each layer are dependent on nodes in the other layer to a cascading-failure process on a multiplex network~\cite{Son2012Percolation}. Bianconi et al.~\cite{Bianconi2014Mutually} reported that one can achieve a similar reduction to finding a mutually-connected component on a multiplex network for cascading failures on a networks of network that is diagonal and layer-coupled (see Section~\ref{general}) as long as, for every pair of layers, there exists a path of dependency edges between a node in one layer and a node in the other layer. Cellai et al.~\cite{Cellai2013Percolation} studied the emergence of a \emph{giant viable cluster} (i.e., a \emph{mutually-connected giant component}) for multiplex networks with overlap, and Hu et al. \cite{hu2013percolation} examined a similar phenomenon.  (See Section \ref{interlayer} for a discussion of overlap between layers in multiplex networks.)

Parshani et al. defined a similar cascade process for multiplex networks in which nodes can be adjacent to each other via connectivity edges and/or dependency edges~\cite{Parshani2011Critical}. In their node-percolation process, nodes are labeled as unoccupied if they are not in the GCC that is formed by occupied nodes of the connectivity layer (i.e., a layer whose edges are connectivity edges). One then labels nodes as unoccupied if they are adjacent to an unoccupied node in the dependency layer (i.e., a layer whose edges are dependency edges).  One then repeats these two processes sequentially until one reaches a stationary state. Parshani et al. placed a configuration-model network in the connectivity layer; in the dependency layer, they placed a network in which a fraction $q$ of the nodes have degree $1$ and the remainder of the nodes have degree $0$.\footnote{For their numerical simulations, Parshani et al.~\cite{Parshani2011Critical} relaxed the restriction that the intra-layer degree of a node in the dependency layer cannot exceed $1$.  References~\cite{Bashan2011Percolation,Bashan2011Combined} subsequently considered this more general situation analytically.} Prior to starting the above sequential percolation process, they initially removed a fraction $p$ of the nodes uniformly at random.  In their study, Parshani et al. derived a transcendental equation for the critical value $q_c$ of the parameter $q$ --- which, analogously to the phase diagram in Fig.~\ref{fig:cascade}f for a similar process, has a corresponding critical value $p_c$ of the parameter $p$ --- and they obtained a closed-form solution for $q_c$ and $p_c$ when they used an ER network in the connectivity layer. For the ER example, they showed that the phase transition for the GCC size as a function of $p$ is second-order for $q < q_c$ but first-order for $q > q_c$.  One can altermatively fix $p$ and use $q$ as a control parameter. Reference~\cite{Li2013Critical} examined the same cascade process on a network in which one can adjust the overlap between dependency edges and connectivity edges. They reported that increasing the overlap fraction can reduce the vulnerability of a network to the removal of nodes uniformly at random.

Very recently, Min and Goh~\cite{Min2014Multiple} examined a variant of the notion of a viable cluster using multiplex networks.  In their setting, some fraction of the nodes are ``source'' nodes. A node is ``viable'' if, in each layer, there exists an intra-layer path of viable nodes that connect that node to a source node.  The mean final fraction (over an ensemble of networks) of viable nodes is called the system's \emph{viability}, which depends both on the network's initial state and on the updating procedure. Note that one specifies the ``state'' of a network by specifying both its structure (i.e., its adjacencies) and the state of each of its nodes. One way to reach an stationary state of the network is via a ``cascade of activations'': one first considers only source nodes as viable and then recursively sets unviable nodes to be viable if they meet the viability condition. One can reach an stationary state with a different value of the viability via a ``cascade of deactivations'': one first considers all nodes as viable and then recursively sets nodes that do not satisfy the viability condition as unviable. The system exhibits hysteresis, which one can observe by plotting the viability $V$ versus a network diagnostic (such as mean intra-layer degree), so one might need to add more edges to the system than were previously removed to be able to recover from the random (e.g., uniformly at random) failure of edges.  In the limit as the number of source nodes becomes $0$, the process studied in Ref.~\cite{Min2014Multiple} reduces to the examination of mutually-connected giant components (i.e., the standard notion of viable clusters). Baxter et al.~\cite{baxter2013} independently studied processes that are akin to those in \cite{Min2014Multiple}. In fact, the cascade of activations in that we described above amounts to the ``weak bootstrap percolation'' that was examined in Ref.~\cite{baxter2013} in the context of multiplex networks. In weak bootstrap percolation, some fraction of nodes is initially set to be invulnerable, all invulnerable nodes are considered to be active, and rest of the nodes are inactive. One then recursively sets inactive nodes to be active if they are adjacent to at least one active node in each layer. The ``weak pruning percolation'' of Ref.~\cite{baxter2013} is similar to the cascade of deactivations, but it is not exactly the same.

\subsubsection{Compartmental Spreading Models and Diffusion.} \label{spreading}

\begin{figure}[!htp]
\begin{center}
\includegraphics[width=1\linewidth]{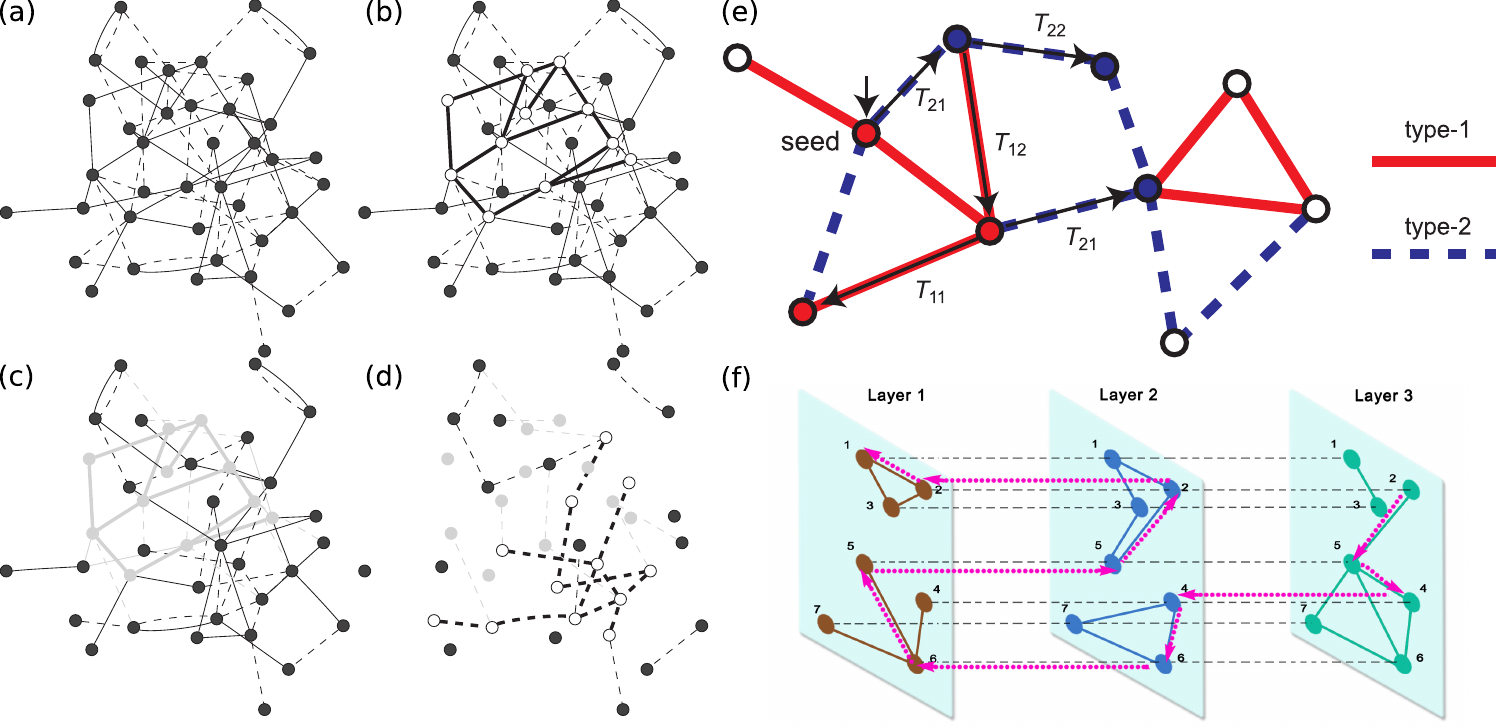}
\end{center}
\caption{(a--d) Illustration of a pair of interacting SIR processes on a two-layer network from Ref.~\protect\cite{Funk2010Interacting}.  (a) One epidemic occurs in each layer; the solid curves are the intra-layer edges for one layer, and the dashed curves are the intra-layer edges for the other layer. (b) Thick solid curves represent ``occupied'' edges in a single-layer ``giant connected component'' (GCC)  \protect\cite{Newmanbook} that represents a large outbreak in one layer (which we call the ``first'' layer). (c) The nodes in that component have recovered; one can either assume that they are now immune and remove them from the other (``second'') layer or suppose that they are only partially immune and posit that they have a reduced probability of being infected by the second epidemic. (d) The thick dashed curves represent edges in a single-layer GCC in the second layer.  [In panels (a)--(d), we have reprinted a figure from Ref.~\protect\cite{Funk2010Interacting} with permission from the authors and the publisher. Copyright (2010) by the American Physical Society.] (e) Schematic of a two-layer network from Ref.~\protect\cite{Min2013Layercrossing} that illustrates SIR dynamics on two-layer networks for which there is an overhead to cross layers.  The transmissibilities $T_{ij}$ are determined both by the type of the incoming infection channel (which is represented by node color) and the type of the outgoing infection channel (i.e., the type of edge).  White disks are susceptible nodes, red disks are recovered nodes that were infected via a type-1 edge, and blue disks are recovered nodes that were infected via a type-2 edge.  [The illustration in panel (e) was created by Byungjoon Min and appears in Ref.~\protect\cite{Min2013Layercrossing}. We are using it with permission from the authors of that paper.] (f) Illustration of a random walk on a multiplex network. The pink dashed arrows indicate the path of a random walker, who uses inter-layer edges to reach nodes that are in different components of the intra-layer networks. [The illustration in panel (f) was adapted by M. De Domenico from a figure in Ref.~\protect\cite{DeDomenico2013Random}. We are using it with permission from the authors of that paper.]
}
\label{fig:spreading}
\end{figure}

It is traditional to study spreading processes using compartmental models on completely mixed populations~\cite{ccc,AndersonMay}.  In such a model, each compartment describe a state (e.g., ``susceptible'', ``infected'', or ``recovered''), and there are parameters that represent transition rates for changing states. However, disease-spreading and information-spreading processes typically take place on networks, so it is important to study the effects of network structure on such processes \cite{Newmanbook,Barrat08}. One places a compartmental model on a network, so each node can be in one of several epidemics states (e.g., ``susceptible'' or ``infected''), and the nodes have either continuous or discrete update rules that govern how the states change. 

The susceptible-infected-recovered (SIR) model\footnote{Alternatively, the ``R'' can stand for ``removed''.} is one of the simplest and most popular compartmental models, and its steady state can be related to a bond-percolation process~\cite{Mollison77,Grassberger82}.  (Note, however, that one needs to be careful about the precise connection between SIR dynamics and percolation \cite{lfd2013}.) For many large-scale spreading processes, one important step towards making models more realistic is to study them on a \emph{metapopulation} structure instead of on a single network. In such a structure, each node is a completely mixed population that is adjacent to other populations via the edges in a network. Epidemic processes on networks of networks (i.e., interconnected networks, node-colored networks, etc.) are natural extensions of completely mixed metapopulation models~\cite{Dickison2012Epidemics,Hindes2013Epidemic}. Several authors have studied an SIR model on two-layer networks. For example, focal ideas have included the correlations of the intra-layer degrees of nodes that are adjacent across layers~\cite{Wang2011Effects}, the strength of coupling (i.e., the number of inter-layer edges) between the two networks~\cite{Dickison2012Epidemics}, and the identification of influential spreaders by computing $k$-shells~\cite{Zhao2014Identifying}.

SIR models have also been studied on multiplex networks. For example,  Min et al. \cite{Min2013Layercrossing} studied SIR spreading on various types of two-layer multiplex networks~\cite{Lee2012Correlated,Min2013Network} with a layer-crossing \emph{overhead}. Such an overhead is a cost for an infection (or information) to change layers (see Fig.~\ref{fig:spreading}e), and this idea is very important for multilayer networks in general \cite{Cozzo2013Clustering}.  References \cite{Lee2012Correlated,Min2013Network} illustrated that there are spreading processes on multiplex networks that cannot be reduced to such processes on aggregated networks. Shai et al.~\cite{Shai2012Effect} demonstrated that an SIR model on positively degree-correlated multiplex networks with two layers can result in a lower epidemic threshold than on negatively degree-correlated networks --- a similar result was later reported by Ref.~\cite{Zhao2013Multiple} --- but that the opposite can be true for a ``constrained'' SIR model in which each node is only allowed to interact with a random subset of its neighbors.  (For each time step, the nodes in this subset are determined uniformly at random, and the subsets at different time steps are independent of each other.)  They consider networks in which each layer is an ER graph as well as ones in which each network is a BA network. SIR dynamics have also been studied on multiplex networks that are not node-aligned.  For example, Ya\u{g}an et al.~\cite{Yagan2012Conjoining} considered a network that contains one ``physical'' layer that includes all nodes and other layers that contain only some subset of these nodes.  Buono et al.~\cite{Buono2013Epidemics} examined SIR dynamics on a network that contains two layers with an equal number of nodes, but only some fraction of the nodes exist in both layers.\footnote{Note that these articles do not use the term ``node-aligned''. Ya\u{g}an et al. called their networks ``overlay networks'', and Buono et al. used the monicker ``partially overlapped multiplex networks''.} 
 
Jo et al.~\cite{Jo2006Immunization} defined a variant of the standard SIR model on a two-layer network in which normal infection spreading occurs in one layer and information spreading occurs in the other layer.  They allowed information spreading to immunize nodes, which could then proceed directly from the ``S'' compartment to the ``R'' compartment. Funk et al. \cite{Funk2009,Funk2010jtb} also investigated an SIR model that was coupled to information-spreading dynamics.  (Additionally, see Ref.~\cite{funk-review2010} for a review article on modelling the influence of human behavior on the spread of diseases.) In their model, better-informed nodes have a reduced susceptibility. Funk and Jansen~\cite{Funk2010Interacting} studied the interplay of two pathogens that spread according to an SIR processes on two-layer networks that can have arbitrary inter-layer degree correlations and overlaps. In their model, nodes that have experienced the first epidemic (which occurs on the ``first'' layer) have a reduced susceptibility to the second pathogen (which occurs on the ``second'' layer). We illustrate the mechanism that they studied in panels (a)--(d) of Fig.~\ref{fig:spreading}.  They illustrated that epidemic sizes are smaller in the second layer when there are positive inter-layer degree correlations than is the case if there are no such correlations or negative correlations. They also considered scenarios in which either epidemic can render the nodes immune to the other epidemic, but they always assumed that the epidemics are not taking place concurrently. Marceau~et al.~\cite{Marceau2011Modeling} built on this work and introduced an SIR model in which the two epidemics unfold simultaneously and can interact dynamically instead of only ``sequentially''. 

The susceptible-infected-susceptible (SIS) spreading model has been studied both on interconnected networks with a given joint degree distribution between layers~\cite{SaumellMendiola2012Epidemic} and for more general situations (e.g., in which the node sets are disjoint, but all inter-layer and intra-layer adjacencies are allowed)~\cite{Sahneh2012Effect} by generalizing the result for monoplex networks \cite{wang2003radius} --- though one needs to be very careful about how such ``results'' have been stated in the literature \cite{gleeson2011} --- that the epidemic threshold is given approximately by the inverse of the spectral radius of the contact network's adjacency matrix. Reference~\cite{Wang2013Effect} also started from the expression that relates an adjacency matrix's spectral radius to the epidemic threshold of an SIS spreading model. However, they took a perturbative approach: they assumed that the spectral radii of the intra-layer and inter-layer supra-adjacency matrices are both known, and they studied the effect of network structure on the spreading process by deriving perturbation approximations for the spectral radius of the supra-adjacency matrix for the entire interconnected network (which we recall is equal to the sum of the intra-layer and inter-layer supra-adjacency matrices). An SIS spreading process was studied using a contact-contagion formulation by Cozzo et al.~\cite{Cozzo2013Contactbased}, who calculated an epidemic threshold that corresponds to the inverse of the spectral radius of the supra-matrix for the contact probabilities. All of the above studies of SIS models noted a decrease in the epidemic threshold upon the introduction of inter-layer adjacencies.  

Several other authors have also examined SIS dynamics and generalized versions of such dynamics. For example, some studies have considered SIS models in which two competing infections spread on two layers of a multiplex network~\cite{Wei2012Competing,Sahneh2013May}.  Sanz et al. \cite{sanz2014} studied interacting epidemics using an extension of the SIS model in which the transitions from infection to recovery and from recovery to infection in one disease depend on the state of each node with respect to the other disease. Additionally, Ref.~\cite{cuenda2013} used SIS dynamics to study malware spreading on a multiplex network in which a node can only be in a single state across all layers (i.e., the inter-layer spreading is infinitely fast) but the spreading process occurs independently (with different contagion rates) on each of the intra-layer networks.

Reference~\cite{clara2013} studied interacting SIS dynamics on multiplex networks in which one layer has literal SIS dynamics on real social contacts and the other layer has SIS-like awareness dynamics (of information about the disease) on virtual social contacts.  Similarly, Sahneh et al.~\cite{Sahneh2012Optimal} studied an SAIS model, which includes an additional ``alert'' state, on a multiplex network in which one layer is responsible for the spreading of an infection and the other is responsible for information dissemination. They also developed a mean-field model that can handle an arbitrary number of node states and layers~\cite{Sahneh2013Generalized}. Lima et al. \cite{manlio2013} considered various types of disease-spreading process that are coupled to information-spreading processes and incorporated human-mobility and mobile-phone data from a large set of individuals in the Ivory Coast. Shai et al.~\cite{Shai2013Coupled} studied a process in which SIS dynamics are coupled to a process that rewires intra-layer edges between susceptible and infected nodes on an interconnected (i.e., node-colored) network.

G\'omez et al.~\cite{Gomez2013Diffusion} examined diffusion in node-aligned multiplex networks. One can examine such a diffusion process analytically by calculating the eigenvalues of a combinatorial supra-Laplacian matrix. G\'omez et al. derived the asymptotic behavior of the eigenvalues of the combinatorial supra-Laplacian of a two-layer network for both strong coupling and weak coupling between layers.  (Two layers are ``strongly coupled'' if the weights of the inter-layer edges are large, and they are ``weakly coupled'' if those weights are small.) De Domenico et al.~\cite{DeDomenico2013Random} introduced several types of random walks on multiplex networks with heterogeneous coupling strengths between nodes.  In Fig.~\ref{fig:spreading}f, we illustrate an example of a random walk on a multiplex network. They demonstrated that the time that is needed for a random walker to reach most of the nodes in a multiplex network depends on the topology of the intra-layer networks, the inter-layer connection strengths, and the type of the random walk. This time can either lie between the times that are necessary to cover the each of the layers separately (i.e., the walk is ``intra-diffusive'') or it can be smaller than each of those times (i.e., the walk is ``super-diffusive'').

The results in Ref.~\cite{Radicchi2013Abrupt} on the algebraic connectivity of the combinatorial supra-Laplacian also have important implications for diffusion on multiplex networks, as how closely the separate layers interact can be very important for diffusive processes (as well as for other dynamics). Reference~\cite{Siudem2013Diffusion} studied diffusion in ``weakly-coupled`` interconnected networks in which there are only a few inter-layer edges. They were able to separate the diffusion into a fast process that takes place inside of the layers and a slow process that takes place between layers.

The susceptible-infected (SI) model is a percolation process in which infected nodes stay that way forever \cite{Newmanbook}. Reference~\cite{Javarone2013Competitive} studied a variant of the SI model in which there are several different infections (i.e., ``lexical innovations'') that compete against each other in a node-colored network in which one layer represents the media and the other represents a social network. Another type of spreading model, which can be used for applications such as information spreading and behavior adoption, is a so-called ``complex contagion'' \cite{centola2007}. Such processes have also been studied on multiplex networks (see Section \ref{otherdynamics}).

\subsubsection{Coupled-Cell Networks}\label{coupledcell}

It is possible to classify the temporal evolution of coupled systems of ordinary differential equations by using \emph{coupled-cell networks} \cite{ashwin1999,Golubitsky2005Patterns,Stewart2003Symmetry} (see Section \ref{othernets}), which can exhibit rich (and sometimes rather surprising) bifurcations and have received a lot of attention in the dynamical-systems community. In coupled-cell networks, a node is associated with a dynamical system, and two nodes have the same ``color'' if they have the same state space and an identical dynamical system. The edges in these networks represent the couplings between the dynamical systems, and two edges have the same ``color'' if the couplings are equivalent.  If two nodes have the same color and the edges that are incident to those nodes all have the same color, then the differential equations at each node are identical. Additionally, for a given network, the formalism of coupled-cell networks automatically gives a set of ``admissible'' vector fields.  (That is, which vector fields are sensible to consider is a natural byproduct of the structure that has been defined.) A coupled-cell network is \emph{homogeneous} if all of the nodes have the same color and also have the same number of incoming edges of each color (i.e., it is a special case of a multiplex network), and it is \emph{regular} if all of the edges have the same color (i.e., it is a monoplex network). 

Most studies of coupled-cell networks have been concerned with systems that have a small number of nodes. A major thread has entailed the derivation of generic results that relate the qualitative behavior of a dynamical system to the topology of the corresponding coupled-cell network~\cite{Golubitsky2006Nonlinear}.  For a monoplex network, synchrony-breaking bifurcations in coupled-cell networks are related to the eigenspace of the network's adjacency matrix. Similarly, the structure of synchrony-breaking bifurcations in a homogeneous coupled-cell network constructed from a Cartesian product of two graphs is related to a tensor product of the eigenspaces of the adjacency matrices of the two original graphs~\cite{Golubitsky2009Bifurcations}.

The careful mathematical study of coupled-cell networks has also yielded many interesting (and mathematically rigorous) results in addition to the aforementioned bifurcation phenomena. Indeed, in order to conduct a generic bifurcation analysis in a coupled-cell network, it is first necessary to classify its ``patterns of synchrony''\footnote{Consider an $n$-node coupled-cell network, and suppose that ${\bf x}^* = (x_1^*,\ldots,x_n^*)$ is a hyperbolic equilibrium point of an admissible vector field.  A ``pattern of synchrony'' is a coloring of the nodes such that two nodes $i$ and $j$ have the same color if and only if $x_i^* = x_j^*$.  See Ref.~\cite{gucken} to recall concepts like ``hyperbolicity'' from the theory of dynamical systems.} \cite{Stewart2003Symmetry,Golubitsky2005Patterns}. Perturbing the system leads to a new system with the same pattern of synchrony if and only if the coloring satisfies a certain combinatorial condition (which does not depend explicitly on the vector field under consideration) \cite{Golubitsky2005Patterns}.  More generally, the formalism of coupled-networks has led --- often after many years of effort \cite{golubitsky2012} --- to results such as a mathematically rigorous classification of ``phase-shift synchrony''\footnote{Given a hyperbolic period-$T_0$ solution ${\bf x}(t) = (x_1(t),\ldots,x_n(t))$ of an admissible vector field on a coupled-cell network, a pair of nodes $i$ and $j$ exhibit ``phase-shift synchrony'' if $x_j(t) = x_i(t + \vartheta_{ij} T_0)$ for some phase-shift constant $\vartheta_{ij} \in S^1 = [0,1)$.} of periodic solutions in (admissible) dynamical systems (i.e., without needing to worry about the specific form of a vector field) in terms of combinatorial conditions on network structure.

As we have illustrated above, the study of coupled-cell networks has been very useful for illuminating rich bifurcation phenomena, and additional studies of such phenomena --- as well as spiritually similar studies of rich, generic phase-transition phenomena that can occur in multilayer networks \cite{Radicchi2013Abrupt,Radicchi2013Driving} --- are important research directions.

\subsubsection{Other Types of Dynamical Systems.}\label{otherdynamics}

Obviously, there are numerous types of dynamics that can occur on multilayer networks, and particularly popular processes like percolation cascades and compartmental spreading models are not the only ones that have received attention. Multilayer structures have important consequences for dynamical processes in general, and such effects (and the magnitude of their importance) can differ for different processes.

The effects of multilayer organization have been studied for several additional types of dynamical processes.  For example, Ref.~\cite{Scotti2013Social} examined ecological dynamics on a multilayer network that includes both social and landscape effects. Additionally, in evolutionary game theory, the coupling of multiple networks can increase cooperation in games such as the Prisoner's Dilemma~\cite{GomezGardenes2012Evolutionary,GomezGardenes2012Evolution,Santos2014Biased}.  For example, it has been reported that one can couple two networks in which the nodes play Prisoner's Dilemma games against their neighbors such that it is optimal to have a nontrivial number (i.e., neither $0$ nor the maximum number) of inter-layer edges to facilitate cooperative behavior~\cite{Jiang2013Spreading,Wang2013Optimal}. It has also been illustrated that the coupling of network layers can increase cooperation in other game-theoretic models (e.g., in a public-goods game in a coupled pair of square lattices~\cite{Wang2012Evolution,Wang2013Interdependent,Szolnoki2013Information}).

The Watts threshold model~\cite{Watts2002Simple}, which is a complex-contagion process for the adoption of ideas, has also been generalized for multiplex networks~\cite{Yagan2012Analysis,Brummitt2012Multiplexityfacilitated}. For example, there exist means for multiplex networks to be more vulnerable to global adoption cascades than monoplex networks~\cite{Brummitt2012Multiplexityfacilitated}. Multiplex networks can also have other important effects on dynamical processes.  For example, it has been reported that multiplex random Boolean networks can be stable\footnote{``Stability'' was measured by initializing the system from two nearby states and measuring the Hamming distance between them after a long time.} for parameter values for which a single layer of the multiplex network in isolation would be unstable~\cite{Cozzo2012Stability}. Kuramoto phase oscillators have also been studied on two-layer networks, for which the coupling via inter-layer edges includes a delay~\cite{Louzada2013Breathing}.  In fact, a perspective based on a network of networks --- where oscillators in a population are coupled to each other via some network structure and then these populations are in turn coupled to each other in a network --- was used several years ago (before networks of networks became popular), with an eye towards applications in neuroscience, to examine synchronization of coupled nonlinear oscillators \cite{kurths2006,kurths2007,barreto2008pre,so2008chaos}.

Brummitt et al.~\cite{Brummitt2010Sandpile,Brummitt2012Suppressing} studied a sandpile model on interconnected (i.e., node-colored) networks in order to model cascading failures on a pair of coupled power grids. In their study, they reported that there is an optimal fraction of interconnected node pairs between the two networks that minimizes the largest cascades. Reference~\cite{Tan2014Traffic} studied routing and traffic congestion in interconnected networks.  That is, they considered a process in which nodes send ``packets'' to other nodes that are routed through the network such that each node can only forward a number of packets, corresponding to its capacity, to its (intra-layer or inter-layer) neighbor nodes at each time step. Reference~\cite{Tan2013Cascading} examined a routing model in which the load on a node is equal to its value of geodesic betweenness centrality (where the paths through the network can traverse either inter-layer and intra-layer edges). At each time step, nodes whose betweenness values exceed their capacity fail; they are then removed from the network, and the betweenness values are recalculated. Zhang et al.~\cite{Zhang2013Robustness} examined a similar dynamical process on two-layer interdependent networks.  However, in their study, paths can only traverse intra-layer edges, and a node's failure in one layer can cause a node that depends on that node in the other layer to fail in a matter that is reminiscent of the cascading failures in Ref.~\cite{Buldyrev2010Catastrophic} (see Section~\ref{cascades}). Morris and Barthelemy~\cite{Morris2012Transport} also examined a similar phenomenon (but without cascading failures) in the context of two-layer transportation systems. In subsequent work~\cite{Morris2013Interdependent}, they studied cascading failures in two-layer electrical networks in which one layer represents a power grid and the other layer represents a control network that monitors the edges in the power lines.

\subsubsection{Control and Dynamics.}\label{control}

The study of dynamical systems has intricate connections with control theory, in which one examines dynamical systems with feedback and inputs.  A control-theoretic perspective can be very insightful for the investigation of dynamical systems on networks \cite{mario-review2010,motter-expo,cowan2012,hill2009}, and it will certainly also be important for dynamical systems on multilayer networks.  For example, one might desire to achieve a desired state --- such as the synchronization (or some other desired behavior) of oscillators that are associated with each node --- so how should one apply low-cost decentralized control strategies to ensure that this occurs as as rapidly as possible?

Indeed, many of the concepts that we have discussed above can be viewed through the lense of control theory. For example, a control-theoretic perspective has been used to examine the temporal evolution of two-layer networks in which a ``control network'' is used to influence an ``open-loop network'' (which, by definition, does not include feedback by itself) \cite{mario-review2010}.  It is also useful to view ``pinning control'' \cite{porfiri2008}, in which one controls only a small fraction of network nodes directly in order to try to influence the dynamics of other nodes, in the context of interdependent networks.  For instance, Wu et al. \cite{wu-chaos2012} recently examined pinning control and synchronization in networks that are both node-colored and edge-colored. A population of controllers interacts with a population of followers, and it is important to distinguish between follower-follower edges and controller-follower edges. 

As Ref.~\cite{prescott2013} illustrated in the context of biochemical systems, one can also use ideas from control theory to decompose networks into different layers \cite{Chiang2007Layering}.  This is somewhat reminiscent of community detection, but the layers need not be based on trying to find densely connected sets of nodes.

\section{Conclusions and Outlook}\label{conc}

The study of multilayer networks --- and of frameworks like multiplex networks and interconnected networks in particular --- has become extremely popular, and it's easy to see why. Most real and engineered systems include multiple subsystems and layers of connectivity, and developing a deep understanding of multilayer systems necessitates generalizing ``traditional'' network theory.  Ignoring such information can yield misleading results, so new tools need to be developed. One can have a lot of fun studying ``bigger and better'' versions of the diagnostics, models, and dynamical processes that we know and (presumably) love --- and it is very important to do so --- but the new ``degrees of freedom'' in multilayer systems also yield new phenomena that cannot occur in single-layer systems.  Moreover, the increasing availability of empirical data for fundamentally multilayer systems amidst the current data deluge also makes it possible to develop and validate increasingly general frameworks for the study of networks.  

In the present article, we discussed the history of research on multilayer networks and related frameworks, and we reviewed the exploding body of recent work in this area.  Numerous similar ideas have been developed in parallel, and the literature on multilayer networks has rapidly become extremely messy.  Despite a wealth of antecedent ideas in subjects like sociology and engineering, many aspects of the theory of multilayer networks remain immature, and the rapid onslaught of papers (especially from the physics community) on various types of multilayer networks necessitates an attempt to unify the various disparate threads and to discern their similarities and differences in as precise a manner as possible.  Towards this end, we presented a general framework for studying multilayer networks and constructed a dictionary of terminology to relate the numerous disparate existing notions to each other. We then provided a thorough discussion to compare, contrast, and translate between related notions such as multilayer networks, multiplex networks, interdependent networks, networks of networks, and many others.  We also discussed existing data sets (including some old ones) that can be represented as multilayer networks and can thus be used to test new ideas in this area.

We reviewed attempts to generalize single-layer network diagnostics, methods, models, and dynamical systems to multilayer settings.  An important theme that has developed in the literature is the importance of multiplexity-induced correlations (and its analogs in multilayer structures more generally), which can have important ramifications for dynamical processes on networks.  For example, such correlations can have significant effects on the speed of transmission of diseases and ideas as well as on the robustness of systems to failure.  Multilayer structures induce new degrees of freedom --- such ``new physics'' is already prominent even for multilayer networks with only a single aspect --- but they remain poorly understood.  We already know that these new degrees of freedoms have important consequences for dynamical processes, and it is crucial to develop a precise understanding of the situations and processes for which such effects (and the magnitude of their importance) are qualitatively different.\footnote{Naturally, empirical data is very helpful --- indeed, it is crucial --- for the study of multilayer networks, though an important limitation of most existing data sets is that it is much easier to reliably estimate the values of intra-layer edge weights than those of inter-layer edge weights. At the moment, transportation networks seem to be the best candidates for multiplex networks in which it is currently possible to reliably estimate the weights of inter-layer edges.  It should usually be much easier to estimate the weights of inter-layer edges for node-colored networks.}  However, just as the trickle of monoplex-network data sets eventually became a flood, we expect to see many more interesting multilayer data sets in the near future. These, in turn, will help scholars to develop new theories, methods, and diagnostics for gaining insight into multilayer networks.

The study of multilayer networks is very exciting, and we look forward to what the next several years will bring.

\section*{Acknowledgements}

All authors were supported by the European Commission FET-Proactive project PLEXMATH (Grant No. 317614). AA also acknowledges financial support from the ICREA Academia, Generalitat de Catalunya (2009-SGR-838), the James S.\ McDonnell Foundation, and FIS2012-38266. JPG acknowledges funding from Science Foundation Ireland (grants 11/PI/1026 and 09/SRC/E1780). YM was also supported by MINECO through Grants FIS2011-25167 and by DGA (Spain). MAP acknowledges a grant (EP/J001759/1) from the EPSRC.  We thank Sergey Buldyrev, Moses Boudourides, Lidia A. Braunstein, Ron Breiger, Javier Buld\'{u}, Kathleen Carley, Davide Cellai, Emanuele Cozzo, Valentin Danchev, Manlio De Domenico, Mario di Bernardo, Peter Erd\"{o}s, Terrill Frantz, Marty Golubitsky, Peter Grindrod, Adam Hackett, Shlomo Havlin, Des Higham, Vincent Jansen, Hang-Hyun Jo, J\'{a}nos Kert\'{e}sz, Dan Larremore, Ian McCulloh, Jim Moody, Peter Mucha, Callum Oakley, Tom Prescott, Ioannis Psorakis, Garry Robins, Joaqu\'{i}n Sanz, Tom Snijders, Gene Stanley, and Alvin Wolfe for helpful comments and suggestions.  We also thank the members of the COSNET Lab and participants in University of Oxford's Networks Journal Club for their feedback and the referee (and/his her team of graduate students and postdoctoral scholars) for helpful comments.

Copyright notice for panel (f) of Fig.~\ref{fig:cascade} and panels (a)--(d) of Fig.~\ref{fig:spreading}: 
``Readers may view, browse, and/or download material for temporary copying purposes only, provided these uses are for noncommercial personal purposes. Except as provided by law, this material may not be further reproduced, distributed, transmitted, modified, adapted, performed, displayed, published, or sold in whole or part, without prior written permission from the American Physical Society.''

\section{Appendix: Glossary and Notation} \label{glossary}

\begin{center}
\begin{tabular}{l|l}
Term/symbol            & Explanation\\
\hline
Multilayer network     & General term for a network with multiple layers (see Section~\ref{general})\\
Multiplex network      & Multilayer network with diagonal couplings (see Section~\ref{multiplex}) \\
Monoplex network       & Network with a single layer; aka: ``singleplex'', ``uniplex'', ``singlex'', ``simplex'' network\\ 
Node-layer tuple       & Tuple that specifies both node and layer; it is a node in a supra-graph (see Section~\ref{general}) \\
Aspect                 & A ``dimension'' of layers (see Section~\ref{general})\\
Supra-adjacency matrix & Matrix representation of a multilayer network (see Section~\ref{supra})\\
Adjacency tensor       & Tensor representation of a multilayer network (see Section~\ref{reps})\\
Node-aligned           & All nodes are shared between all layers (see Section~\ref{general}) \\
Layer-disjoint         & Each node is present only in a single layer (see Section~\ref{general})\\
Diagonal couplings     & Inter-layer edges are only between nodes and their counterparts (see Section~\ref{general})\\
Layer-coupled          & Diagonal couplings that are independent of the nodes (see Section~\ref{general})\\
Categorical couplings  & Diagonal couplings for which all possible inter-layer edges are present (see Section~\ref{general})\\
Ordinal couplings      & Diagonal couplings with inter-layer edges only between nodes in neighboring layers \\ & (see Section~\ref{temporal})\\ \hline
$\aspects$             & Number of aspects (i.e., the ``dimensionality'' of the layers) \\
$\nnodes$              & Number of nodes in one layer in a node-aligned network \\
$\nlayers$             & Number of layers in a network with a single aspect\\
$\layervector$         & Sequence of sets of layers\\
$\layers$              & Set of layers in a network with a single aspect\\
${\bf A}$              & Adjacency matrix\\
${\bf W}$              & Weighted adjacency matrix\\
$\mathcal{A}$          & Adjacency tensor\\
$\mathcal{W}$          & Weighted adjacency tensor\\
$\nodeindex[1],\nodeindex[2]$                  & Nodes (and node indices)\\
$\layerindex[1],\layerindex[2]$& Layers (and layer indices)\\
$\layerindexvector[1],\layerindexvector[2]$      & Sequences of layer indices (e.g., $\layerindexvector[1]=\layerindex[1]_1, \dots, \layerindex[1]_{\aspects}$ or $\layerindexvector[1]=\layerindex[1]_1 \dots \layerindex[1]_{\aspects}$) \\
\end{tabular}
\end{center}

\bibliography{multislicerev2}

\end{document}